\documentclass[twocolumn,showpacs,preprintnumbers,amsmath,amssymb]{revtex4}

\usepackage{graphicx}
\usepackage{dcolumn}
\usepackage{bm}

\newenvironment{Aeqno}%
{}{}

\begin{document}

\title{Three-component multiplicity distribution,
oscillation of \\combinants and properties of clans
in $pp$ collisions at the LHC}

\author{I. Zborovsk\'{y}}
 \email{zborovsky@ujf.cas.cz}
\affiliation{%
Nuclear Physics Institute of the CAS, 250 68 \v{R}e\v{z}, Czech Republic
}%

\begin{abstract}
Multiplicity distributions of charged particles measured by the ATLAS Collaboration 
in proton-proton collisions 
at $\sqrt{s}=8$ and 13~TeV 
are analyzed in the framework of a weighted superposition of
three negative binomial distribution functions.
The examination of the experimental data confirms the existence of a narrow peak at low multiplicities 
found earlier at $\sqrt{s}=0.9$ and 7~TeV. 
The peak is described by the third separate component of the total distribution.
The energy dependence of the multiplicity characteristics of the three-component scenario is studied.  
It is demonstrated that the third multiplicity component  
might be responsible for oscillations of combinants
observed lately in the analysis of the multiplicity distributions measured 
in proton-proton collisions at the LHC. 
Some consequences of the three-component description of the data for the clan parameters are analyzed. 
\end{abstract}

\maketitle

\subsection{Introduction}
\label{sec:A}

The multiplicity of produced particles is an important characteristic of the final state 
in high energy proton-proton collisions.
The shapes of the charged-particle multiplicity distributions (MDs) measured with high accuracy 
can provide valuable information on various processes in these interactions.
The distributions reflect correlations in the system encoded in an integrated form.
At lower collision energies, the two-particle correlations are dominant and a single negative 
binomial distribution (NBD) \cite{Giov1} gives a satisfactory description of the MD up to the ISR energies.

At higher collision energies, a shoulder structure of the MD at high multiplicities \cite{UA5} 
and oscillation 
of the ratio $H_q=K_q/F_q$ of the cumulant to factorial moments \cite{Dremin0} have been found.  
No pattern of this kind appears in a phenomenological fit by a single NBD \cite{Dremin1}.
It has been proposed \cite{GiovUgo1} to describe the observed shoulder structure as a weighted 
superposition of soft events (without mini-jets) and semi-hard events (with mini-jets), 
each of the NBD type. The suggestion was emphasised by the interesting and remarkable observation
that the oscillations of $H_q$ vs. rank $q$ obtained from data is reproduced
by the $K_q$ over $F_q$ ratio calculated from the superposition of two NBDs.
The appearance of the second multiplicity component under the tail of the MD is
connected with cumulants each of 
which involves an infinite ``cumulative" sum over ${\it all}$ multiplicity probabilities~$P(n)$. 

The extrapolation of the weighted superposition of the two defined classes 
of events from the TeV to the multi-TeV energy domain
and a study of the properties of particle clans raised intriguing questions concerning the onset 
of a third class of events in the tail of the distribution, both
in the full phase space \cite{GiovUgo2} and in the limited  windows in pseudorapidity \cite{GiovUgo3}. 
Later study \cite{IZ} of the MDs measured 
at $\sqrt{s}= 7$~TeV indicated  
existence of the third multiplicity component at low $n$.
In this region, near the maximum of $P(n)$,
there is a suitable tool, the method of combinants \cite{Gyulassy}, to be used  
in searching for new phenomena in the behavior of the MDs.
The combinants ${\cal C}(i)$ have similar additive properties as cumulants 
which are expressible as an infinite sum over all probabilities.
The latter are therefore convenient for an analysis of the total shape of the MDs. 
On the other hand, the combinants can be expressed as a finite combination of the ratios $P(n)/P(0)$. 
The combinants are thus extremely suitable
for study of the final set of the multiplicity frequencies which make up the 
experimental data at low multiplicities.

In this paper we study in terms of a three-component description the data on MD of charged particles produced  
in $pp$ collisions at the LHC at new high energies, $\sqrt{s}=8$ and 13~TeV. 
Each component is represented by a single NBD and parametrized by two parameters, $\bar{n}$ and $k$. 
The total distribution is weighted sum of three NBDs.
The paper extends an earlier analysis \cite{IZ} performed with data on MD measured  
at $\sqrt{s}=0.9$ and 7~TeV.  
The study of the ATLAS measurements \cite{ATLAS1,ATLAS2,ATLAS3,ATLAS4}
show that the two-component description of the MD is unsatisfactory. Within the
multi-component NBD parametrization, the structure of the data indicates the necessity
of a third component in the region of low multiplicities.
We demonstrate that, unlike the two-component fits to the MDs measured at the LHC, 
the phenomenological fits with the third NBD component at low $n$ 
can give oscillations of the combinants multiplied by their rank discovered lately \cite{WW1} 
in multiplicity data measured by the CMS \cite{CMS} and ALICE \cite{ALICE8} Collaborations. 
Although the sensitivity of the oscillations to the systematic uncertainties 
of the measurements is large, the natural applicability and power of the method of combinants 
in the analysis of the fine structure of the MDs observed in $pp$ collisions at the LHC
is of interest in the search for and study of new phenomena emerging in the global characteristics
of particle production.  

The second part of the paper is dedicated to a study of the obtained results within
the clan structure analysis of data on MD. 
The qualifying assumption is that each of the components of the weighted superposition 
used to parametrize the experimental data has the NBD form.
The emergence of the peak in the MD at low $n$, described by the third NBD, has a serious impact    
on the clan parameters of the first and the second component
of the total distribution.
The parameters changed dramatically in comparison with the two-NBD analyses 
considered usually in different scenarios of soft and semi-hard events.
The most appealing observation is the increase of the average number of clans 
in the high-$n$ (semi-hard) component with the center-of-mass 
energy $ \sqrt{s} $. This is in sharp contrast with two-NBD parametrizations of the multiplicity data. 
The contribution of the third NBD to the low multiplicity part of the charged particle distribution 
has a serious consequence 
for the average number of particles per clan in the first and the second component. 
It turns out that the dominant component with largest probability is characterized with few clans 
containing many particles. We illustrate that the second, high-$n$ component consists 
of large number of clans containing less particles.  
The properties of clans of the three-NBD parametrization of the MDs measured at the LHC and some  
corresponding mechanisms of particle production are discussed.

\subsection{Weighted superposition of NBDs}
\label{sec:B}

The data on MD provide information on various processes underlying the production 
mechanism contain information as regards correlations in the system and can serve as a test for
probing the dynamics of the interaction.
The particle production  
is based on quantum chromodynamics (QCD). 
The multi-particle processes responsible for the final multiplicity  
involve also soft scales, 
including hadronization, 
which remain beyond the reach of perturbative QCD.
The complex phenomena that influence MD are therefore hard to describe in detail and one has 
to rely on phenomenology.
One of the most successful distribution functions used in describing the probability distributions 
of produced particles in $pp/\bar{p}p$ collisions is the two-parameter NBD, 
\begin{equation}
P(n,\bar{n},k)=\frac{\Gamma(n+k)}{\Gamma(k)\Gamma(n+1)}
\left[\frac{\bar{n}}{k+\bar{n}}\right]^n
\left[\frac{k}{k+\bar{n}}\right]^k,
\label{eq:r1}
\end{equation}
where $\bar{n}$ is the average multiplicity and $k$ characterizes the width of
the distribution.
The history and genesis of NBD can be found e.g. in \cite{NBDhistory}.
An important application of this distribution in the particle production is 
connected with cascading mechanisms.
A frequently used and popular interpretation of NBD comes 
from the clan model \cite{GioVanHove} where a particle emits additional particles in a  
self-similar branching pattern. The clans are produced independently and contain particles of the same ancestry. 
The Poisson distribution is the distribution of clans consisting of single particles. It is obtained for $k=\infty$.
A superposition of NBDs was exploited by different approaches
to multiple production in hadron collisions.
It is based on the decomposition 
\begin{equation}
P(n)=
\sum_{i=1}^N \alpha_i P(n,\bar{n}_i,k_i), \ \ \ \
\sum_{i=1}^N \alpha_i = 1,
\label{eq:r2}
\end{equation}
where $P(n,\bar{n}_i,k_i)$ is given by (\ref{eq:r1}).
In this paper we study high statistic multiplicity data \cite{ATLAS2,ATLAS3,ATLAS4} 
obtained by the ATLAS Collaboration at the LHC 
in the framework of a weighted superposition of three NBDs.
The aim is to achieve a detailed description of the data including the shape
of the distributions in the region of maximal values of $P_n$.
Specifically, we consider the function (\ref{eq:r2}) for $N=3$ 
and extract values of the eight independent parameters ($\bar{n}_i$, $k_i$,  $\alpha_2$ and $\alpha_3$)  
corresponding to the ATLAS data 
with different transverse momentum cuts
at the energies $\sqrt{s}=8$ and 13~TeV. 
The results are compared with our previous analysis \cite{IZ}
at $\sqrt{s}=0.9$ and 7~TeV and with a two-NBD parametrization of the same data.

\subsection{Analysis of new ATLAS data at midrapidity} 
\label{sec:2}

The multiplicity distributions of charged particles produced in $pp$ collisions 
have been measured at the LHC by different experiments, in different kinematic regions, 
at different energies and for different classes of events.
The experimental data accumulated by the ATLAS Collaboration are exceptional, for they have small systematic errors 
and are based on the analysis of an extremely large number of events.
The ATLAS data at $\sqrt{s}=8$~TeV \cite{ATLAS2} and $\sqrt{s}=13$~TeV \cite{ATLAS3,ATLAS4} 
exploit a similar methodology to that used at lower centre-of-mass energies \cite{ATLAS1}.
The analyzed charged particle MDs were obtained in the same 
phase-space regions with application of the same multiplicity cuts 
in the same pseudorapidity window $|\eta|<2.5$.
The recorded number of events for the conditions $p_T > 100$~MeV/c and $n_{ch}>1$
exceeds 9 million for both energies, $\sqrt{s}=8$ and 13~TeV.
A similar large number of events was considered in the data sample with the
higher transverse momentum cut $p_T > 500$~MeV/c and $n_{ch}>0$.
The measurements in both regions give a severe restriction on models 
of multiparticle production in $pp$ collisions at high energies.
This concerns in particular the weighted superposition of two NBDs, which 
is usually attributed to the classification of events into soft and semi-hard events
with respect to the momentum 
transfer in parton-parton scatterings.

\subsubsection{MDs at low transverse momentum}
\label{sec:C.1}

\begin{figure}
\includegraphics[width=78mm,height=78mm]{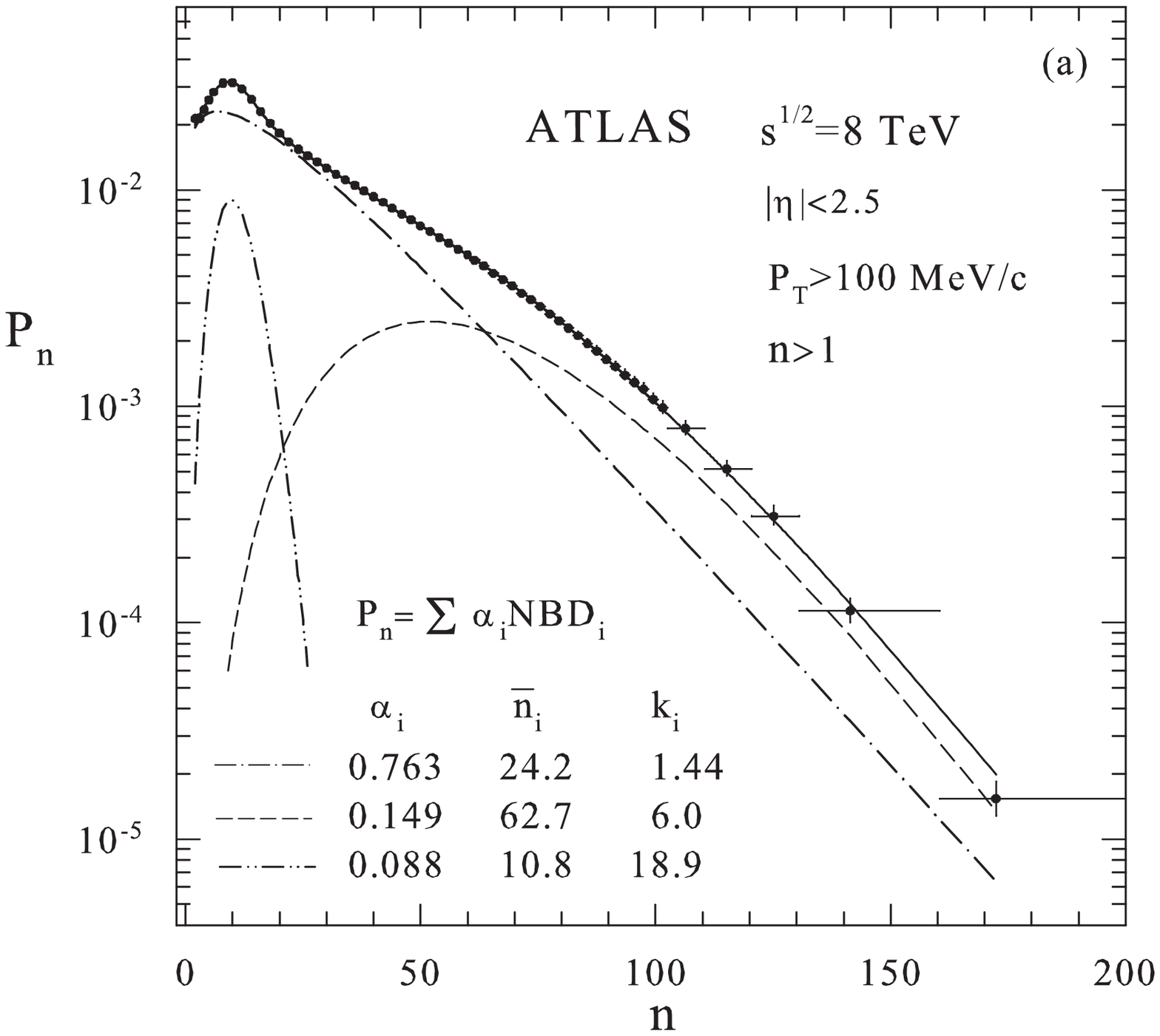}
\vskip -0.6cm
\includegraphics[width=78mm,height=78mm]{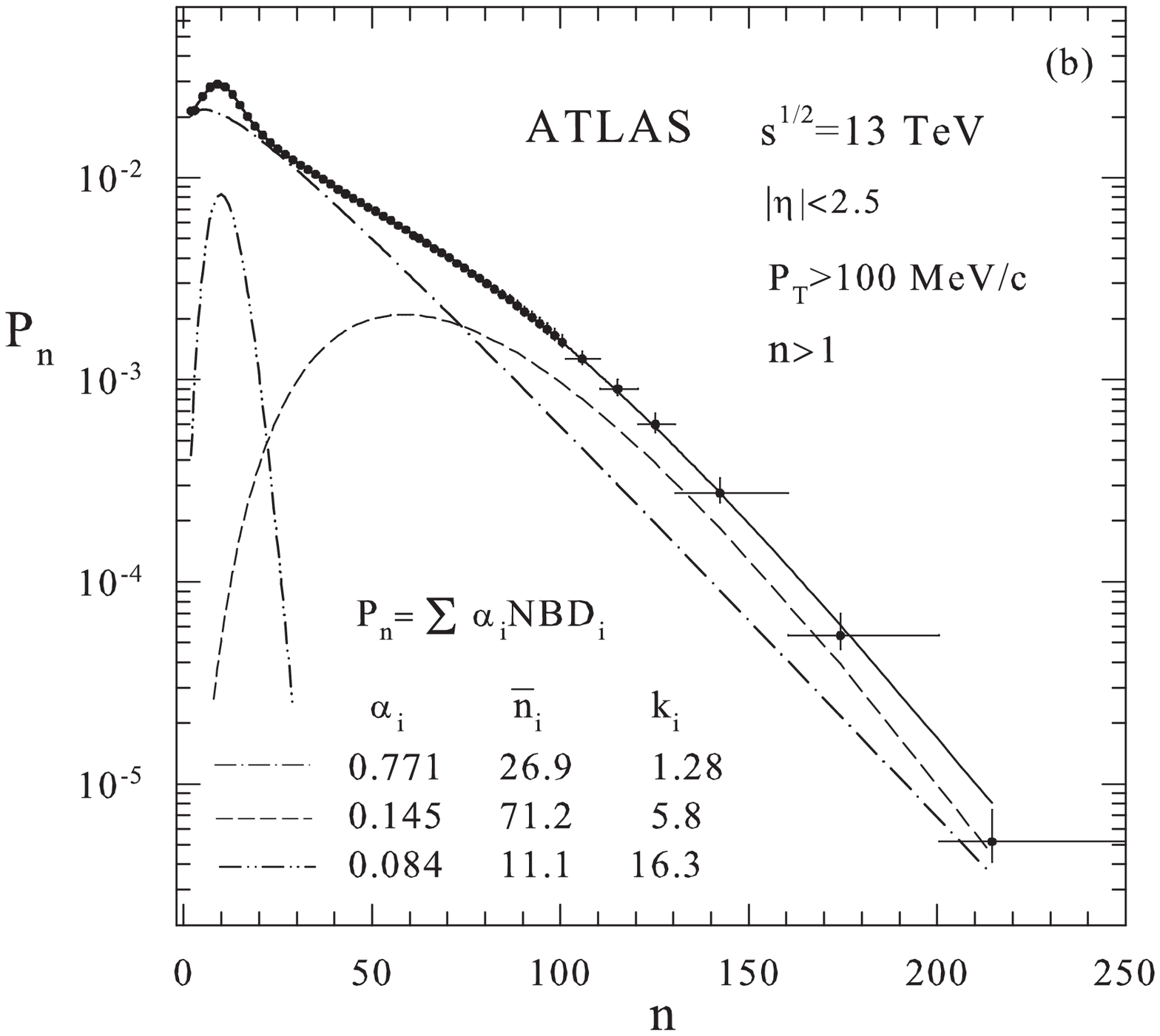}
\vskip -0.7cm
\caption{
MD of charged particles
measured by the ATLAS Collaboration \protect\cite{ATLAS2,ATLAS4}
in the pseudorapidity interval $|\eta|<2.5$ for $p_T > 100$~MeV/c, $n>1$
at {\bf a} $\sqrt s=8$~TeV and {\bf b} $\sqrt s=13$~TeV.
The error bars include both the statistical and the systematic uncertainties summed in quadrature.
The solid lines represent a superposition of three NBDs 
with the parameters from Tables \ref{tab:1} and \ref{tab:2}, respectively.
The dash-dot, dash and dash-dot-dot lines show
single components of the total MD  
}
\label{fig:1}       
\end{figure}
 
The most inclusive  phase-space region covered by the ATLAS measurements corresponds to
the conditions $p_T > 100$~MeV/c and  $n_{ch}>1$.
We have fitted the data \cite{ATLAS2,ATLAS4} on MD measured 
in the interval $ |\eta|<2.5$ in this region with a weighted superposition of three NBDs.
The decomposition of the total MD at the energies $\sqrt s=8$~TeV and 13~TeV
is shown in Fig. \ref{fig:1}a,~b, respectively.
The experimental data are indicated by symbols and the fitted three-component 
function (\ref{eq:r2}) is depicted by the solid line.
The dash-dot, dash and dash-dot-dot lines represent  
single NBD components of the total distribution.  
The corresponding parameters and values of $\chi^2$ are quoted in Tables~\ref{tab:1} and \ref{tab:2}. 
Terms of the fitting procedure are explained in Appendix A.

\begin{table}
\caption{
The parameters of the superposition of three and two NBDs  
obtained from fits to  data \protect\cite{ATLAS2} on MD measured by the ATLAS
Collaboration in the pseudorapidity window $|\eta|<2.5$ with the cut $p_T>100$~MeV/c, $n>1$ at 
$ \sqrt{s}=8$~TeV.
The parameter values were obtained by minimization of Eq.~(\ref{eq:a1})
}
\label{tab:1}       
\begin{tabular*}{\columnwidth}{@{}l@{\extracolsep{\fill}}lll@{}} 
\hline\noalign{\smallskip}
{\it i} &  $\ \ \ \alpha_i$  & \ \ \ $\bar{n}_i$ & \ \ \ $k_i$  \\             
\noalign{\smallskip}\hline\noalign{\smallskip}
 1 &  0.763$^{+0.041}_{-0.055}$  &  24.2$^{+1.4  }_{-1.9  }$ &\ \,1.44$^{+0.06 }_{-0.04 }$   \\ \noalign{\smallskip} 
 2 &  0.149$^{+0.046}_{-0.032}$  &  62.7$^{+1.7  }_{-2.3  }$ &\  \,6.0$^{+0.8  }_{-0.6  }$   \\ \noalign{\smallskip}
 3 &  0.088$^{+0.009}_{-0.009}$  &  10.8$^{+0.2  }_{-0.2  }$ &    18.9$^{+7.8  }_{-4.5  }$   \\ \noalign{\smallskip}
   & \multicolumn{3}{c}{$\chi^2/dof$ = 10.2/(85-8)}  \\
\noalign{\smallskip}\hline\noalign{\smallskip}
 1 &  0.440\,$\pm$\,0.019   &   11.90\,$\pm$\,0.25  & \ 2.58\,$\pm$\,0.10 \\ 
 2 &  0.560\,$\pm$\,0.019   &   42.78\,$\pm$\,0.76  & \ 2.72\,$\pm$\,0.12 \\ \noalign{\smallskip} 
   & \multicolumn{3}{c}{$\chi^2/dof$ = 119/(85-5)}  \\
\noalign{\smallskip}\hline
\end{tabular*}
\end{table}

\begin{table}[h!]
\caption{
The parameters of the superposition of three and two NBDs  
obtained from fits to  data \protect\cite{ATLAS4} on MD measured by the ATLAS
Collaboration in the pseudorapidity window $|\eta|<2.5$ with the cut $p_T\!>\!100$~MeV/c, $n\!>\!1$ at 
$ \sqrt{s}\!~=\!~13$~TeV.
The parameter values were obtained by minimization of Eq.~(\ref{eq:a1})
}
\label{tab:2}       
\begin{tabular*}{\columnwidth}{@{}l@{\extracolsep{\fill}}lll@{}} 
\hline\noalign{\smallskip}
{\it i} &  $\ \ \ \alpha_i$  & \ \ \ $\bar{n}_i$ & \ \ \ $k_i$  \\             
\noalign{\smallskip}\hline\noalign{\smallskip}
 1 &  0.771$^{+0.036}_{-0.048}$  &  26.9$^{+1.5  }_{-2.0  }$ &\ \,1.28$^{+0.04 }_{-0.03 }$ \\ \noalign{\smallskip} 
 2 &  0.145$^{+0.041}_{-0.029}$  &  71.2$^{+1.4  }_{-2.2  }$ &\  \,5.8$^{+0.8  }_{-0.6  }$ \\ \noalign{\smallskip}
 3 &  0.084$^{+0.007}_{-0.007}$  &  11.1$^{+0.1  }_{-0.2  }$ &    16.3$^{+3.5  }_{-2.5  }$ \\ \noalign{\smallskip}
   & \multicolumn{3}{c}{$\chi^2/dof$ = 18.9/(86-8)}  \\
\noalign{\smallskip}\hline\noalign{\smallskip}
 1 &  0.337\,$\pm$\,0.013   &   10.78\,$\pm$\,0.10 &    \ 2.55\,$\pm$\,0.11  \\ 
 2 &  0.663\,$\pm$\,0.013   &   43.62\,$\pm$\,0.51 &    \ 1.96\,$\pm$\,0.06  \\ \noalign{\smallskip} 
   & \multicolumn{3}{c}{$\chi^2/dof$ = 265/(86-5)}  \\
\noalign{\smallskip}\hline
\end{tabular*}
\end{table}

One can see from Fig.~\ref{fig:1} 
that the decomposition into three components is similar at both energies and reveals identical 
features to those found \cite{IZ} in the same class of events at lower $\sqrt s$.
The dominant component with the largest probability gives the main contribution $ \alpha_1\bar{n}_1 $ 
to the total average multiplicity.
The other two components contribute to the high- and low-multiplicity region, respectively.
The average multiplicities of the first and the second component, $\bar{n}_1$ and  $\bar{n}_2$,
increase with energy. This results in broadening of the total distribution.
The average multiplicity $\bar{n}_3 \simeq 11$ of the third component 
is nearly energy independent.
Within the errors quoted in Table~\ref{tab:1}, 
the probabilities $ \alpha_i $ show a weak energy dependency as well.
The values of the parameters $k_i$ increase with decreasing probabilities $\alpha_i$.
The first NBD with the largest probability $\alpha_1$ is characterized by the smallest parameter $k_1$.
The NBD component under the peak of the distribution at low multiplicities
is narrow with large value of $k_3$.

\begin{figure}
\includegraphics[width=78mm,height=78mm]{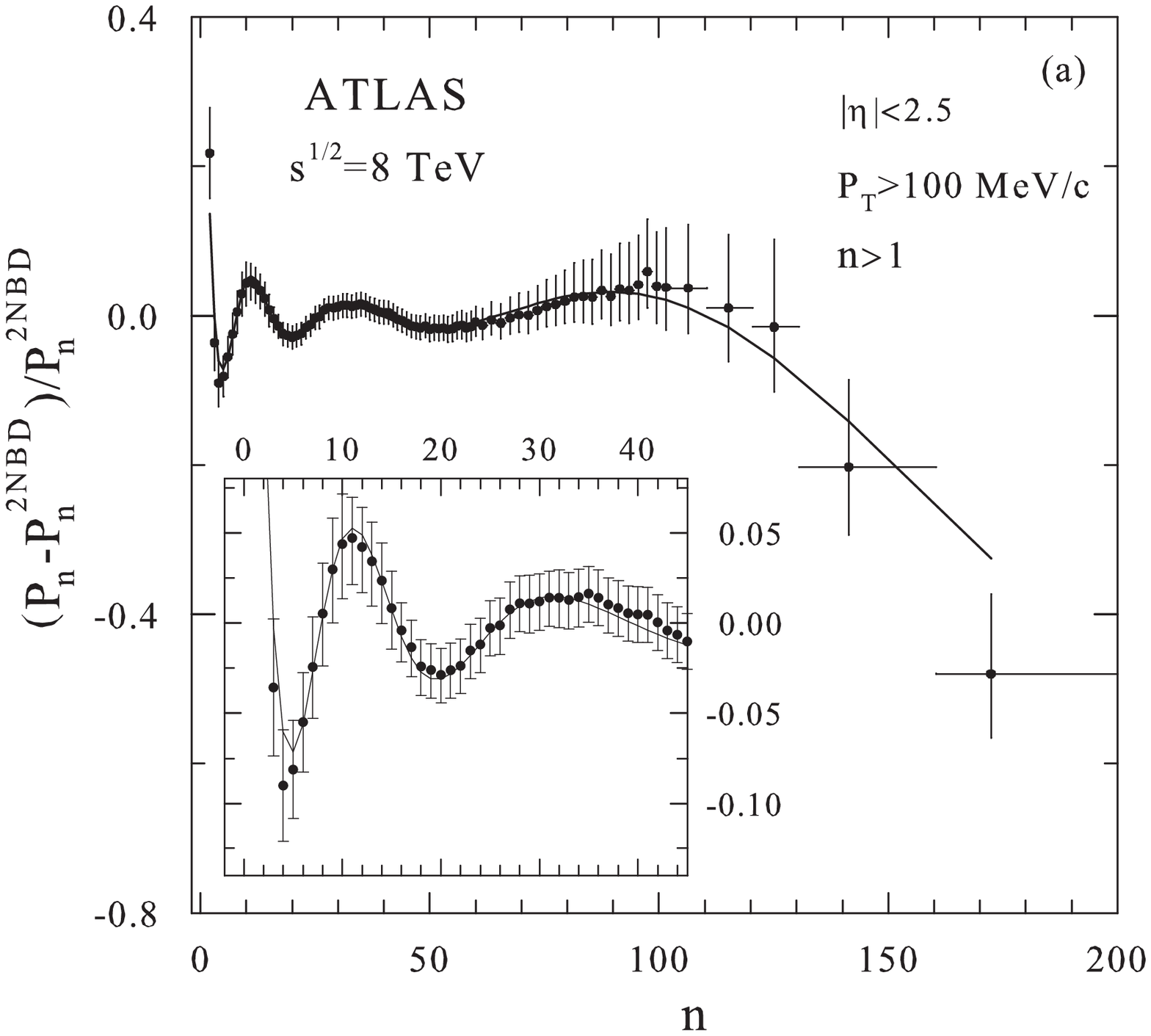}
\vskip -0.7cm
\includegraphics[width=78mm,height=78mm]{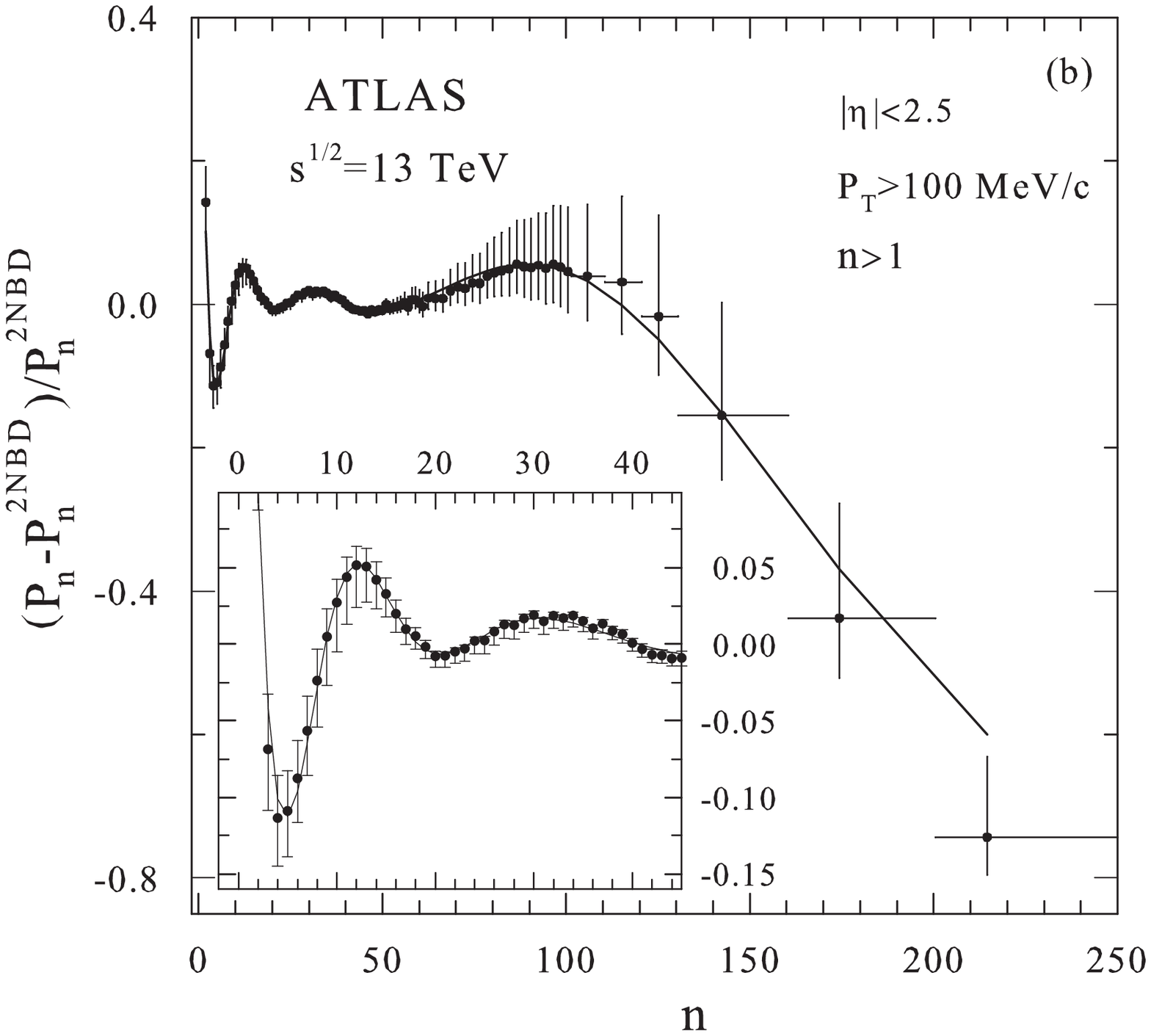}
\vskip -0.7cm
\caption{
Normalized residues of the MDs relative to superposition of two NBDs $(\mathrm{P_n^{2NBD}})$
with the parameters listed in Tables~\ref{tab:1} and \ref{tab:2}, respectively.
The symbols correspond to data on MDs measured by the ATLAS Collaboration \protect\cite{ATLAS2,ATLAS4}
in the interval $|\eta|<2.5$ with the cut $p_T > 100$~MeV/c, $n>1$
at {\bf a} $\sqrt s=8$~TeV and {\bf b} $\sqrt s=13$~TeV.
The error bars include both the statistical and the systematic uncertainties summed in quadrature.
The solid lines represent a three-NBD 
superposition with the parameters from Tables \ref{tab:1} and \ref{tab:2}, respectively.
The inserts show the detailed behaviour of the residues at 
low~$n$  
}
\label{fig:2}       
\end{figure}

We have fitted the ATLAS data with a weighted superposition of two NBDs.
The obtained values of the parameters for the two-component hypothesis  
are given in Tables~\ref{tab:1} and \ref{tab:2}.
Figure \ref{fig:2} shows the relative residues of data with respect to the two-NBD fits.
The points correspond to the measurements of the ATLAS Collaboration
in the pseudorapidity interval $|\eta|<2.5$ for $p_T > 100$~MeV/c, $n>1$.
The inserts show the detailed structure of the residues at low multiplicities.
The solid lines are given by the three-component description of the ATLAS data 
with the parameters quoted in Tables \ref{tab:1} and \ref{tab:2}. 
One can see that the two-NBD approximation of the measured MDs is unsatisfactory at
both energies. 
The corresponding values of $ \chi^2/dof $ are too large, 
especially at the energy $\sqrt{s}=13$~TeV.
The high statistic ATLAS data measured with relatively small systematic uncertainties 
manifest a distinct peak around $n\sim 11$. 
The description of the peak clearly seen in the residues in Fig.~\ref{fig:2}
was obtained by the third negative binomial component with $\bar{n}_3 \simeq 11$, 
similar to the demonstration in \cite{IZ} at lower $\sqrt s$. 
The third component is depicted by the dash-dot-dot lines in Fig.~\ref{fig:1}.

\subsubsection{MDs with the cut $p_T>$500 MeV/c}
\label{sec:C.2}

The ATLAS Collaboration presented data on MD produced in $pp$ collisions
in the separate phase-space region defined by the conditions $p_T > 500$~MeV/c and $n_{ch}>0$
at the energies $\sqrt s=8$~TeV \cite{ATLAS2} and 13~TeV \cite{ATLAS3}. 
The distributions were measured in the window $|\eta|<2.5$. 
They differ from earlier analyses of primary charged particles  
in that charged particles with a life time 30~ps $<\tau <300$~ps were considered as secondary 
particles and thus excluded.   

\begin{figure*}
\begin{center}
\includegraphics[width=78mm,height=78mm]{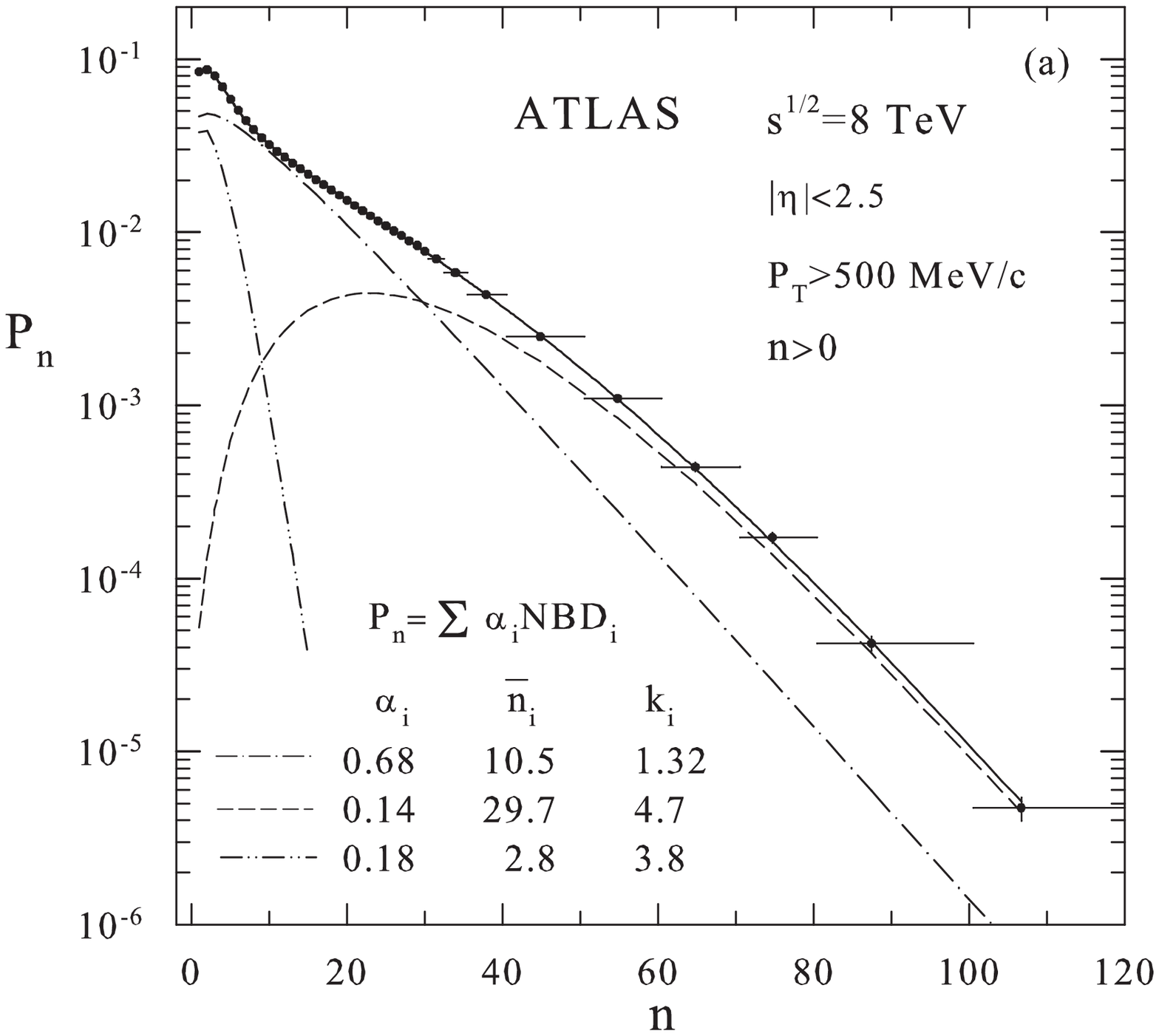}
\includegraphics[width=78mm,height=78mm]{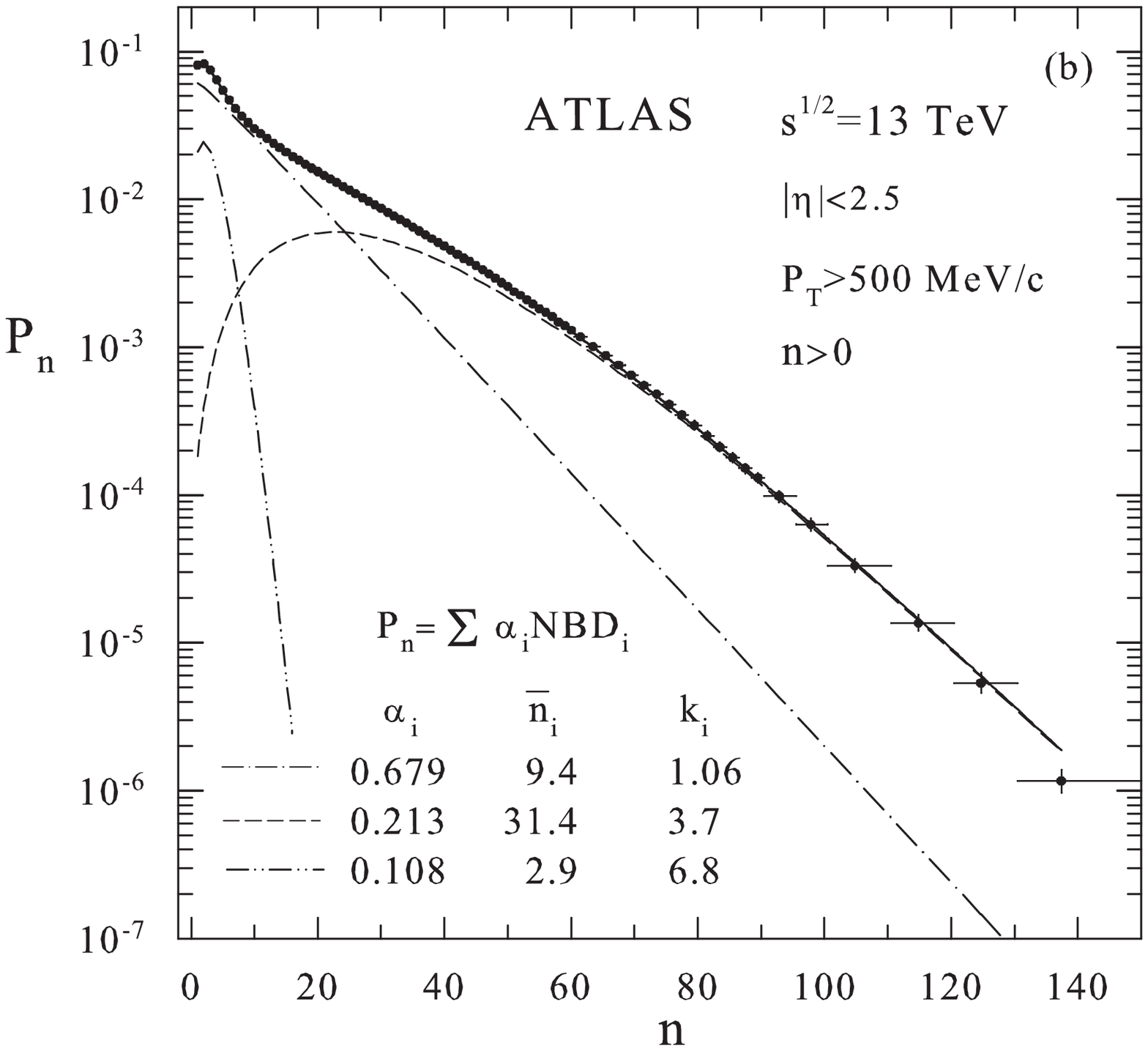}
\vskip -1.0cm
\includegraphics[width=78mm,height=78mm]{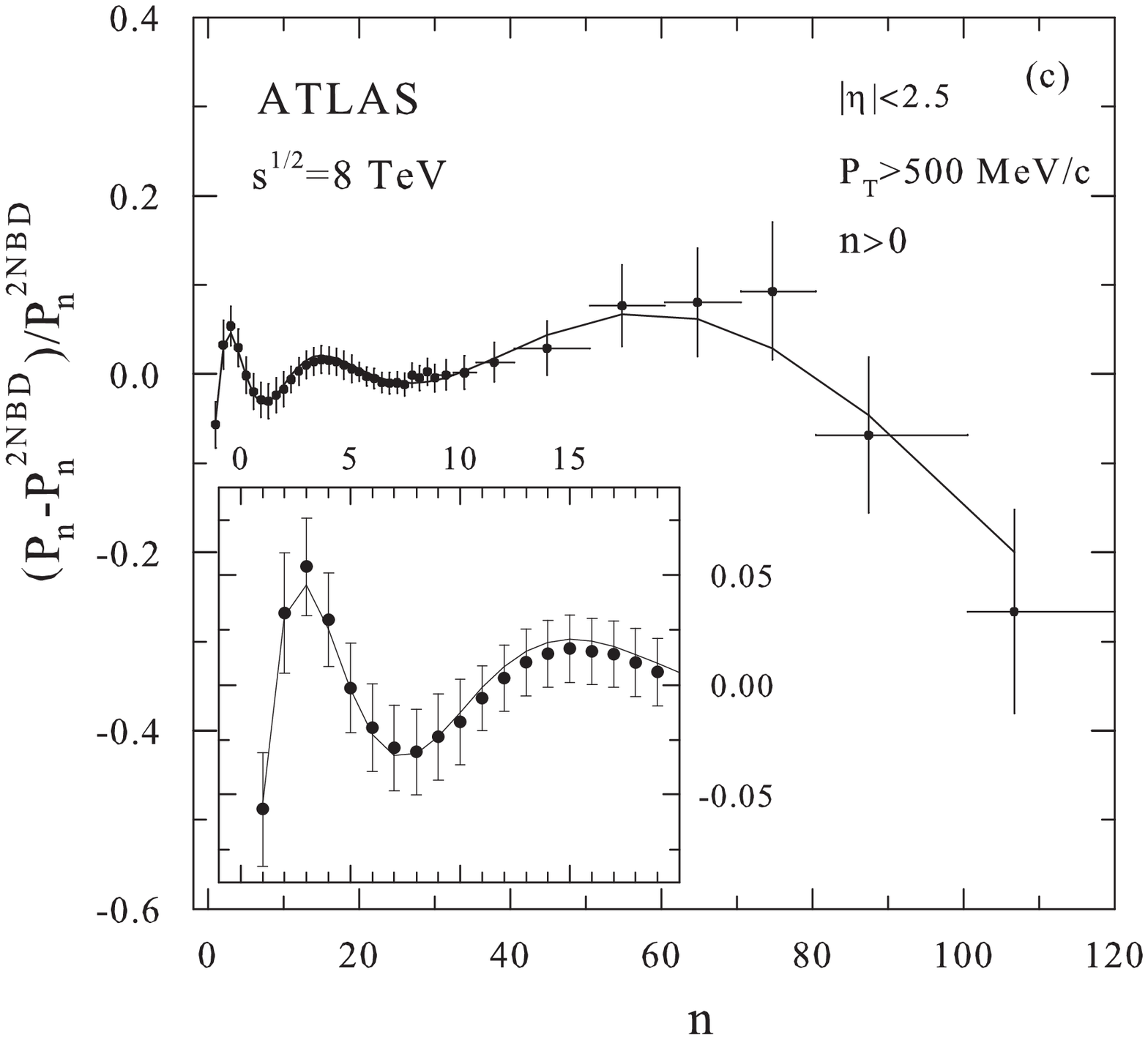}
\includegraphics[width=78mm,height=78mm]{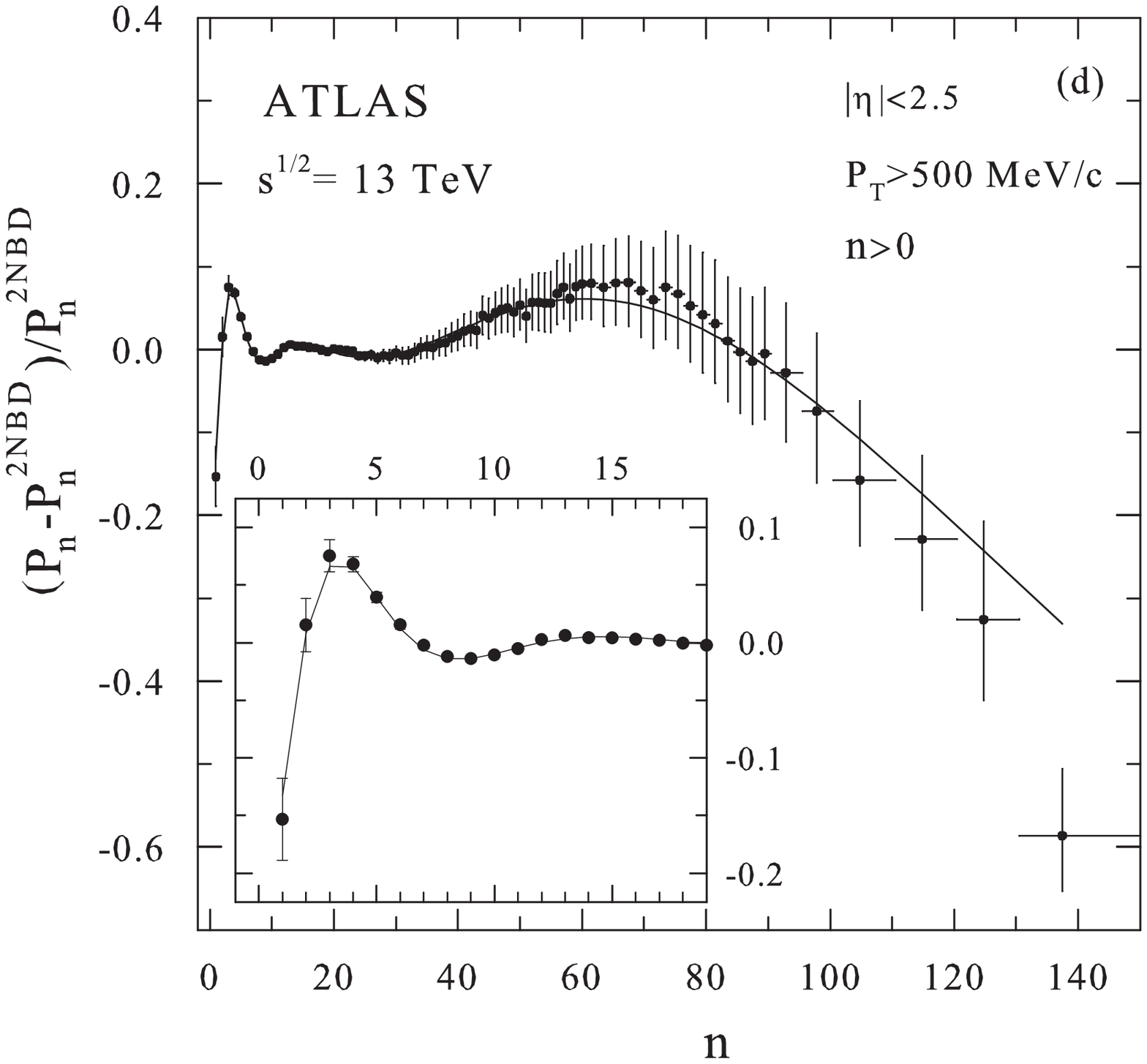}
\end{center} 
\vskip -1.2cm
\caption{
MD of charged particles
measured by the ATLAS Collaboration \protect\cite{ATLAS2,ATLAS3}
in the pseudorapidity interval $|\eta|<2.5$ for $p_T > 500$~MeV/c, $n>0$
at {\bf a} $\sqrt s=8$~TeV and {\bf b} $\sqrt s=13$~TeV.
Normalized residues of the data on MDs relative to weighted superposition 
of two NBDs $(\mathrm{P_n^{2NBD}})$
with the parameters listed in Table~\ref{tab:3} and  Table~\ref{tab:4} at {\bf c}  $\sqrt s=8$~TeV and 
{\bf d} $\sqrt s=13$~TeV, respectively. 
The error bars include both the statistical and the systematic uncertainties summed in quadrature.
The lines have the same meaning as in Figs. \ref{fig:1} and \ref{fig:2}
}
\label{fig:3}       
\end{figure*}

The ATLAS data at $\sqrt s=8$ and 13~TeV together with three-NBD fits are shown
in Fig.~\ref{fig:3}a, b, respectively.
The symbols and the lines have the same meaning as in Fig.~\ref{fig:1}.
The corresponding parameters and values of $\chi^2/dof$ are given 
in Tables \ref{tab:3} and \ref{tab:4}.
One can see some similarities when comparing the description of the most inclusive and 
the $p_T$-cut data shown in Figs. \ref{fig:1} and \ref{fig:3}, respectively.   
The component with the largest probability $\alpha_1$ has the smallest parameter $k_1$.
The parameters $k_i$ increase as the probabilities $\alpha_i$ decrease.
The average multiplicity $\bar{n}_3 \simeq 3$ of the third component  
is nearly energy independent.
The probability $\alpha_3$ of the third component is non-zero also in the $p_T$-cut data sample.  
On the other hand, there are differences between the description of data shown
in Figs. \ref{fig:1} and \ref{fig:3}, respectively.
With the imposed $p_T$ cut, the value of $\bar{n}_3$ becomes significantly smaller 
relative to the average multiplicities of the first and the second NBD component.
The multiplicity $\bar{n}_1$ does not increase with energy but shows signs of saturation.   
The contribution $\alpha_2\bar{n}_2$ of the second component to the total average multiplicity 
increases rapidly with $\sqrt{s}$ and becomes dominant at $\sqrt s=13$~TeV.
A decreasing tendency of the probabilities $\alpha_i$ with the index $i$
(seen in minimal $p_T$-biased data)
is well visible for $p_T>500$~MeV/c at $\sqrt s=13$~TeV only.
A similar description applies to the hierarchy of the parameters $k^{-1}_i$ characterizing the widths 
of the single NBD components.

\begin{table}
\caption{
The parameters of the superposition of three and two NBDs  
obtained from fits to  MDs \protect\cite{ATLAS2}  measured by the ATLAS
Collaboration in the pseudorapidity window $|\eta|<2.5$ with the cut $p_T>500$~MeV/c, $n>0$ at 
$ \sqrt{s}=8$ TeV.
The parameter values were obtained by minimization of Eq. (\ref{eq:a1})
}
\label{tab:3}       
\begin{tabular*}{\columnwidth}{@{}l@{\extracolsep{\fill}}lll@{}} 
\hline\noalign{\smallskip}
{\it i} &  $\ \ \ \alpha_i$  & \ \ \ $\bar{n}_i$ & \ \ \ $k_i$  \\             
\noalign{\smallskip}\hline\noalign{\smallskip}
 1 &   0.68$^{+0.15 }_{-0.36 }$  &   10.5$^{+2.0  }_{-2.4  }$   &  1.32$^{+1.64 }_{-0.25 }$ \\ \noalign{\smallskip}
 2 &   0.14$^{+0.17 }_{-0.06 }$  &   29.7$^{+3.1  }_{-5.4  }$   &   4.7$^{+1.3  }_{-1.3  }$ \\ \noalign{\smallskip}
 3 &   0.18$^{+0.19 }_{-0.09 }$  &\ \,2.8$^{+0.2  }_{-0.2  }$   &   3.8$^{+10.9 }_{-1.5}$   \\ \noalign{\smallskip}
   & \multicolumn{3}{c}{$\chi^2/dof$ = 4.5/(39-8)}  \\
\noalign{\smallskip}\hline\noalign{\smallskip}
 1 &  0.548\,$\pm$\,0.030   &\ \,4.47\,$\pm$\,0.24  &    1.38\,$\pm$\,0.08  \\ 
 2 &  0.452\,$\pm$\,0.030   &   20.05\,$\pm$\,0.71  &    2.47\,$\pm$\,0.16  \\ \noalign{\smallskip} 
   & \multicolumn{3}{c}{$\chi^2/dof$ = 43.8/(39-5)}  \\
\noalign{\smallskip}\hline
\end{tabular*}
\end{table}
\begin{table}
\caption{
The parameters of the superposition of three and two NBDs  
obtained from fits to MDs \protect\cite{ATLAS3} measured by the ATLAS
Collaboration in the pseudorapidity window $|\eta|<2.5$ with the cut $p_T>500$~MeV/c, $n>0$ at 
$ \sqrt{s}=13$ TeV.
The parameter values were obtained by minimization of Eq. (\ref{eq:a1})
}
\label{tab:4}       
\begin{tabular*}{\columnwidth}{@{}l@{\extracolsep{\fill}}lll@{}} 
\hline\noalign{\smallskip}
{\it i} &  $\ \ \ \alpha_i$  & \ \ \ $\bar{n}_i$ & \ \ \ $k_i$  \\             
\noalign{\smallskip}\hline\noalign{\smallskip}
 1 &  0.679$^{+0.059}_{-0.073}$  &\ \,9.4$^{+0.7  }_{-0.7  }$   &  1.06$^{+0.15 }_{-0.09 }$ \\ \noalign{\smallskip} 
 2 &  0.213$^{+0.042}_{-0.035}$  &   31.4$^{+1.2  }_{-1.4  }$   &   3.7$^{+0.3  }_{-0.3  }$ \\ \noalign{\smallskip}
 3 &  0.108$^{+0.031}_{-0.024}$  &\ \,2.9$^{+0.1  }_{-0.1  }$   &   6.8$^{+4.6  }_{-2.2  }$ \\ \noalign{\smallskip}
   & \multicolumn{3}{c}{$\chi^2/dof$ = 46.5/(81-8)}  \\
\noalign{\smallskip}\hline\noalign{\smallskip}
 1 &  0.532\,$\pm$\,0.013   &\ \,4.37\,$\pm$\,0.11  &    1.01\,$\pm$\,0.02 \\ 
 2 &  0.468\,$\pm$\,0.013   &   22.38\,$\pm$\,0.33  &    2.11\,$\pm$\,0.06  \\ \noalign{\smallskip} 
   & \multicolumn{3}{c}{$\chi^2/dof$ = 548/(81-5)}  \\
\noalign{\smallskip}\hline
\end{tabular*}
\end{table}

The normalized residues relative  
to the two-NBD parametrization (see Tables~\ref{tab:3} and \ref{tab:4}) 
of the ATLAS data measured in the pseudorapidity window  $|\eta|<2.5$ 
with $p_T > 500$~MeV/c,
$n>0$ at the energies $\sqrt s=8$~TeV \cite{ATLAS2} and 13~TeV \cite{ATLAS3}
are depicted in Fig. \ref{fig:3}c,~d, respectively.
Both data show sizeable discrepancies with respect to the weighted superposition
of two~NBDs. 
The values of $\chi^2/dof=43.8/34$ at $\sqrt s=8$~TeV  and $\chi^2/dof=548/76$ 
at $\sqrt s=13$~TeV for the two-NBD fits are too large, especially at the higher energy.

The residues manifest  
a clear peak visible at low multiplicities around $n\!\simeq\!3$.
The description of the peak is obtained by the third negative binomial component 
with $\bar{n}_3\!\simeq\!3$ shown by the dash-dot-dot lines in Fig.~\ref{fig:3}a, b.
The analysis of the new ATLAS data at $ \sqrt{s}=8$ and 13~TeV  
confirms the existence of the peak \cite{IZ} emerging near $n\simeq 3$ at $ \sqrt{s}=7$~TeV. 
Such an effect is negligible in the ATLAS measurements  
with $p_T > 500$~MeV/c at $\sqrt{s}=0.9$~TeV (see Fig. 6(b) in \cite{IZ}).

\subsection{Energy dependence of three-NBD description}
\label{sec:D}

The ATLAS data on MDs in two $p_T$-cut regions  
show a distinct peak at multiplicities where the soft production processes dominate. 
The peaky structure can be well described by a separate component
within the three-component 
parametrization of the experimental data.
A description of the MDs by superposition of 
three NBDs reveals some properties which are compared below for both analyzed kinematic regions.
It concerns the energy dependence of the corresponding parameters obtained 
at lower energies ($\sqrt{s}=0.9$ and 7~TeV) \cite{IZ} 
and in this analysis ($\sqrt{s}=8$ and 13~TeV).

\begin{figure}
\includegraphics[width=78mm,height=78mm]{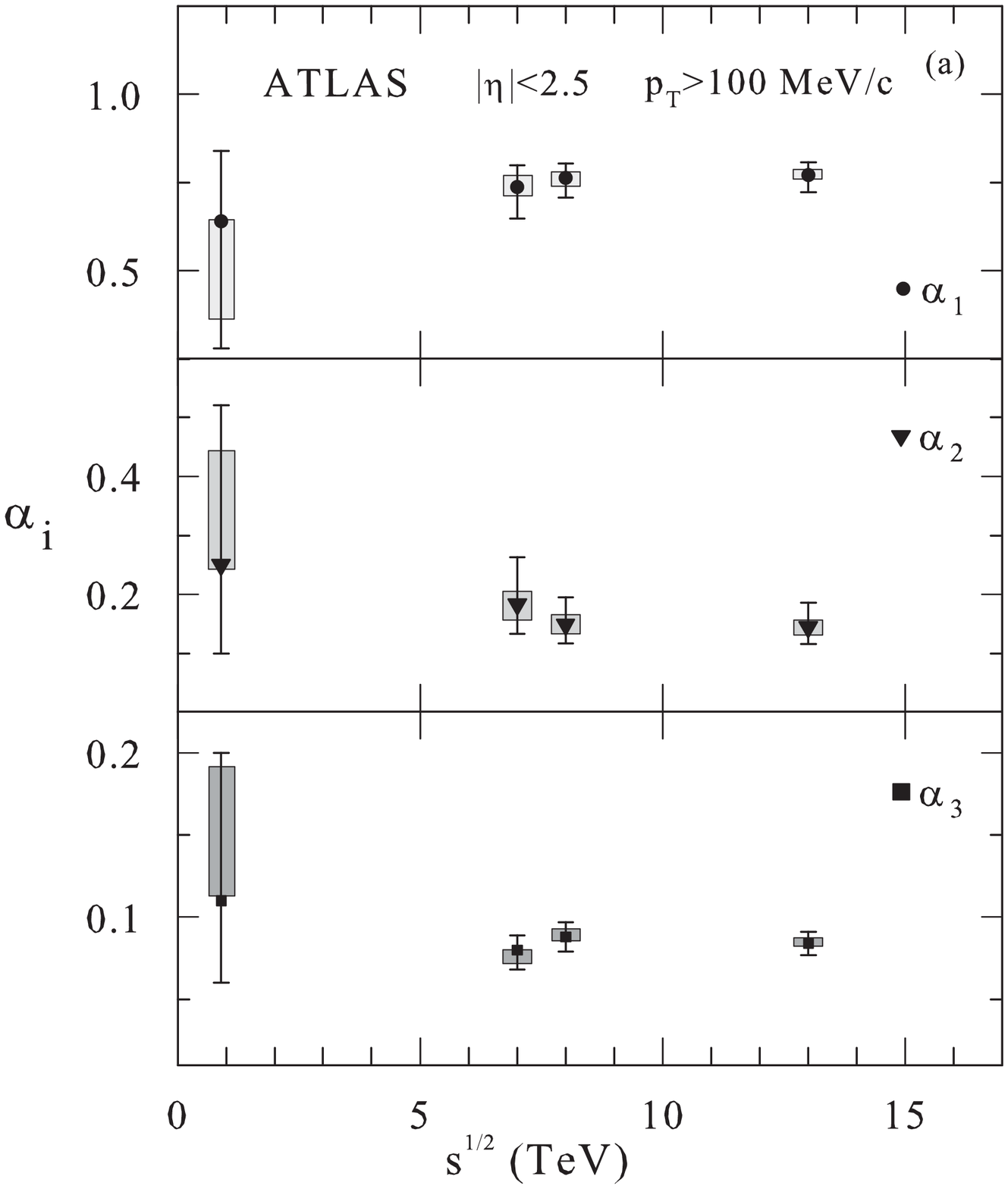}
\vskip -0.2cm
\includegraphics[width=78mm,height=78mm]{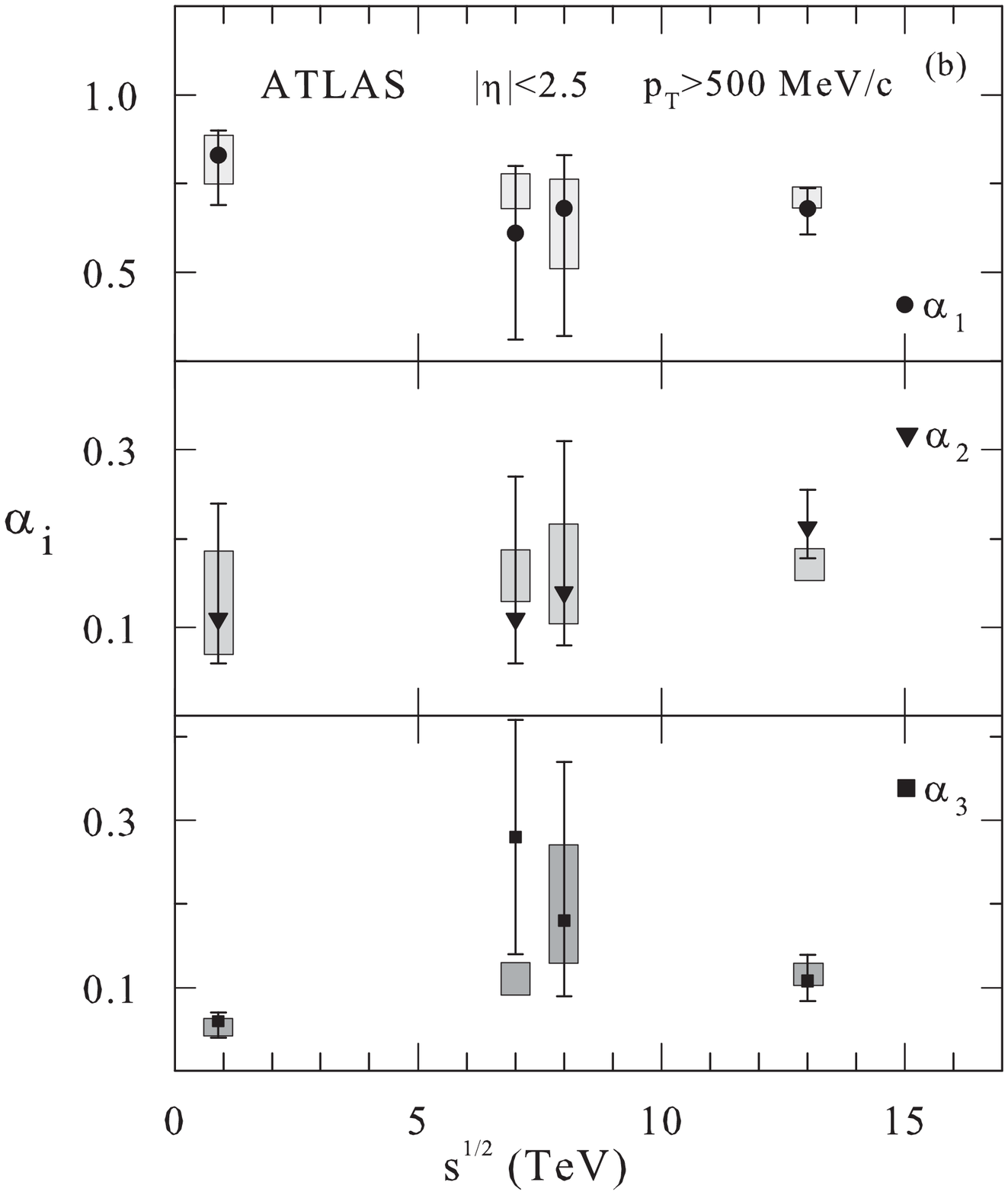}
\vskip 0.5cm
\caption{
Energy dependence of the probabilities $\alpha_i$ of single components of the 
weighted superposition of three NBDs fitted to the charged particle MDs 
\protect\cite{ATLAS1,ATLAS2,ATLAS3,ATLAS4}
measured by the ATLAS Collaboration in the interval $|\eta|<2.5$
for {\bf a} $p_T>100$~MeV/c and {\bf b} $p_T>500$~MeV/c.  
The symbols with error bars and shaded rectangles correspond to the parameter values 
obtained by minimization of 
Eqs. (\ref{eq:a1}) and (\ref{eq:a4}), respectively  
}
\label{fig:4}       
\vskip -0.5cm
\end{figure}

\begin{figure*}
\begin{center}
\includegraphics[width=78mm,height=78mm]{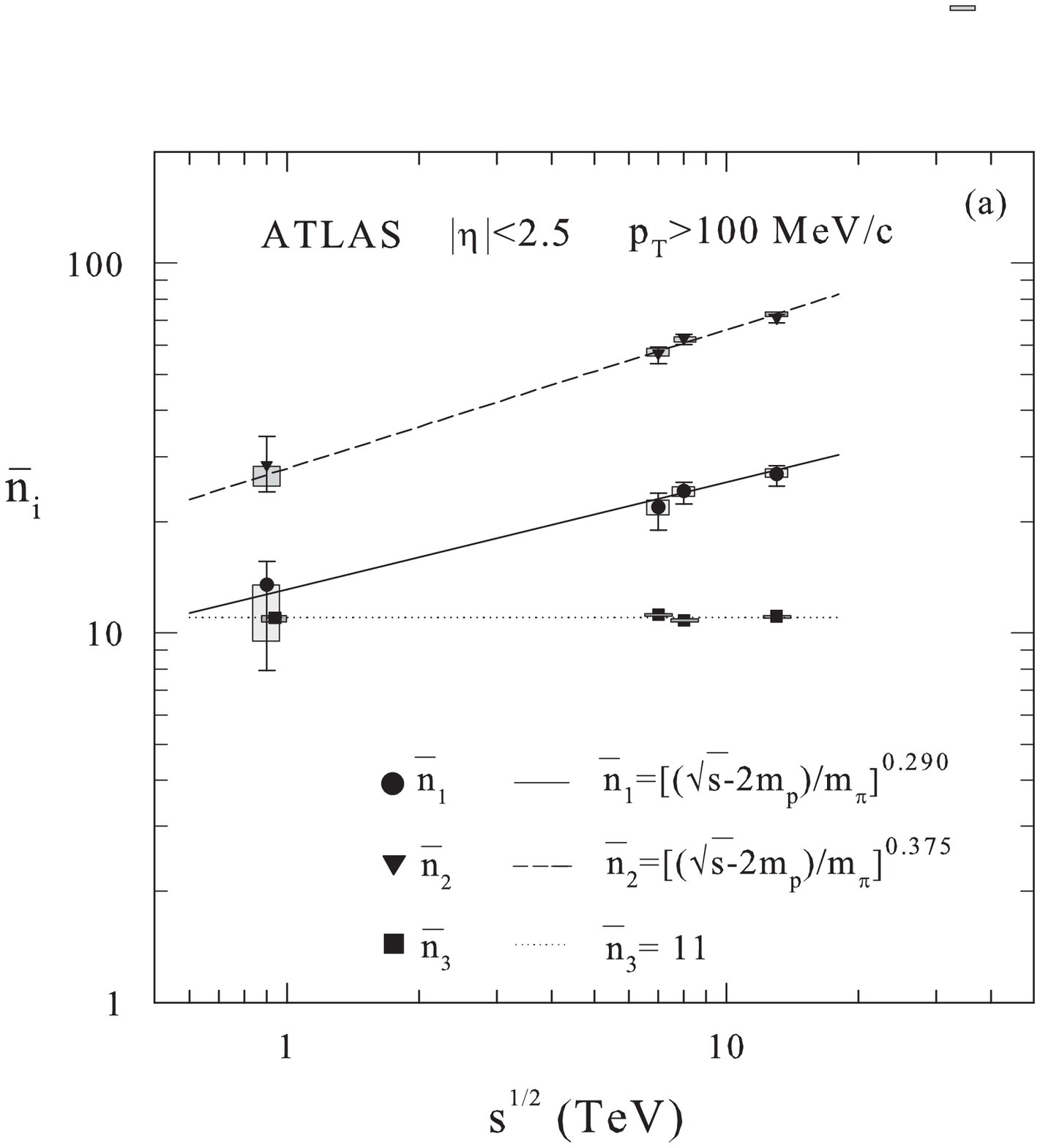}
\includegraphics[width=78mm,height=78mm]{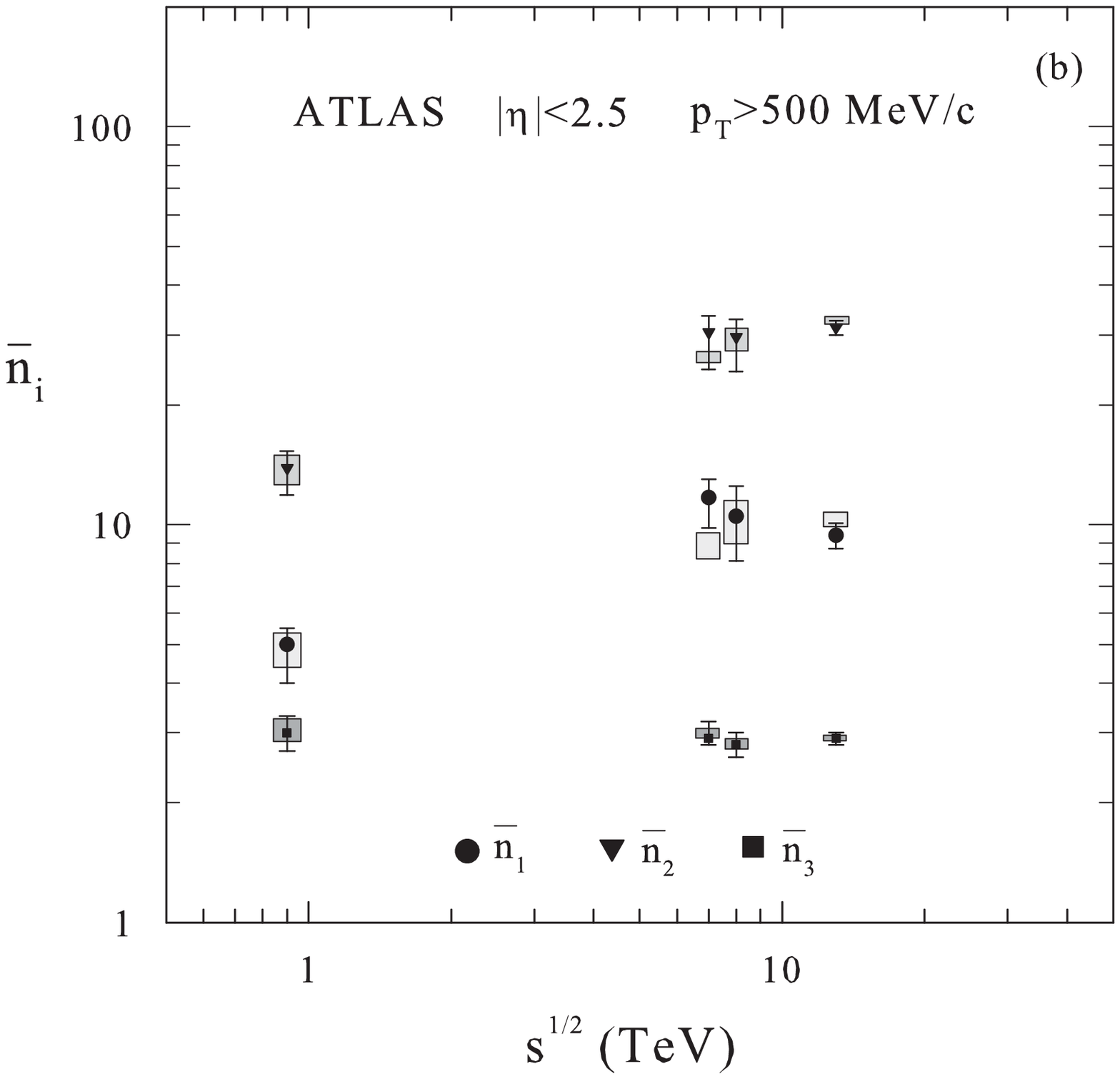}
\end{center} 
\vskip -1.0cm
\caption{
Energy dependence of the average multiplicities $\bar{n}_i$
of single components of the weighted superposition of three NBDs 
fitted to the charged-particle MDs 
\protect\cite{ATLAS1,ATLAS2,ATLAS3,ATLAS4}
measured by the ATLAS Collaboration in the interval $|\eta|<2.5$
for {\bf a} $p_T>100$~MeV/c and {\bf b} $p_T>500$~MeV/c. 
The symbols with error bars and shaded rectangles correspond to the parameter values 
obtained by minimization of 
Eqs. (\ref{eq:a1}) and (\ref{eq:a4}), respectively.  
The lines correspond to the quoted formulas where $m_p$ and $m_{\pi}$
are the proton and pion mass, respectively
}
\label{fig:5}       
\end{figure*}
 
\begin{figure*}
\begin{center}
\includegraphics[width=78mm,height=78mm]{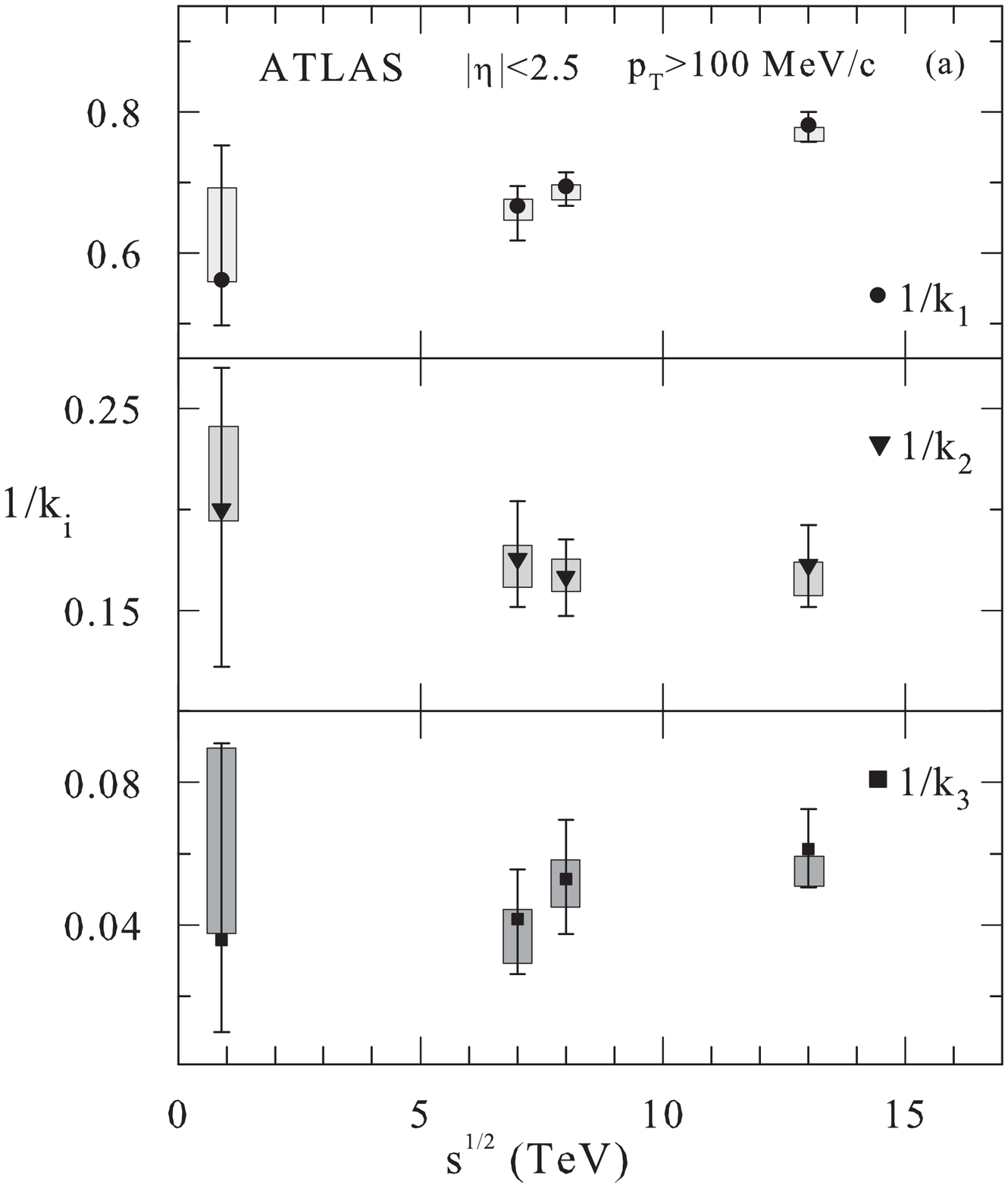}
\includegraphics[width=78mm,height=78mm]{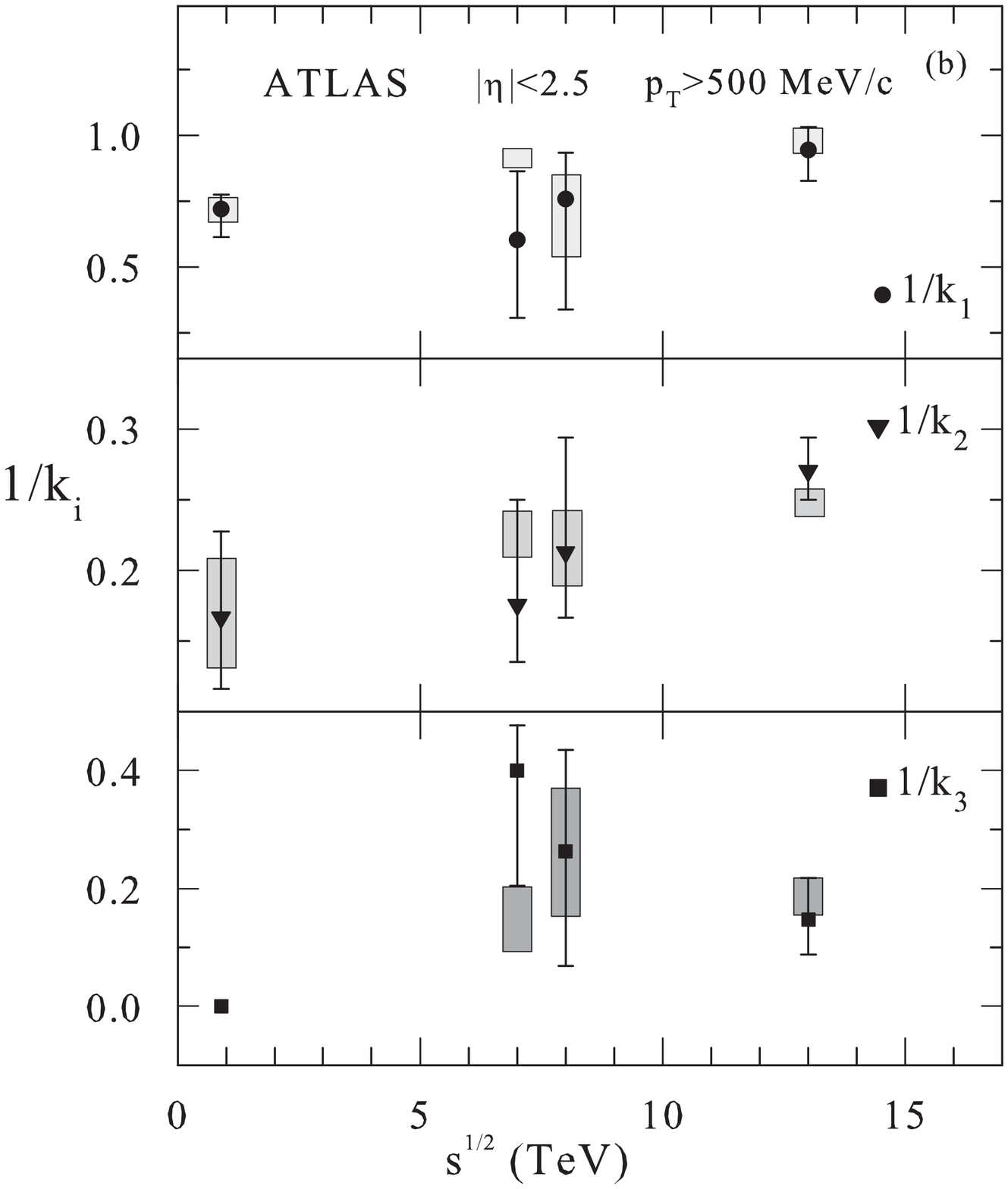}
\end{center} 
\vskip -0.4cm
\caption{
Energy dependence of the inverse values of the parameters $k_i$ 
of single components of the weighted superposition of three NBDs 
fitted to the charged-particle MDs 
\protect\cite{ATLAS1,ATLAS2,ATLAS3,ATLAS4}
measured by the ATLAS Collaboration in the interval $|\eta|<2.5$
for {\bf a} $p_T>100$~MeV/c and {\bf b} $p_T>500$~MeV/c.
The symbols with error bars and shaded rectangles correspond to the parameter values 
obtained by minimization of 
Eqs. (\ref{eq:a1}) and (\ref{eq:a4}), respectively    
}
\label{fig:6}       
\vskip -0.3cm
\end{figure*}

Figure \ref{fig:4} shows the probabilities $\alpha_i$ of single NBD components 
in dependence on $\sqrt{s}$.
The parameters $\alpha_i$ were obtained from fits to the ATLAS data on multiplicities
in the window  $|\eta|<2.5$ 
for (a) an almost inclusive ($p_T>100$~MeV/c) and (b) a transverse momentum cut ($p_T>500$~MeV/c) 
data sample.
The symbols with error bars and shaded rectangles correspond to the parameter values 
obtained by minimization of 
Eqs.~(\ref{eq:a1})~and~(\ref{eq:a4}), respectively. 
Despite considerable errors,
there are some trends visible in the behavior of the probabilities. 
The values of $\alpha_i$ in Fig.~\ref{fig:4}a show a weak or nearly no energy dependence
in the multi-TeV energy 
region.~Errors of their determination allow to trace a hierarchy in relation to the index $i$ 
which is rather stable against $\sqrt{s}$. 
The same ordering of $\alpha_i$  
is obvious in Fig.~\ref{fig:4}b at $\sqrt{s}=13$~TeV only. 
The ordering of 
$\alpha_2$ and $\alpha_3$ for the data with $p_T>500$~MeV/c is due to large errors 
unclear at $\sqrt{s}=$7 and 8~TeV. 
The probability of the third NBD component is non-zero 
in all analyzed cases.

Figure \ref{fig:5} shows the $\sqrt{s}$ dependence of the average multiplicities $\bar{n}_i$ of 
single NBDs in the window  $|\eta|<2.5$ 
for (a) $p_T>100$~MeV/c and (b) $p_T>500$~MeV/c cuts.
The values of $\bar{n}_i$ extracted from the ATLAS data \cite{ATLAS1,ATLAS2,ATLAS3,ATLAS4} 
are depicted in the log-log plot to check their power behavior. 
As seen from Fig.~\ref{fig:5}a, the average multiplicities $\bar{n}_1$ and $\bar{n}_2$ 
demonstrate a power increase with the energy $\sqrt s$. 
Both dependences are well parametrized in terms of 
$Y_{max}\!=\!\ln[(\sqrt{s}\!-\!2m_p)/m_{\pi}]$, which is 
the maximal rapidity of pions in the $pp$ c.m. system.
A similar description seems to be valid for the $\sqrt{s}$ dependence of $\bar{n}_2$ 
shown in Fig.~\ref{fig:5}b, thought 
for the multiplicity $\bar{n}_1$ with $p_T>500$~MeV/c it is problematic to draw such a conclusion. 
The third NBD component at low $n$ reveals remarkable properties.
Its average multiplicity $ \bar{n}_3$ is nearly energy independent
for both $p_T$-cut data samples.

The inverse values of the parameters $k_i$ 
of the weighted superposition of three NBDs are depicted  
as a function of $\sqrt s$ in Fig.~\ref{fig:6}.
They were found from an analysis of the same data as the parameters shown in
Figs.~\ref{fig:4} and \ref{fig:5}.
As seen from Fig.~\ref{fig:6}a, the parameters demonstrate a clear hierarchy 
with the index $i$ for the $p_T>100$~MeV/c data sample.
The width of the dominant multiplicity component characterized by $k^{-1}_1$ shows 
a slight growth with energy. 
The parameter $k^{-1}_2$ of the component under the tail of the distributions reveals 
approximate constancy with respect to $\sqrt{s}$ in the multi-TeV energy region. 
The width of the peak at low multiplicities ($k^{-1}_3$) indicates a similar tendency, though
such a statement is not too conclusive due to errors. 
The same hierarchy of the parameters $k^{-1}_i$ with respect to the index $i$
as in Fig.~\ref{fig:6}a 
is observed for the data sample with $p_T>500$~MeV/c
at $\sqrt{s}=$ 13~TeV only. The ordering of the parameters at lower energies is hard to tell
and their energy dependence is unclear from Fig.~\ref{fig:6}b.

\subsection{Combinants of the MDs}
\label{sec:E}

A general form of the MD is useful to make a characterization 
in terms of its deviations from the Poisson distribution.
The natural logarithm of the Poisson generating functional is given by a linear 
dependence on its argument.
The higher order terms of the power series expansion of the logarithm of 
a general generating functional denote the deviation from the Poisson distribution. 
The expansion coefficients ${\cal C}(i)$, named combinants \cite{Gyulassy}, 
characterize the MD in terms of the generating function
\begin{equation}
G(z)=\exp\left(\sum_{i=1}^{\infty} {\cal C}(i)(z^i-1)\right).
\label{eq:r3}
\end{equation}

The combinants possess the additivity property and are expressible 
as a finite combination of the ratios $P(n)/P(0)$. 
Some of them are permitted to have negative values \cite{Gyulassy}. 
Individual interpretations of the coefficients 
depend on different models of multiparticle production. 
The quantity of interest, named here ``cumulative combinant", is $i{\cal C}(i)$.
In the model of independent boson production with geometric distributions of particles 
in each production mode (e.g. in the mode with an average number of bosons $\overline{n}_i$ 
with momenta in the interval $(p_i, p_i+dp_i)$), 
the cumulative combinant $(i\!+\!1){\cal C}(i\!+\!1)$ 
is the mean number of modes which have an occupation number greater than $i$ \cite{Gyulassy78}.
The cumulative combinants can be rewritten in the form 
$(i\!+\!1){\cal C}(i\!+\!1) = \langle N\rangle C_i$ where 
$\langle N\rangle=G'(z=1)$ is the average multiplicity and
the coefficients $C_i$ are commonly 
used in the  relation 
\begin{equation}
(n+1)P(n+1)=\langle N\rangle \sum_{i=0}^{n}  C_i P(n-i).
\label{eq:r4}
\end{equation}
\vskip -0.1cm
Such a kind of recurrence formula occurs in cascade-stochastic processes \cite{SalehTeich,SchmittMarsan}.
The relation was applied to the parametrization of the MD of charged particles in hadron-hadron collisions 
at high energies \cite{Rusov1,Rusov2}.

\begin{figure}
\includegraphics[width=78mm,height=78mm]{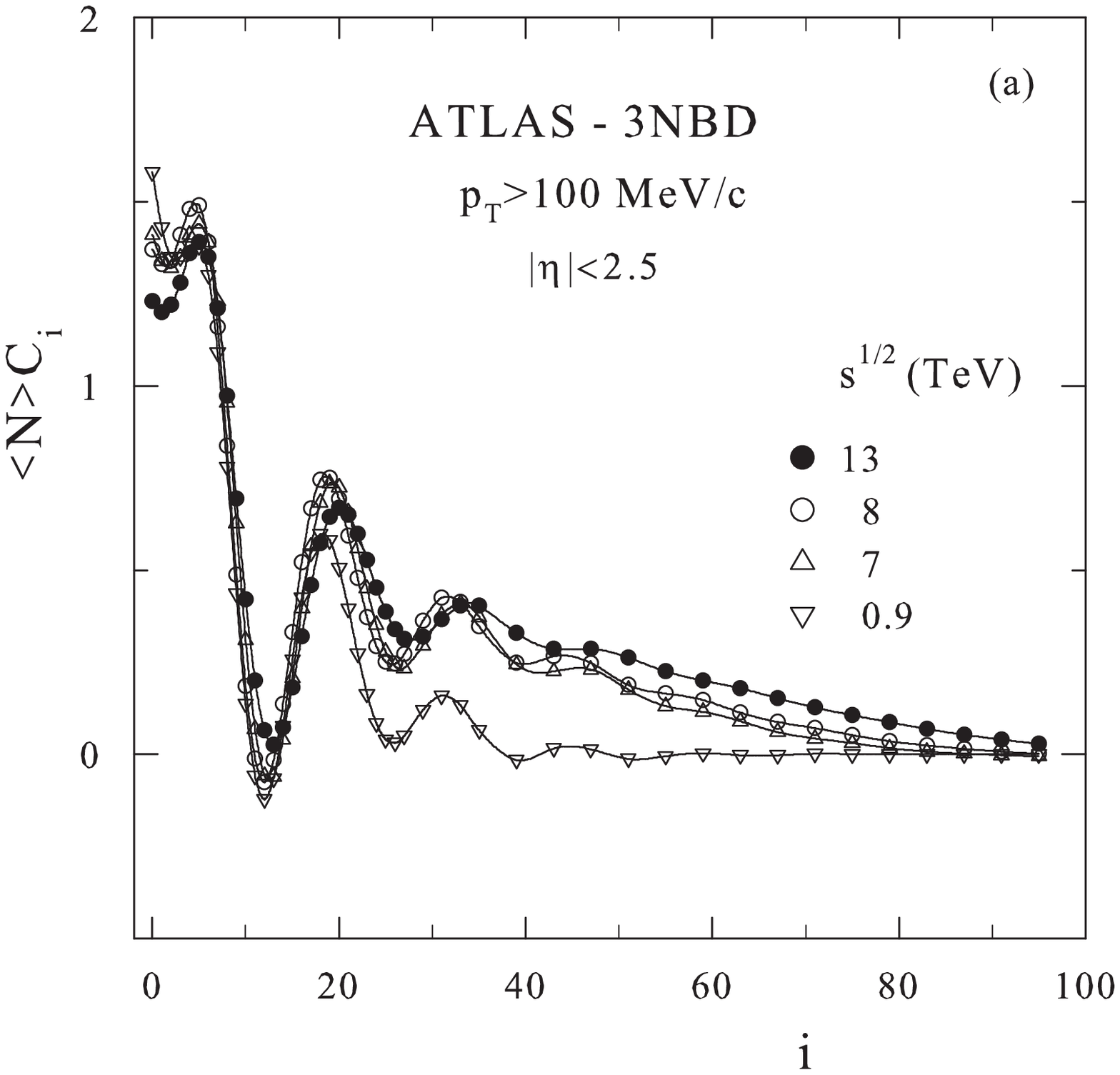}
\vskip -1.0cm
\includegraphics[width=78mm,height=78mm]{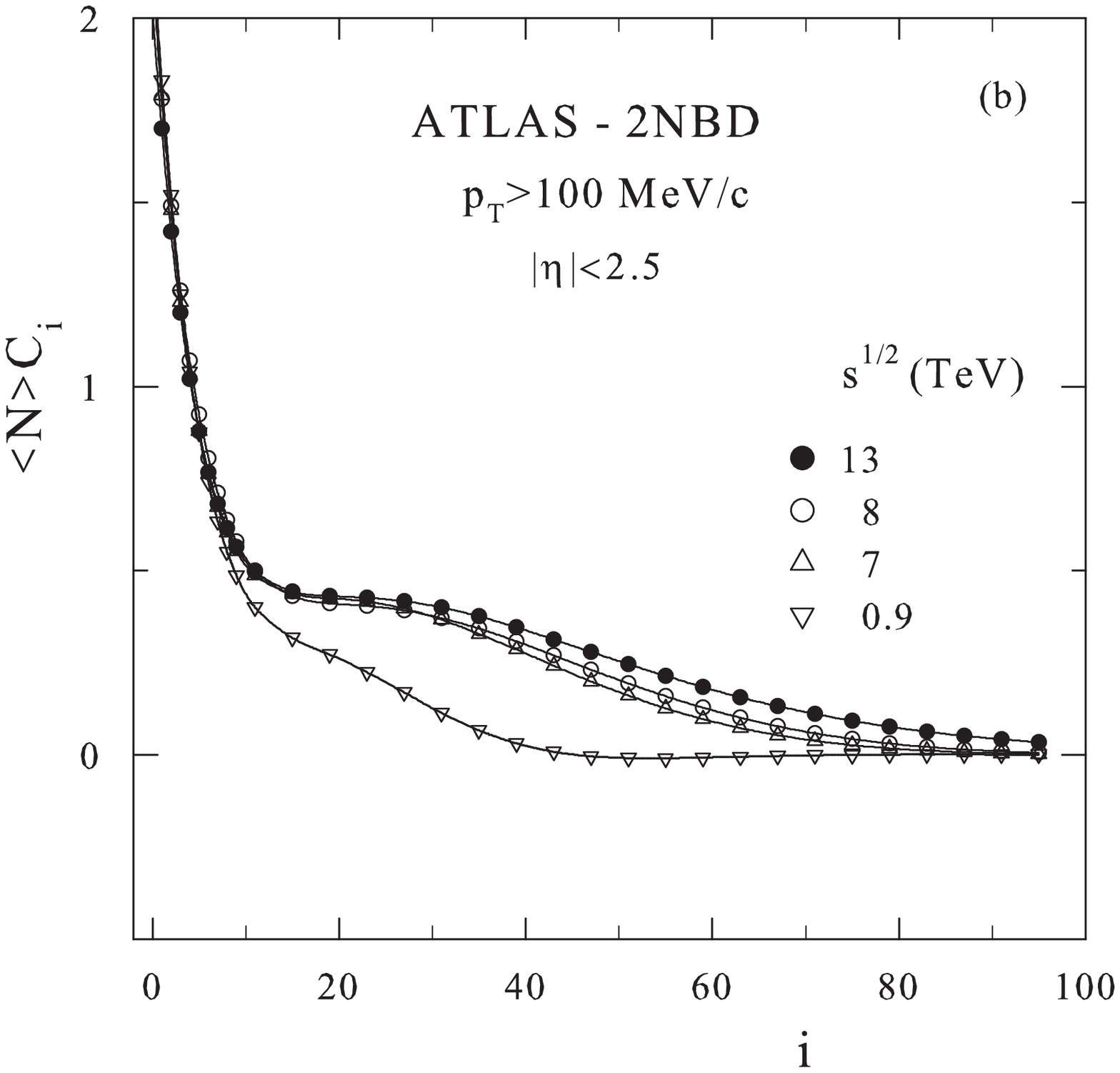}
\vskip -0.7cm
\caption{
Energy dependence of the cumulative combinants $\langle N\rangle C_i$ 
calculated from a weighted superposition of {\bf a} three and {\bf b} two NBDs 
fitted to the charged-particle MDs 
\protect\cite{ATLAS1,ATLAS2,ATLAS4}
measured by the ATLAS Collaboration in the interval $|\eta|<2.5$
for $p_T>100$~MeV/c  
}
\label{fig:7}       
\vskip -0.5cm
\end{figure}

The quantities $\langle N\rangle C_i$ 
calculated  by  Eq. (\ref{eq:r4}) from MDs measured 
by the CMS \cite{CMS} and ALICE \cite{ALICE8} Collaborations in $pp$ collisions 
possess remarkable oscillating properties \cite{WW1}.
Unlike this, 
the two-NBD superposition used to fit the data is not able to account 
for the oscillating structure of $\langle N\rangle C_i$.

\begin{figure*}
\begin{center}
\includegraphics[width=78mm,height=78mm]{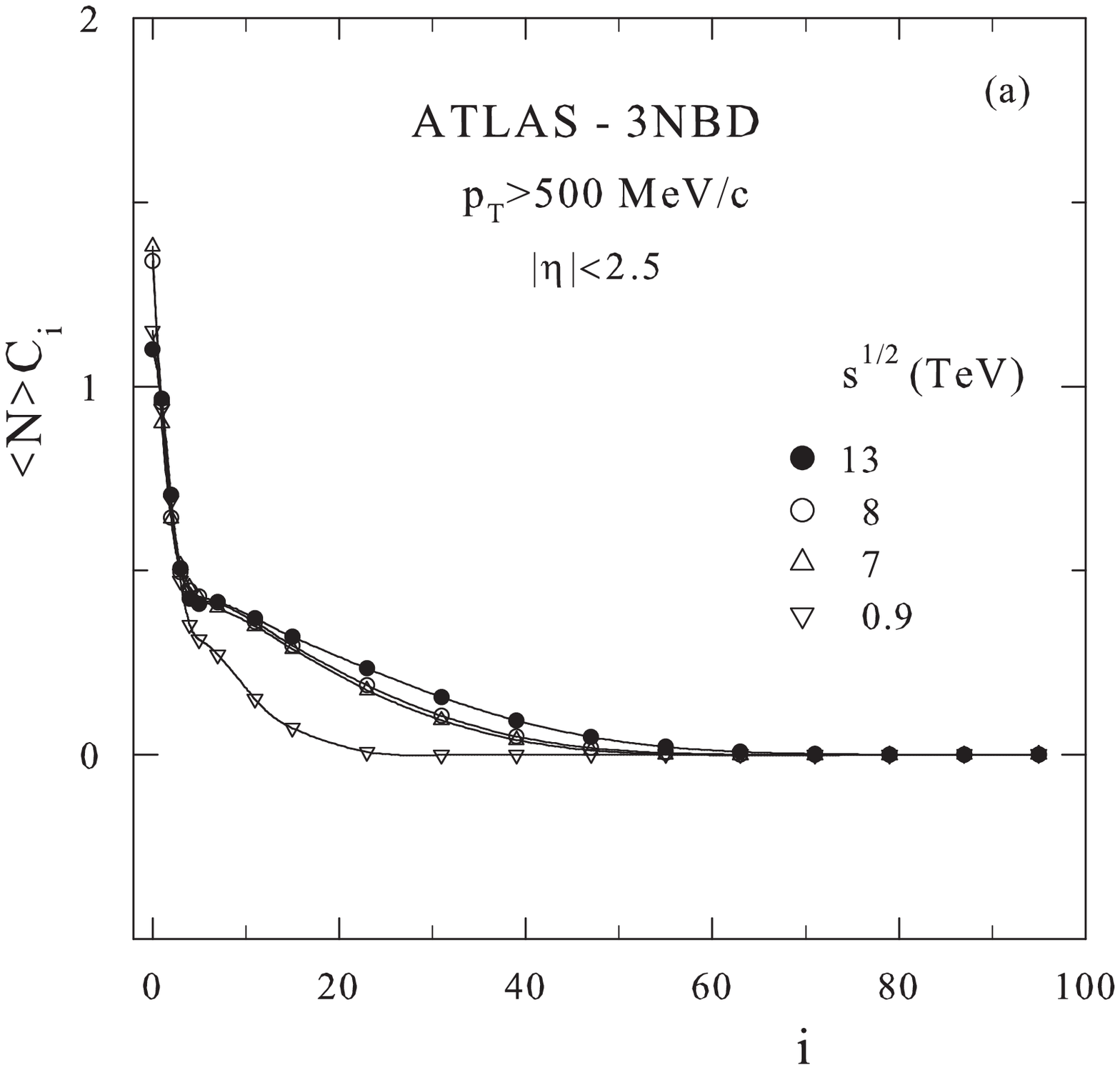}
\includegraphics[width=78mm,height=78mm]{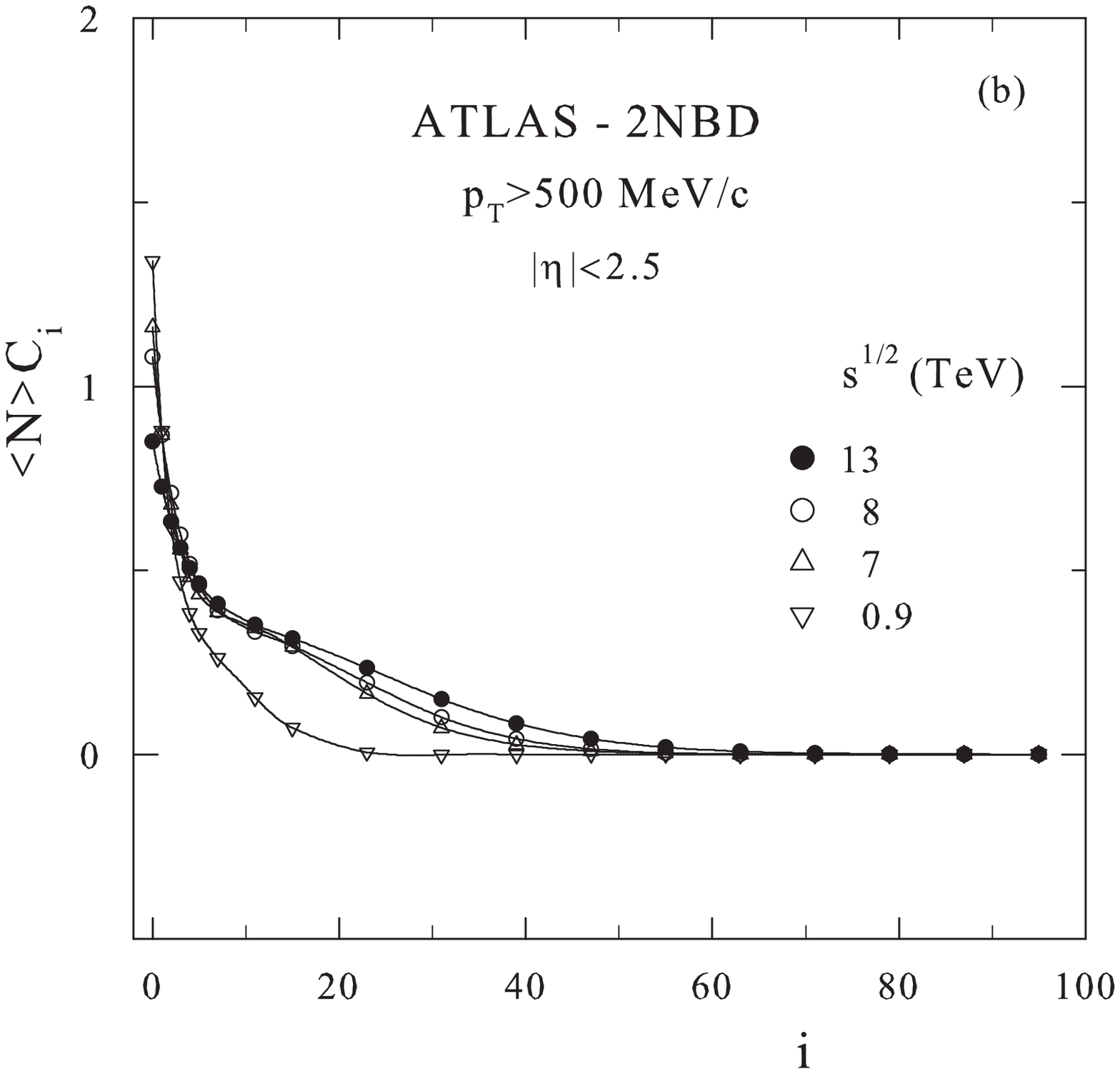}
\end{center} 
\vskip -0.8cm
\caption{
Energy dependence of the cumulative combinants $\langle N\rangle C_i$  
calculated from a weighted superposition of {\bf a} three and {\bf b} two NBDs 
fitted to the charged-particle MDs 
\protect\cite{ATLAS1,ATLAS2,ATLAS3}
measured by the ATLAS Collaboration in the interval $|\eta|<2.5$
for $p_T>500$~MeV/c  
}
\label{fig:8}       
\end{figure*}
\begin{figure*}
\begin{center}
\includegraphics[width=78mm,height=78mm]{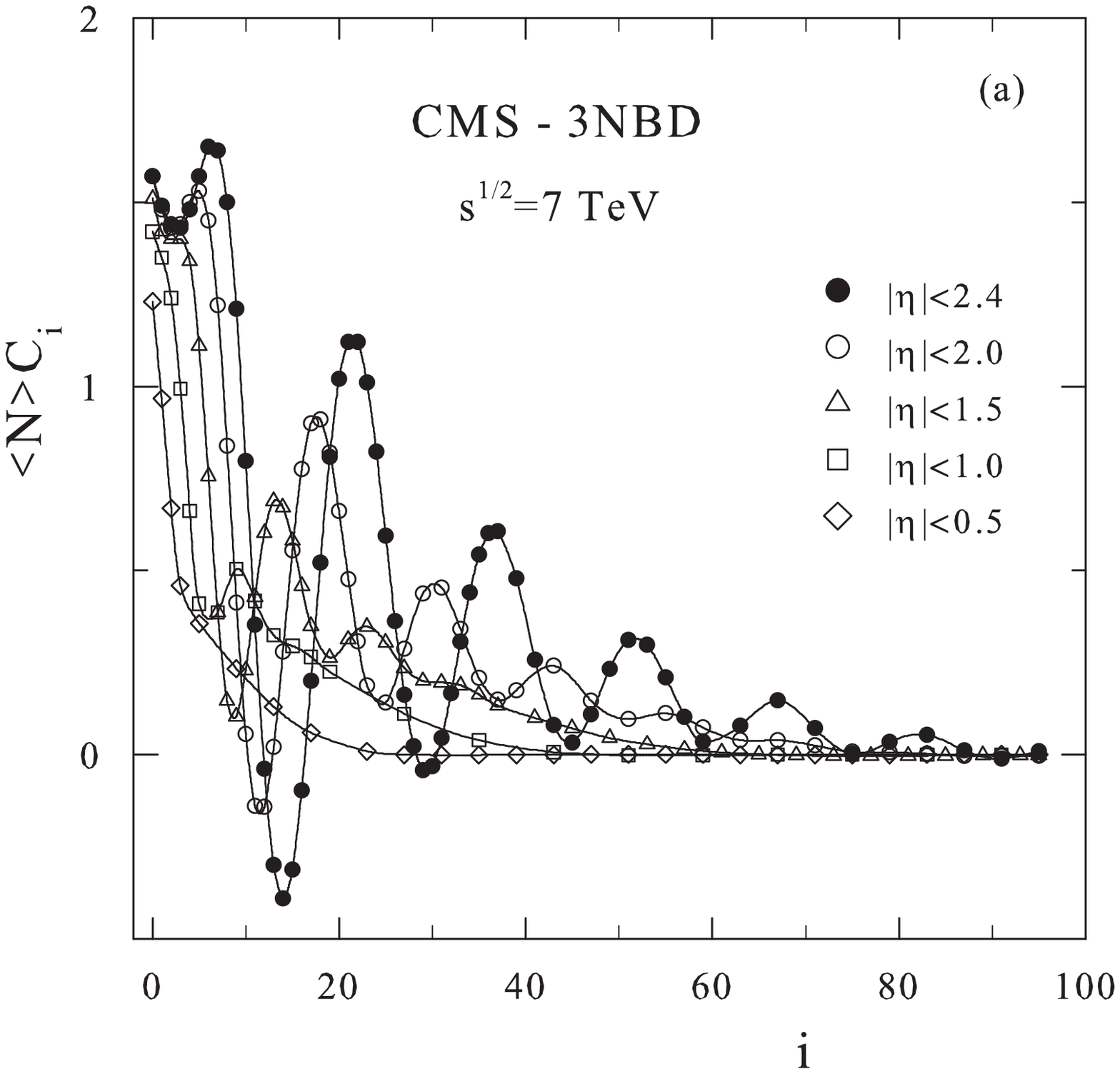}
\includegraphics[width=78mm,height=78mm]{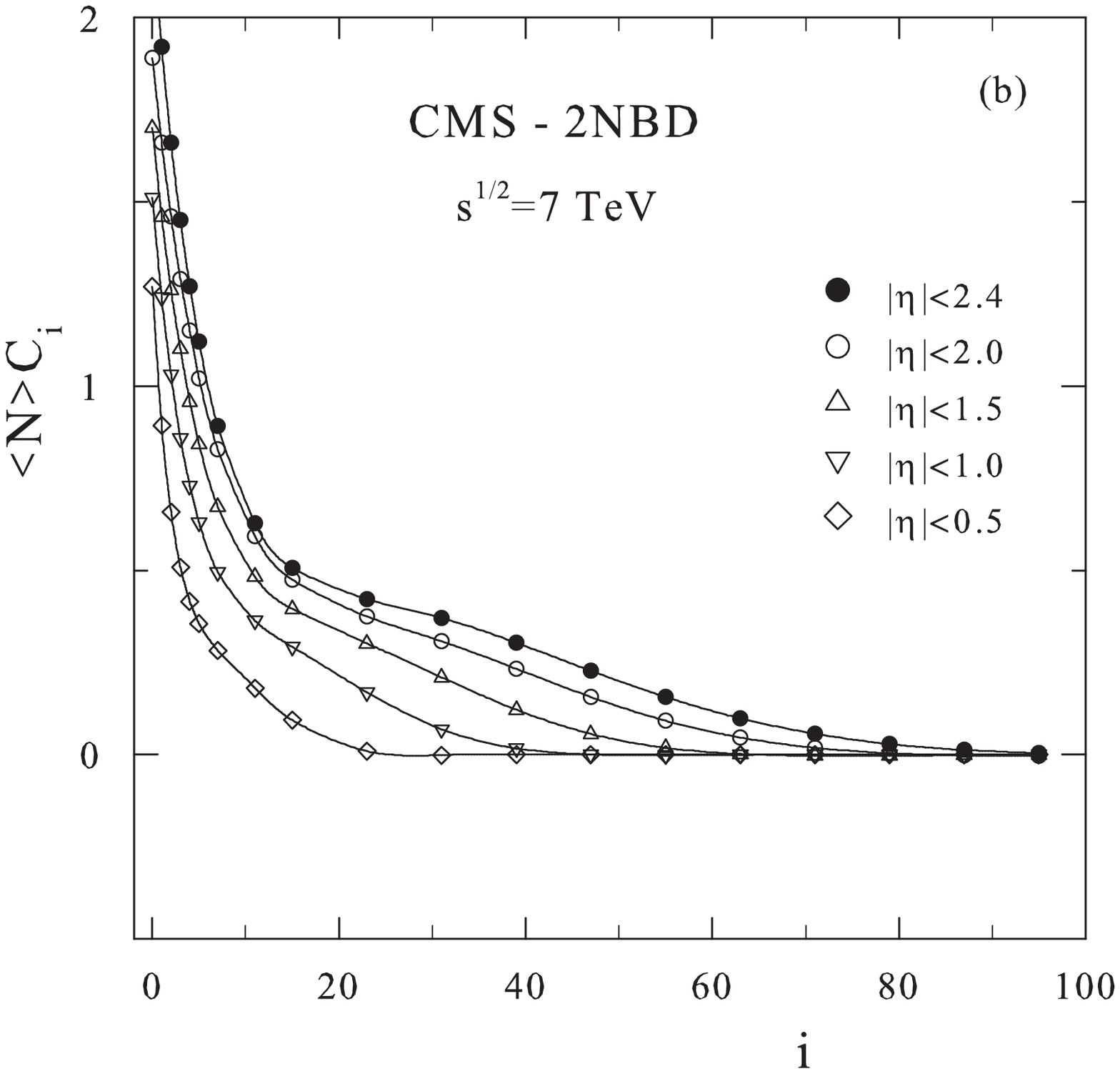}
\end{center} 
\vskip -0.8cm
\caption{
The cumulative combinants $\langle N\rangle C_i$  
calculated from a weighted superposition of {\bf a} three and {\bf b} two NBDs 
fitted to the charged-particle MDs 
\protect\cite{CMS}
measured by the CMS Collaboration at $\sqrt s=$7~TeV for $p_T>0$ 
in different pseudorapidity intervals 
}
\label{fig:9}       
\end{figure*}

Next we show that oscillations of the cumulative combinants
$\langle N\rangle C_i=(i\!+\!1){\cal C}(i\!+\!1)$ 
can arise from the three-NBD superposition fitted to the data with the third 
component, which accounts for the peak at low~$n$.
We calculate the coefficients $\langle N\rangle C_i$ 
from Eq. (\ref{eq:r4}) using 
the values of $P(n)$ obtained from the corresponding fits. 

Figure \ref{fig:7} shows the quantities 
$\langle N\rangle C_i$ in dependence on the rank $i$ 
calculated from a weighted superposition of (a) three  
and (b) two NBDs used to fit the data on MD
measured by the ATLAS Collaboration in the interval $|\eta|<2.5$
with the transverse momentum cut $p_T>100$~MeV/c.
The parameters of the fitted functions 
are taken from Table 1 of \cite{IZ} (for $\sqrt s=$0.9 and 7~TeV)
and from Tables \ref{tab:1} and \ref{tab:2} of the present paper (for $\sqrt s=$8 and 13~TeV).
One can see that two-NBD fits to the ATLAS data give a smooth behavior of the cumulative combinants
in dependence on their rank.
In contrast to this, the three-NBD fits to the ATLAS data are characterized by the oscillations 
of $\langle N\rangle C_i$ at low $i$.
The shape of the first oscillation is nearly energy independent. 
The position of the first minimum corresponds to the energy independent 
value of the average multiplicity $\bar{n}_3\sim 11$ 
of the third NBD component of the total distribution. 

As discussed in \cite{WW1}, the oscillations can be connected with memory 
(or correlations) in the produced multi-particle system. 
The correlation of particles in the minima is weaker. 
The first minimum
corresponds to the production of particles within the third independent NBD component which 
makes the average correlation of the total system weaker. The consecutive oscillations  
represent non-monotonic loss of memory (or correlations) away from a considered multiplicity $n$.   

The cumulative combinants obtained from 
a weighted sum of three and two NBDs used to fit the ATLAS data 
in the interval $|\eta|<2.5$
with the cut $p_T>500$~MeV/c are shown in Fig.~\ref{fig:8}a, b, respectively.
One can see that the oscillations have disappeared for the data sample with 
$p_T>500$~MeV/c  even for the three-NBD description. 
This is despite the fact that a three-NBD fit to the data is still needed to account 
for the peak at low multiplicities (see Fig.~\ref{fig:3}c, d).
The disappearance of the oscillations is caused by the small value of $\bar{n}_3\sim 3$  
(cf. the diminishing of the oscillations for small windows in Fig.~\ref{fig:9}a where  
$\bar{n}_3=2.7$ for $ |\eta|<$0.5).

The data on MD of charged particles measured by the CMS Collaboration 
in the central interaction region \cite{CMS} allow one to study 
the cumulative combinants in dependence on the pseudorapidity window.
As shown in \cite{WW1}, the amplitude and periodicity of the oscillations increase with 
the window size. For small windows they practically vanish.
Figure \ref{fig:9}a demonstrates that similar properties of the cumulative combinants 
can be obtained from three-NBD fits \cite{IZ} to the CMS data. 
As seen from Fig.~\ref{fig:9}b, 
the two-NBD fits \cite{IZ} to the CMS data do not give any oscillations.
The cumulative combinants calculated from weighted superposition of two NBDs used to fit the data
are decreasing functions of the rank $i$. 
In contrast, the combinants calculated from three-NBD fits to the data 
reveal oscillations at low $i$. 
The magnitude of the oscillations decreases with the decreasing window size. 
The larger periodicity is in the larger windows. 
The first minimum
corresponds to production of particles within the third-NBD component, which 
makes the average correlation of the total system weaker. 
Its position (for larger windows) corresponds to the pseudorapidity dependence 
of the average multiplicity $\bar{n}_3$ (see Table 4 of \cite{IZ})
of the third-NBD component of the total distribution. 
Let us note that, for the CMS data, the multiplicity $n=0$ was excluded from the fits. 
Therefore the cumulative combinants shown in Fig.~\ref{fig:9} are calculated  
with the parameters quoted in Tables 4 and 5 of \cite{IZ}
without any correction to the value of $P(0)$ obtained from the fits. 

We have studied the sensitivity of the oscillations of the cumulative combinants 
to changes of the
probabilities $\alpha_i$ of single NBD components of the total distribution.
For this purpose, we used auxiliary parameters $\epsilon_i$ by which 
the corresponding probabilities $\alpha_i$ are gradually turned down. 
In order to keep the total probability equal to unity, 
the attenuation of $\alpha_1$ with $\epsilon_{1} < 1$ is compensated 
by the proportional amplification of $\alpha_2$ and $\alpha_3$, which
is realised as follows:
\begin{equation}
\alpha^{'}_{1}(\epsilon_1)=\alpha_1\epsilon_1,\ \ \
\alpha^{'}_{i}(\epsilon_1)=\alpha_i \frac{(1-\alpha^{'}_{1})}{(1-\alpha_1)},\ \ \
i=2,3.
\label{eq:r5}
\end{equation}
Similar equations were used for weakening of the second and 
the third component by $\epsilon_2$ and $\epsilon_3$, respectively.

\begin{figure}
\includegraphics[width=78mm,height=78mm]{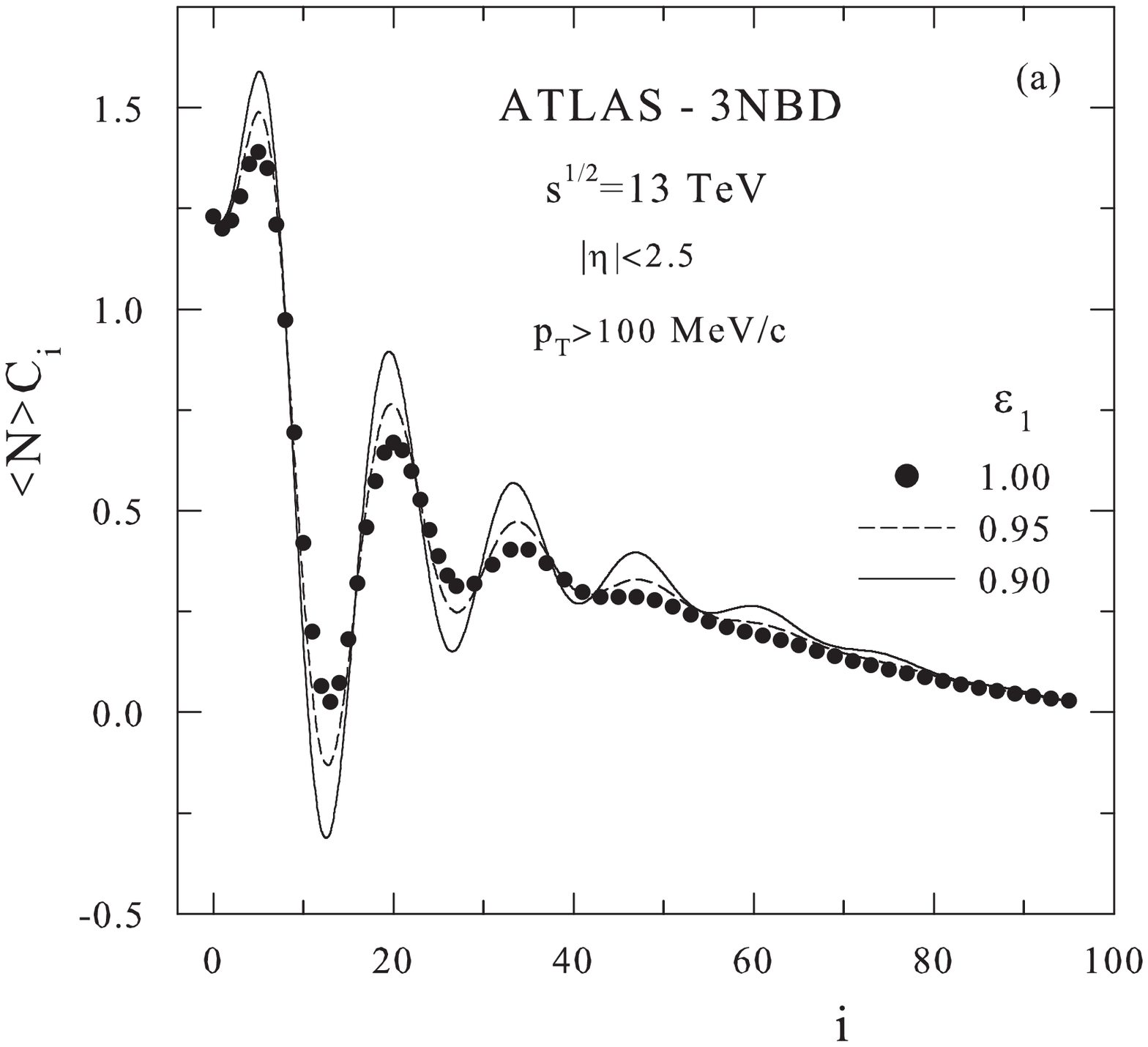}
\vskip -0.7cm
\includegraphics[width=78mm,height=78mm]{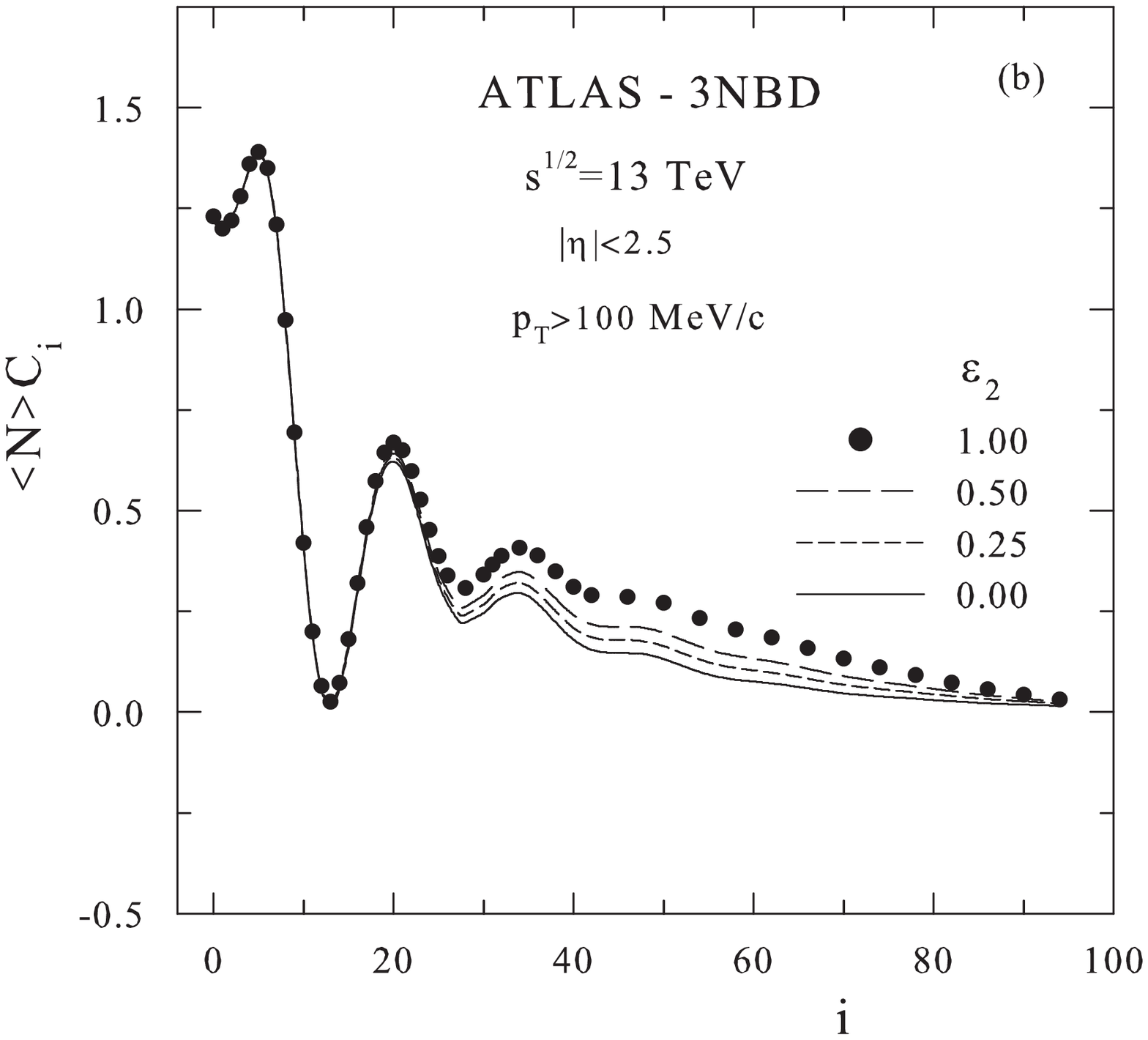}
\vskip -0.7cm
\includegraphics[width=78mm,height=78mm]{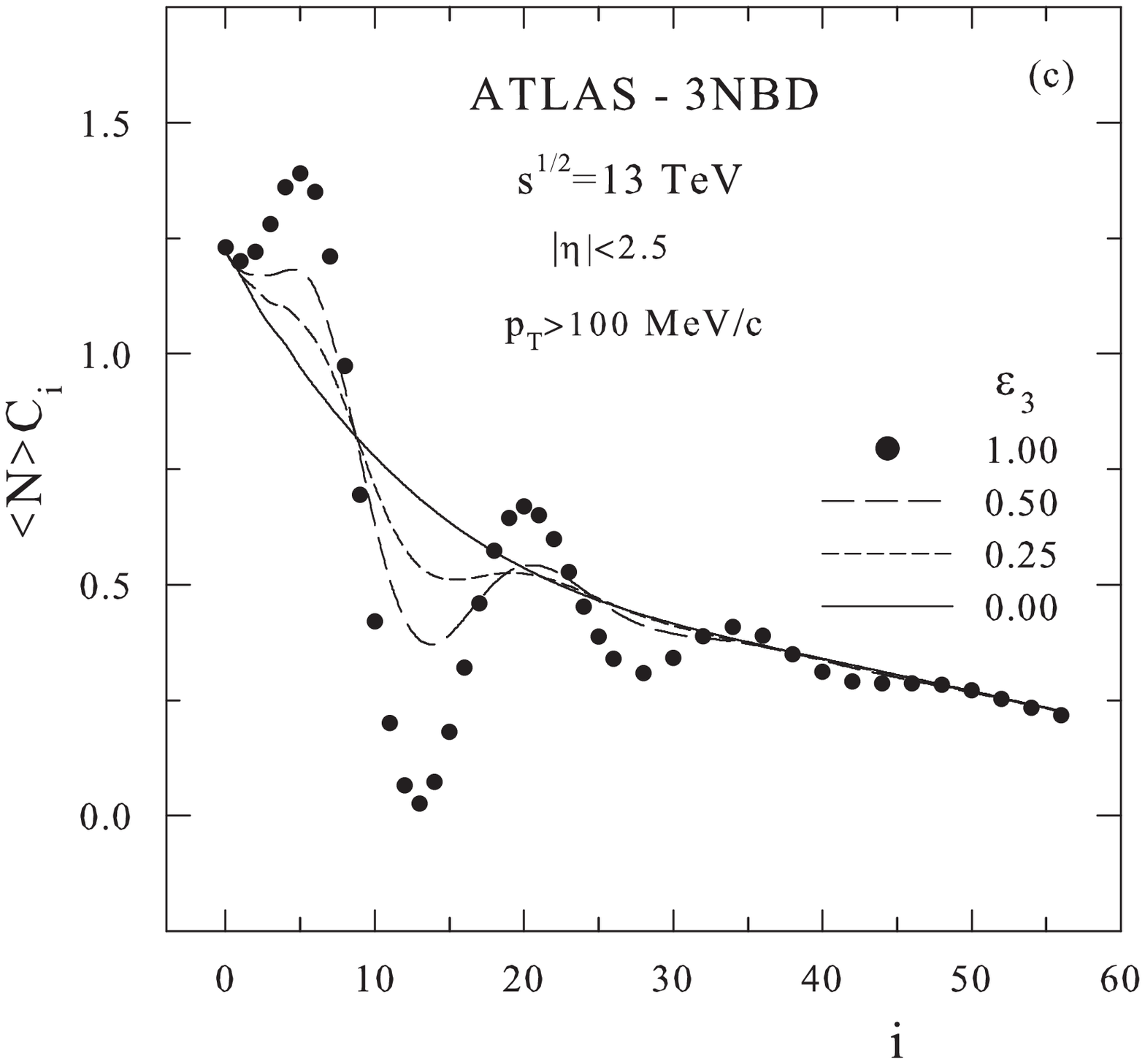}
\vskip -0.6cm
\caption{
The cumulative combinants $\langle N\rangle C_i$  
calculated from three-NBD fit to the ATLAS data \protect\cite{ATLAS4} at $\sqrt s=$13~TeV 
(black circles from Fig.~\ref{fig:7}a).
The modifications of $\langle N\rangle C_i$ (lines) with reduction 
of the {\bf a} first, {\bf b} second, and {\bf c} third NBD component (see text) 
}
\label{fig:10}       
\end{figure}

Figure \ref{fig:10}a demonstrates the increase of the amplitude of oscillations when 
the probability $\alpha_1$ of the main NBD component is reduced by 5 or 10\%. The locations 
of the minima and maxima of the oscillations remain intact as the other parameters were unchanged.
Figure \ref{fig:10}b depicts the sensitivity of the cumulative combinants to the probability 
$\alpha_2$ of the NBD component under the tail of the total distribution. 
One can see that the first wave up to $n\sim 24$ is not affected, even for $\epsilon_2=0$.
The behavior of the cumulative combinants in dependence on $\epsilon_3$ is shown
in Fig.~\ref{fig:10}c. One can see that the amplitudes of the oscillations diminish 
with decreasing probability $\alpha^{'}_{3}$ of the third NBD component at low multiplicities. 
The oscillations disappear completely in the case of $\alpha^{'}_3=0$.

The uncertainties of the quantities $\langle N\rangle C_i$
and, more specifically, correlation of their size with the pattern of their oscillations are sensitive to statistical fluctuations and systematic uncertainties of the raw data to such an 
extent that the wavy structure in $\langle N\rangle C_i$ may be rendered 
insignificant given sufficiently large experimental uncertainties,
as demonstrated above by varying the fitted three-NBD parameters.
This will be further confirmed by direct extraction of 
modified versions of the cumulative combinants from experimental data on MDs in Sect.~\ref{sec:G}.

The natural applicability of the method of combinants 
to the study of fine structure of the MDs  appears to be useful.
This together with the relation of the combinants to the
finite set of multiplicities which make up the data seems to be an appropriate tool 
in justification of the existence of the third multiplicity component emerging  
in $pp$ collisions at the LHC at low
multiplicities.

\subsection{Clan structure of three-NBD description} 
\label{sec:F}

The correlation structures of multi-particle states created in high energy collisions are often studied
in the framework of the clan concept introduced to interpret NBD occurring in 
different reactions over a wide range of energies.    
In hadron-hadron collisions, the properties of clans  have been investigated and discussed 
within a two-component model used to describe the MD of the produced particles 
by a superposition of two NBDs \cite{GiovUgo1,GiovUgo4}.
The first component, connected  with soft events, reveals invariant properties 
as a function of the c.m. energy \cite{CDF}.  
The second component, under the shoulder of the MD at high $n$, is associated with semi-hard processes
including the production of mini-jets.
It is believed that  mini-jets with low $p_T$ are an important part of
particle production at high energies~\cite{Minijets}.

The clan model is based on the decomposition of NBD to Poisson distributed clusters  
or independent bunches of produced particles \cite{GioVanHove,SimakSumbera}.
According to the common definition, a single cluster (clan) contains at least one particle.
The number of particles inside the clan follows a logarithmic distribution.
Among possible interpretations of the logarithmic distribution 
let us mention the time average of time-dependent cascading \cite{GioVanHove}
and a model of cascade processes with collapsing of clans \cite{IZ_clans}.   
The average number of clans, $\bar{N}$, 
and the average number of particles per clan, $\bar{n}_c$, are expressed in terms of the
NBD parameters $\bar{n}$ and $k$ as follows:
\begin{equation}
\bar{N}=k\ln\left(1+\frac{\bar{n}}{k}\right), \ \ \ \ \ \bar{n}_c=\frac{\bar{n}}{\bar{N}}.
\label{eq:r6}
\end{equation}
For a weighted superposition of NBDs, the clan characteristics of a single NBD depend 
on the number of components used for a description of the experimental data.
The analysis of the data in terms of 
two and three NBDs is instructive,
because it allows one to study changes of clan parameters 
with the emergence of a distinct peak observed in MD at the LHC at low $n$.

\begin{figure}
\includegraphics[width=78mm,height=78mm]{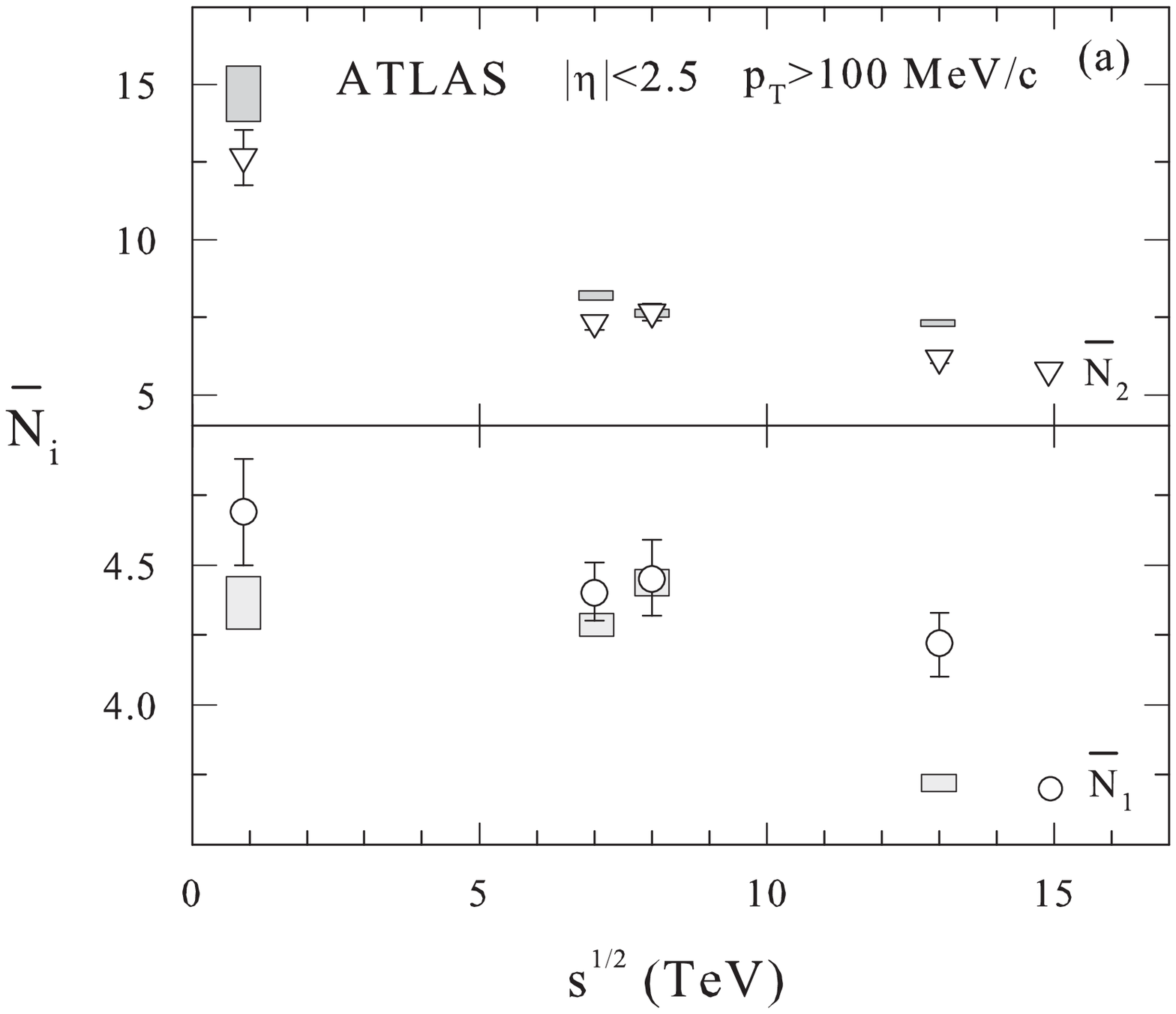}
\vskip -1.8cm
\includegraphics[width=78mm,height=78mm]{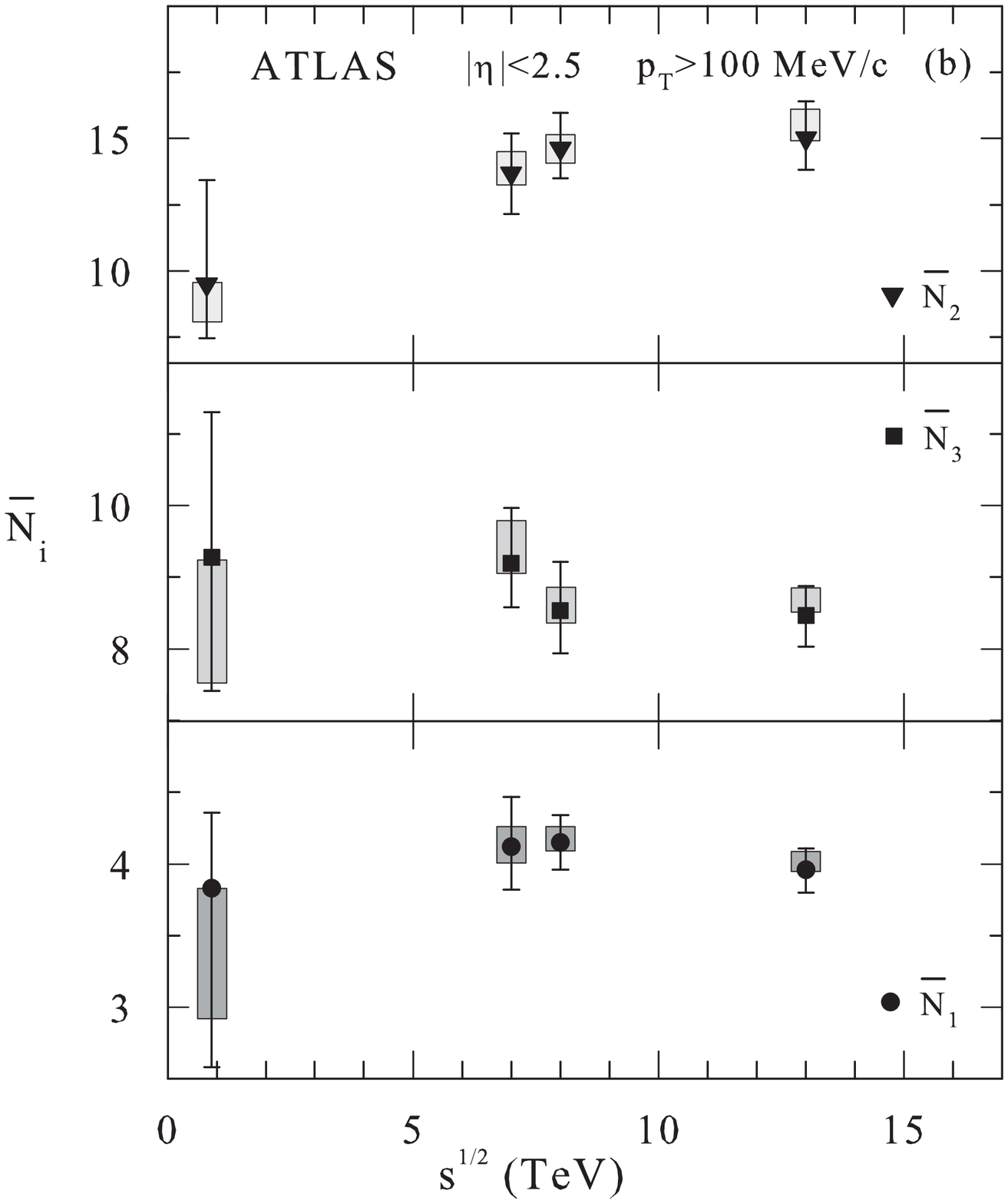}
\caption{
Energy dependence of the average number of clans $\bar{N}_i$ 
calculated from a weighted superposition of {\bf a} two and {\bf b} three NBDs 
fitted to the charged-particle MDs 
\protect\cite{ATLAS1,ATLAS2,ATLAS4}
measured by the ATLAS Collaboration in the interval $|\eta|<~2.5$
for $p_T>100$~MeV/c. 
The symbols with error bars and shaded rectangles correspond to the parameter values 
obtained by minimization of 
Eqs. (\ref{eq:a1}) and (\ref{eq:a4}), respectively   
}
\label{fig:11}       
\end{figure}

\begin{figure}
\includegraphics[width=78mm,height=78mm]{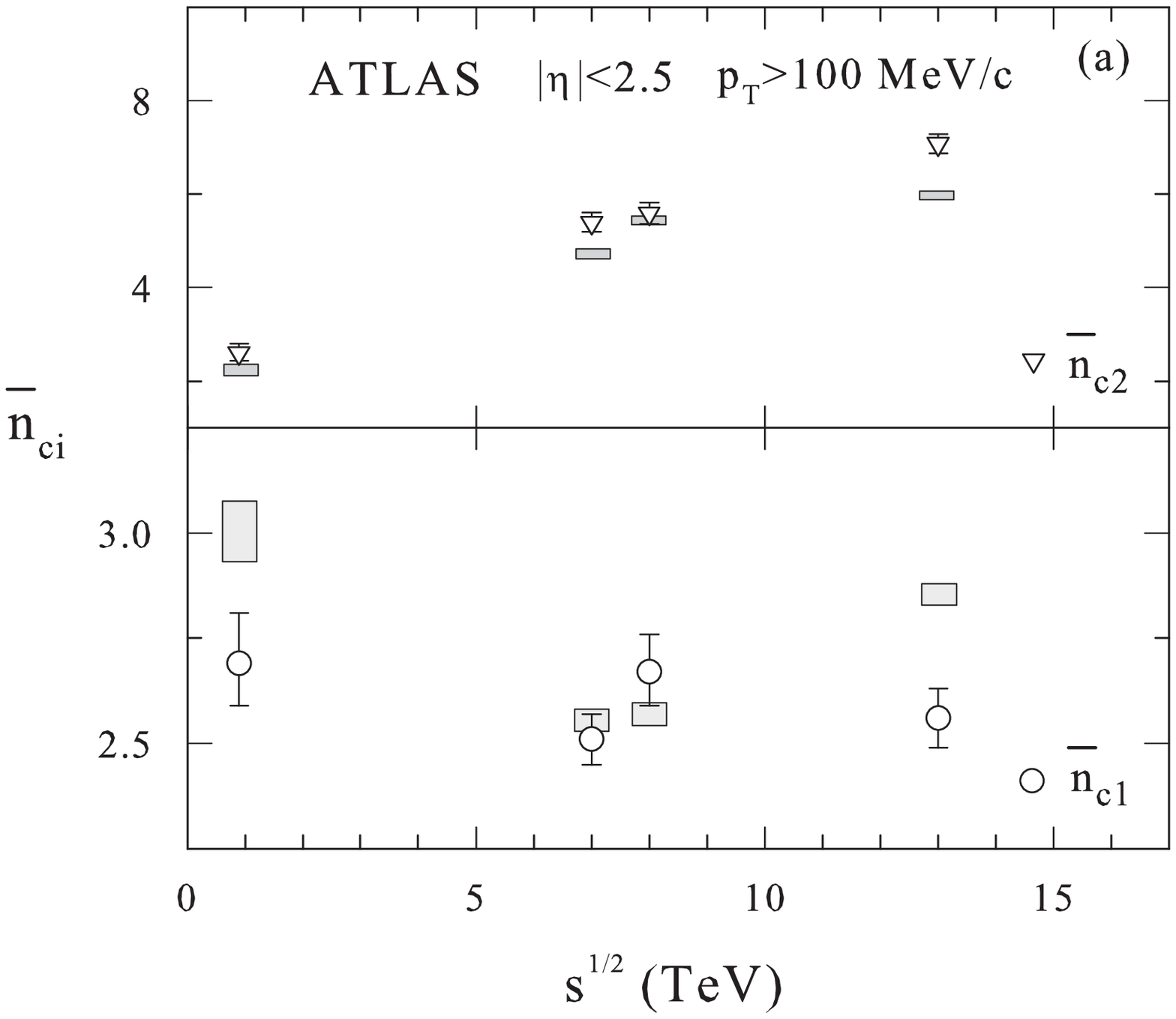}
\vskip -1.8cm
\includegraphics[width=78mm,height=78mm]{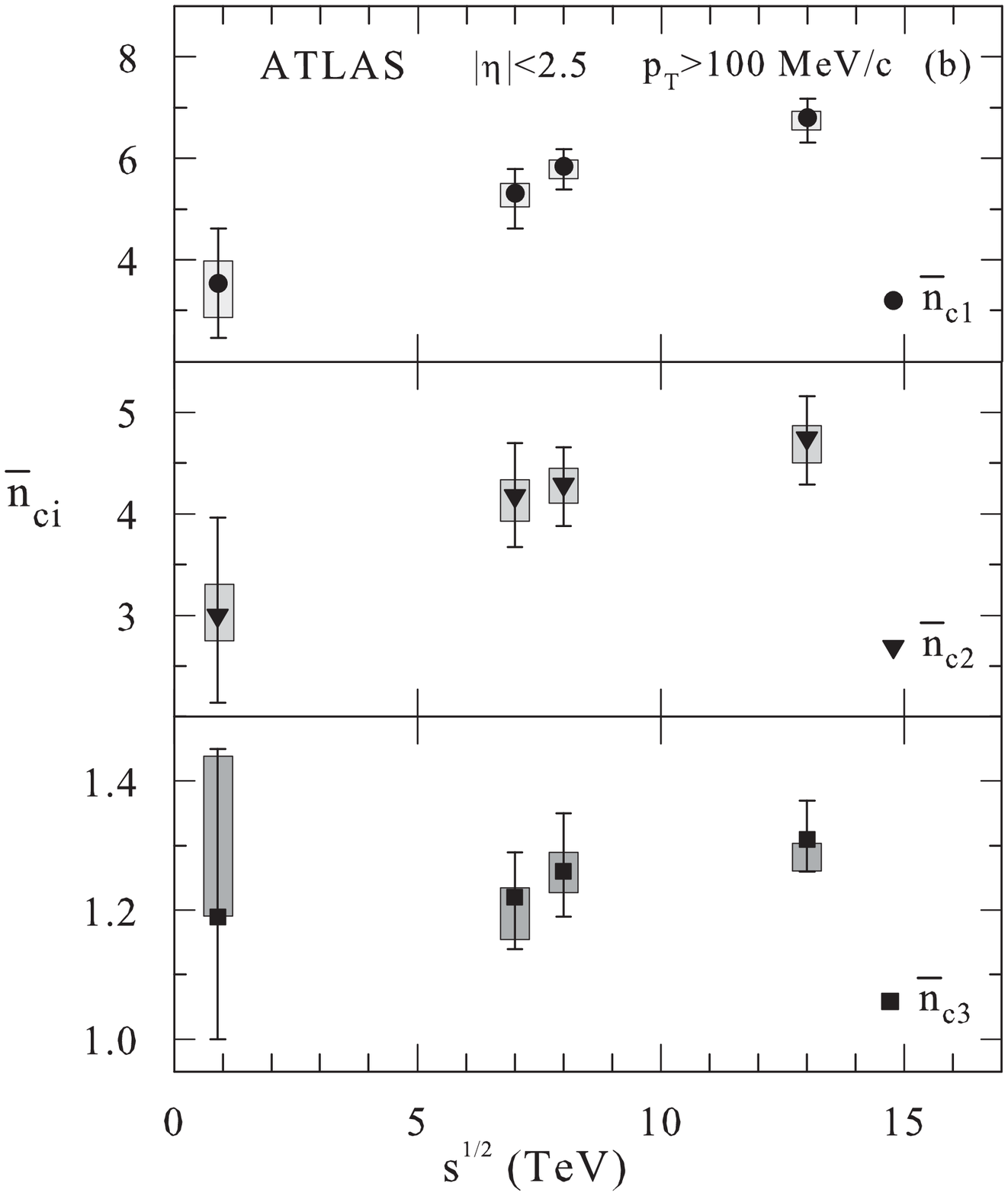}
\caption{
Energy dependence of the average number of particles per clan $\bar{n}_{ci}$ 
calculated from weighted superposition of {\bf a} two and {\bf b} three NBDs 
fitted to the charged-particle MDs 
\protect\cite{ATLAS1,ATLAS2,ATLAS4}
measured by the ATLAS Collaboration in the interval $|\eta|<2.5$
for $p_T>100$~MeV/c.  
The symbols with error bars and shaded rectangles correspond to the parameter values 
obtained by minimization of 
Eqs. (\ref{eq:a1}) and (\ref{eq:a4}), respectively  
}
\label{fig:12}       
\end{figure}

Figure \ref{fig:11}a shows the $\sqrt s$ dependence of the average number of clans calculated 
from a weighted superposition of two NBDs. The parameters were calculated from fits 
to the ATLAS data  
measured in the interval  $|\eta|<2.5$ for $p_T>100$~MeV/c. 
The symbols with error bars and shaded rectangles are given by parameter values 
using Eqs. (\ref{eq:a1}) and (\ref{eq:a4}), respectively. 
As one can see, the average number of clans of the first component, $\bar{N}_1$, 
is approximately constant with energy (except the last shaded rectangle corresponding 
to large value of $\widetilde{\chi}^2/dof$ = 776/81).
A similar observation \cite{Ghosh} follows from the measurements  of the CMS Collaboration \cite{CMS}
in the interval $|\eta|<2.4$.
The most peculiar feature of the two-NBD description is the decrease of the average number of 
clans $\bar{N}_2$ of the semi-hard component with energy (upper panel in Fig. \ref{fig:11}a).  
This property, confirmed by the analysis \cite{Ghosh} of CMS data, 
was discussed \cite{GiovUgo4} within 
different scenarios for soft and semi-hard events.

The $\sqrt s$ dependence of the average number of clans 
is quite different in the three-NBD model. 
As seen from Fig. \ref{fig:11}b,
the values of $\bar{N}_1$ for the dominant component with largest probability $\alpha_1$
are nearly constant at high energies. 
The average number of clans
in the second component, $\bar{N}_2$, increases with energy 
and becomes much larger than $\bar{N}_1$ at high $\sqrt s$. 
This is a new feature, substantially different from the parametrization of the data by two NBDs 
(cf. the upper panel in Fig. \ref{fig:11}a).
Another observation is that the values of $\bar{N}_3$ corresponding 
to the third NBD are relatively high and nearly constant.
In a certain sense, the energy independence of $\bar{N}_1$ and $\bar{N}_3$ 
substitutes the approximate constancy of the average number of clans 
of the soft component in the two-NBD scenario
(cf. the lower panel in Fig. \ref{fig:11}a).

Figure \ref{fig:12} shows the average number of particles per clan, $\bar{n}_{ci}$,
calculated from a weighted superposition of (a) two and (b) three NBDs 
fitted to the ATLAS data in the interval $|\eta|<2.5$
for $p_T>100$~MeV/c.
In the two-NBD scenario (Fig. \ref{fig:12}a), 
$\bar{n}_{c1}$ for the soft component does not change much with energy.
The value of $\bar{n}_{c2}$ grows with energy and becomes larger than $\bar{n}_{c1}$.
Such a trend 
is compensated with the decrease of the average number of the semi-hard clans with $\sqrt s$, 
as is illustrated in the upper panel in Fig. \ref{fig:11}a.

\begin{figure*}
\begin{center}
\includegraphics[width=78mm,height=78mm]{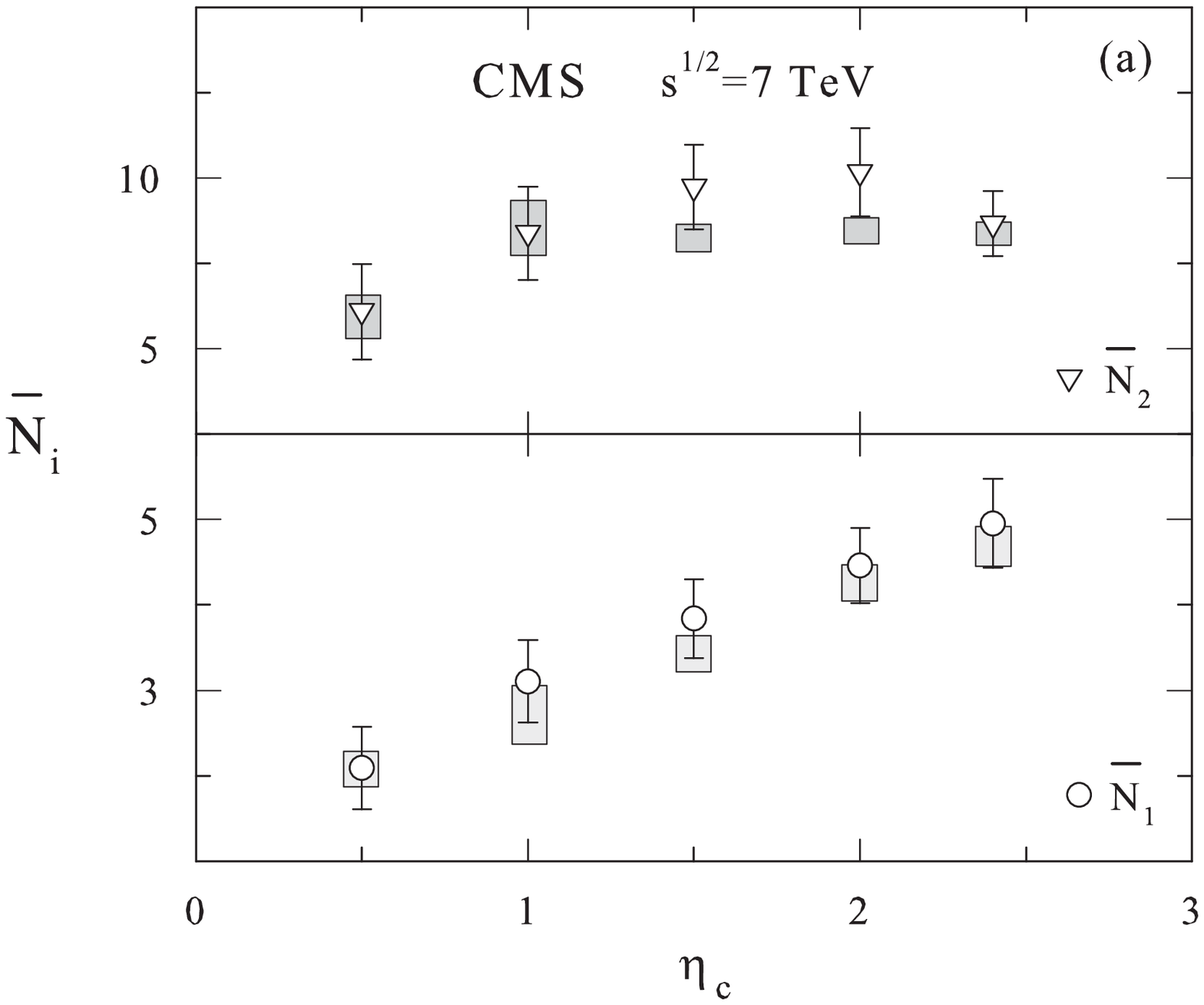}
\includegraphics[width=78mm,height=78mm]{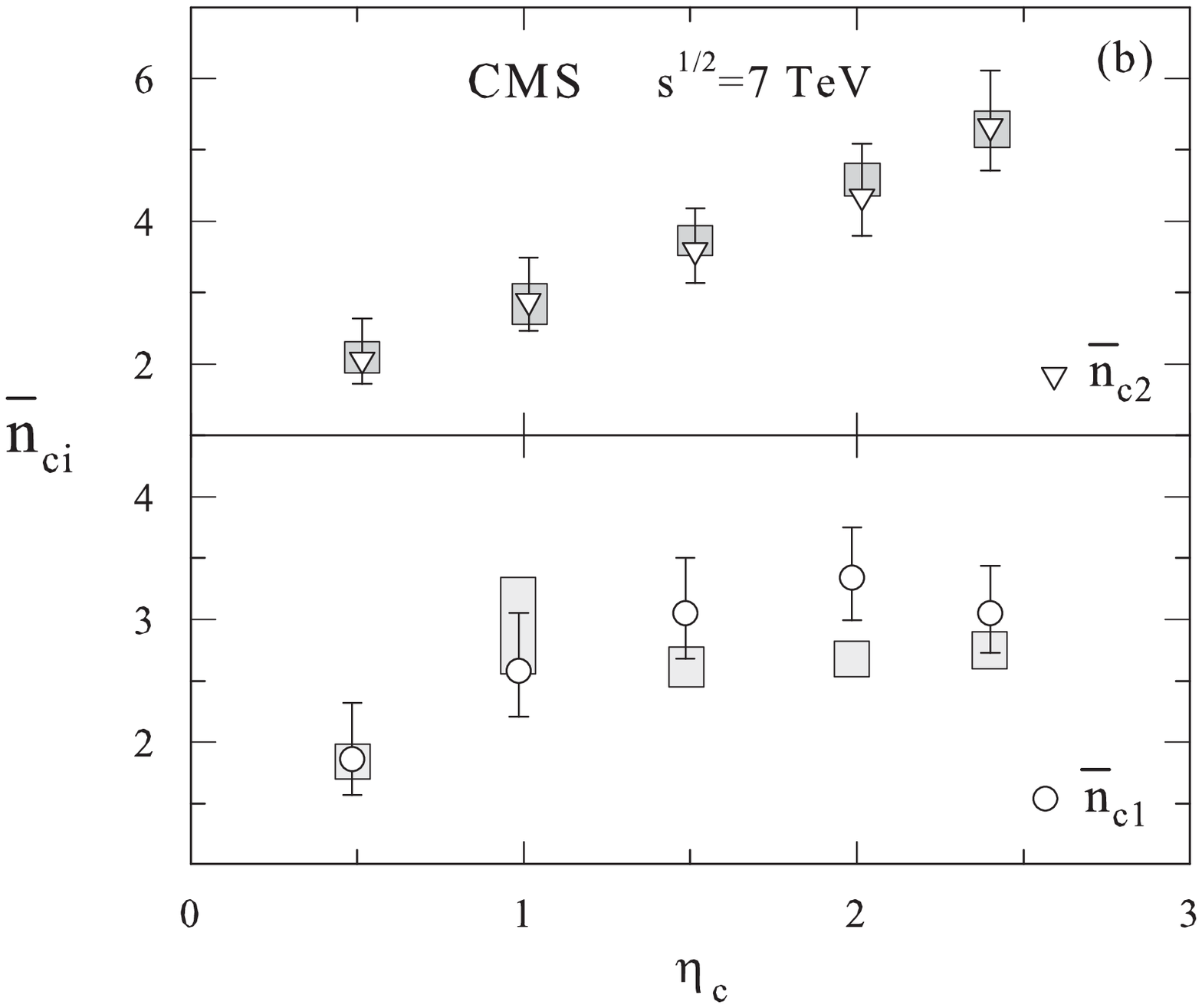}
\end{center} 
\vskip -2.2cm
\caption{
The $\eta_c$ dependence of {\bf a} the average number of clans $\bar{N}_{i}$ and {\bf b} 
the average number of charged particles per clan, $\bar{n}_{ci}$, for a two-NBD description
of the CMS data \protect\cite{CMS}
on MD measured in different pseudorapidity intervals $|\eta|<\eta_c$ at $\sqrt s =$ 7 TeV.
The symbols with error bars and shaded rectangles correspond to the parameter values 
obtained by minimization of 
Eqs. (\ref{eq:a1}) and (\ref{eq:a4}), respectively  
}
\label{fig:13}       
\end{figure*}
\begin{figure*}
\begin{center}
\includegraphics[width=78mm,height=78mm]{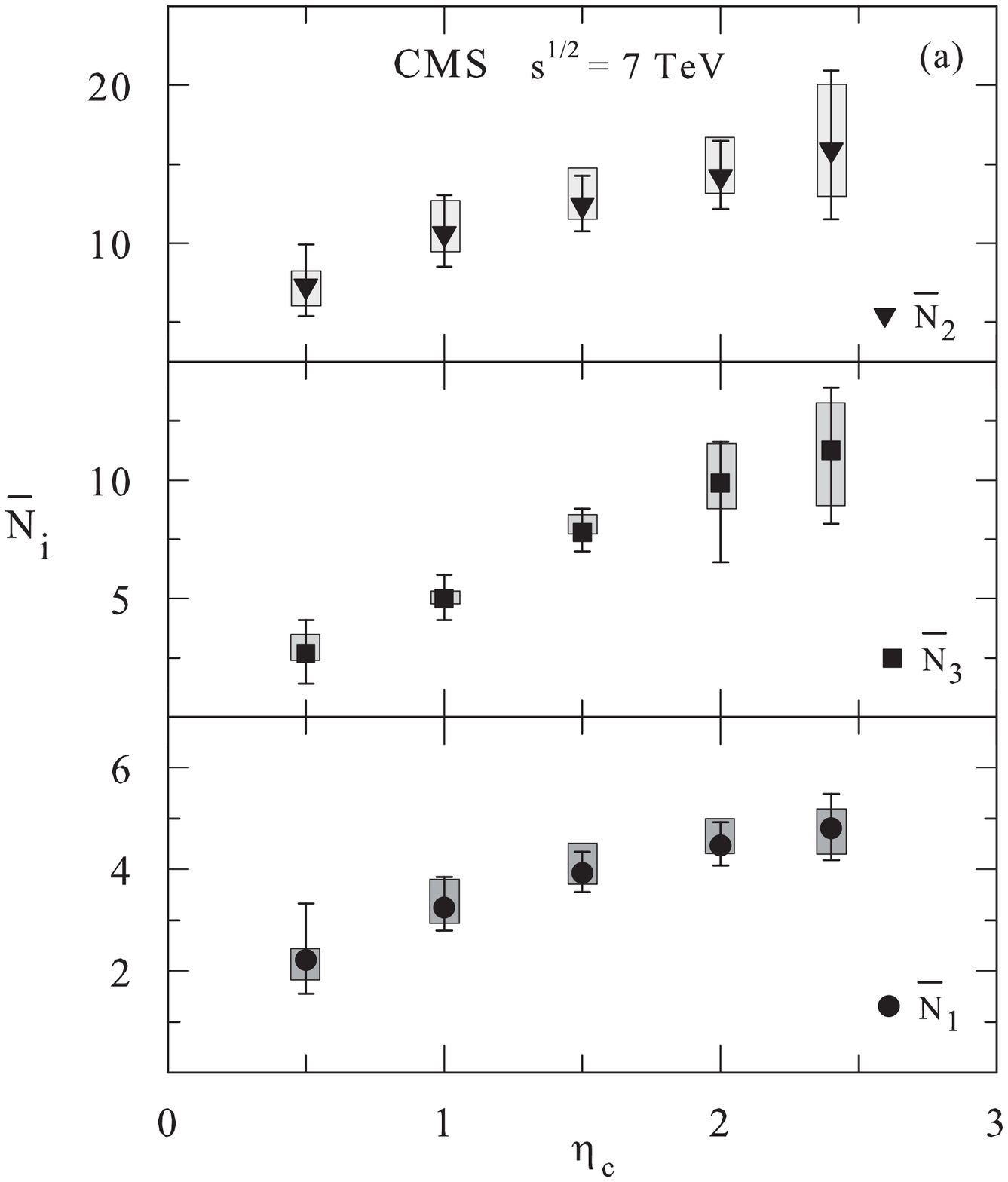}
\includegraphics[width=78mm,height=78mm]{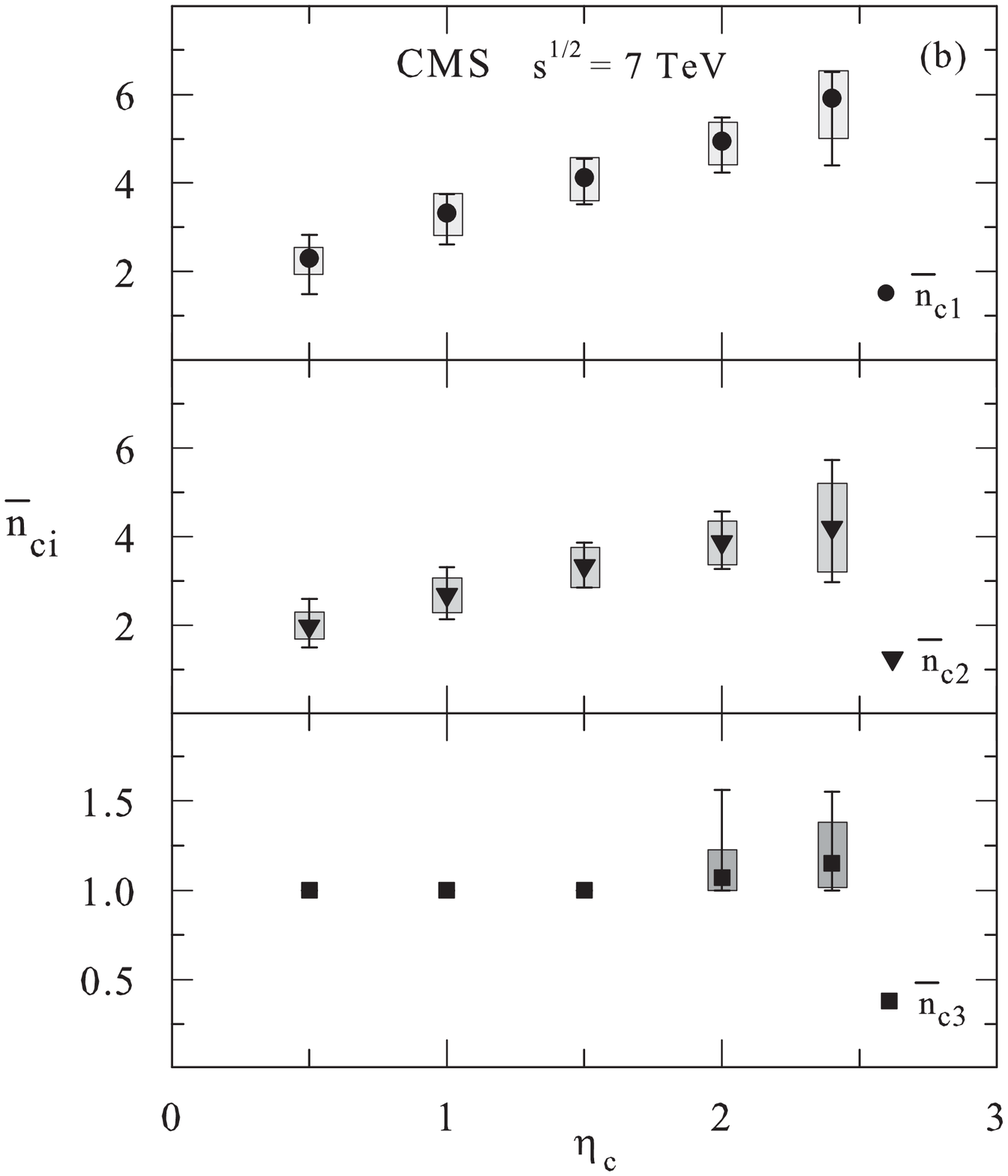}
\end{center} 
\vskip -0.5cm
\caption{
The $\eta_c$ dependence of {\bf a} the average number of clans $\bar{N}_{i}$ and {\bf b} 
the average number of charged particles per clan,~$\bar{n}_{ci}$, for a three-NBD description
of the CMS data \protect\cite{CMS}
on MD measured in different pseudorapidity intervals $|\eta|<\eta_c$ at $\sqrt s =$~7~TeV.
The symbols with error bars and shaded rectangles correspond to the parameter values 
obtained by minimization of 
Eqs. (\ref{eq:a1}) and (\ref{eq:a4}), respectively   
}
\label{fig:14}       
\end{figure*}

The situation becomes completely different for the three-NBD description.
As demonstrated in Fig. \ref{fig:12}b,
the average number of particles  inside clans of the first (dominant) NBD component, $\bar{n}_{c1}$,
increases rapidly with~$\sqrt s$. 
The values of $\bar{n}_{c2}$ of the second 
component under the tail of MD reveal growth with energy as well, but not as rapid as in Fig. \ref{fig:12}a.
The substantial difference is that, contrary to the two-NBD description, $\bar{n}_{c1} > \bar{n}_{c2}$
in the multi-TeV energy domain. 
The average number of particles inside clans of the third NBD component, $\bar{n}_{c3}$, is only slightly
larger than unity and depends weakly on~$\sqrt s$.  

Based on the performed analysis of the ATLAS data with 
weighted superposition of three NBDs, 
we draw the following conclusions concerning the clan properties of the first 
and the second NBD component of the total distribution.
The first component, corresponding to the dominant class of events, 
contains large clans consisting of many particles. 
The number of particles inside the clans grows strongly with energy.
The average number of such clans is small ($ \bar{N}_1\simeq~4$) and independent of $\sqrt s$. 
The small and constant number of clans suggests 
that there are few (or no) ``mini-jet clans" in these events.

The multiplicity component under the tail of the distribution corresponds to  
a class of events in which much more clans are produced than in the first, dominant one.
The average number of clans of the second component increases with energy 
and reaches large values ($ \bar{N}_2\sim~15$) at high $\sqrt s$.
The corresponding number of particles per clan $\bar{n}_{c2}$ is somewhat smaller 
than $\bar{n}_{c1}$ in the high energy region. 
It grows with $\sqrt s$ as well.
The large, increasing number of clans and the growing number of particles per
clan suggest that there are many clans ``mini-jets" in this class of events.
The properties of the second multiplicity component under the tail of the MD are 
characteristic for semi-hard processes with abundant production of mini-jets.

We have studied the pseudorapidity dependence of the clan parameters 
using the data \cite{CMS} 
measured by the CMS Collaboration in the intervals $ |\eta|< \eta_c =$ 0.5, 1.0, 1.5, 2.0, and 2.4 
at $\sqrt s =$ 7~TeV. 
Figure \ref{fig:13} shows (a) the average number of clans 
$\bar{N}_{i}$ and (b) the average number of particles per clan $\bar{n}_{ci}$
calculated from two-NBD description \cite{IZ} of the CMS data in 
dependence on the width of the pseudorapidity window.
The behavior of the parameters calculated from three-NBD fits \cite{IZ} to the same data
is depicted in Fig. \ref{fig:14}.

As one can see from Figs. \ref{fig:13}a and \ref{fig:14}a, 
the $\eta_c$ dependence of the average number of clans of the first component, $\bar{N}_{1}$,
is nearly the same for two- and three-NBD parametrization. 
Despite large errors, a difference is visible in the behavior of $\bar{N}_{2}$ 
for the semi-hard events with mini-jets, 
especially in the large pseudorapidity windows. 
The values of $\bar{N}_{2}$  increased slightly 
for the three-NBD description
and do not show a sign of saturation for larger values of $\eta_c$. 
The average number of clans grows with $\eta_c$ for all three components.
Another difference between the parametrization of the CMS data with two and three NBDs 
is seen in the $\eta_c$ dependence of the average number of particles per clan.
There is some indication that   
clans of the first component contain more particles 
than the clans of the second one ($\bar{n}_{c1}>\bar{n}_{c2}$)
when fitting the data with weighted superposition of three NBDs in larger windows.  
This is in accord with the clan analysis of the ATLAS data shown in Fig. \ref{fig:12}.

The above observations are consistent with the interpretation of $1/k_i$ as an aggregation parameter \cite{GiovUgo4}
\begin{equation}
\frac{1}{k_i} = \frac{P_i(1,2)}{P_i(2,2)},
\label{eq:r7}
\end{equation}
where $P_i(N,m)$ is the probability that $m$ particles belong to $N$ clans.
As seen from Figs. \ref{fig:11}b and \ref{fig:12}b, 
clans of the second component are more numerous and contain
fewer particles  in comparison with the clans of the first component. 
This means that there is 
much less aggregation of particles in the semi-hard events with mini-jets than in the 
first, dominant class of events.
Such a property is reflected 
by the relation $1/k_2<1/k_1$ observed in central pseudorapidity intervals at $\sqrt s =$ 7~TeV
(Fig. 18(a) in \cite{IZ})
and at different energies in the window $|\eta|<$ 2.5 (Fig.~\ref{fig:6}a).

\begin{figure}
\includegraphics[width=78mm,height=78mm]{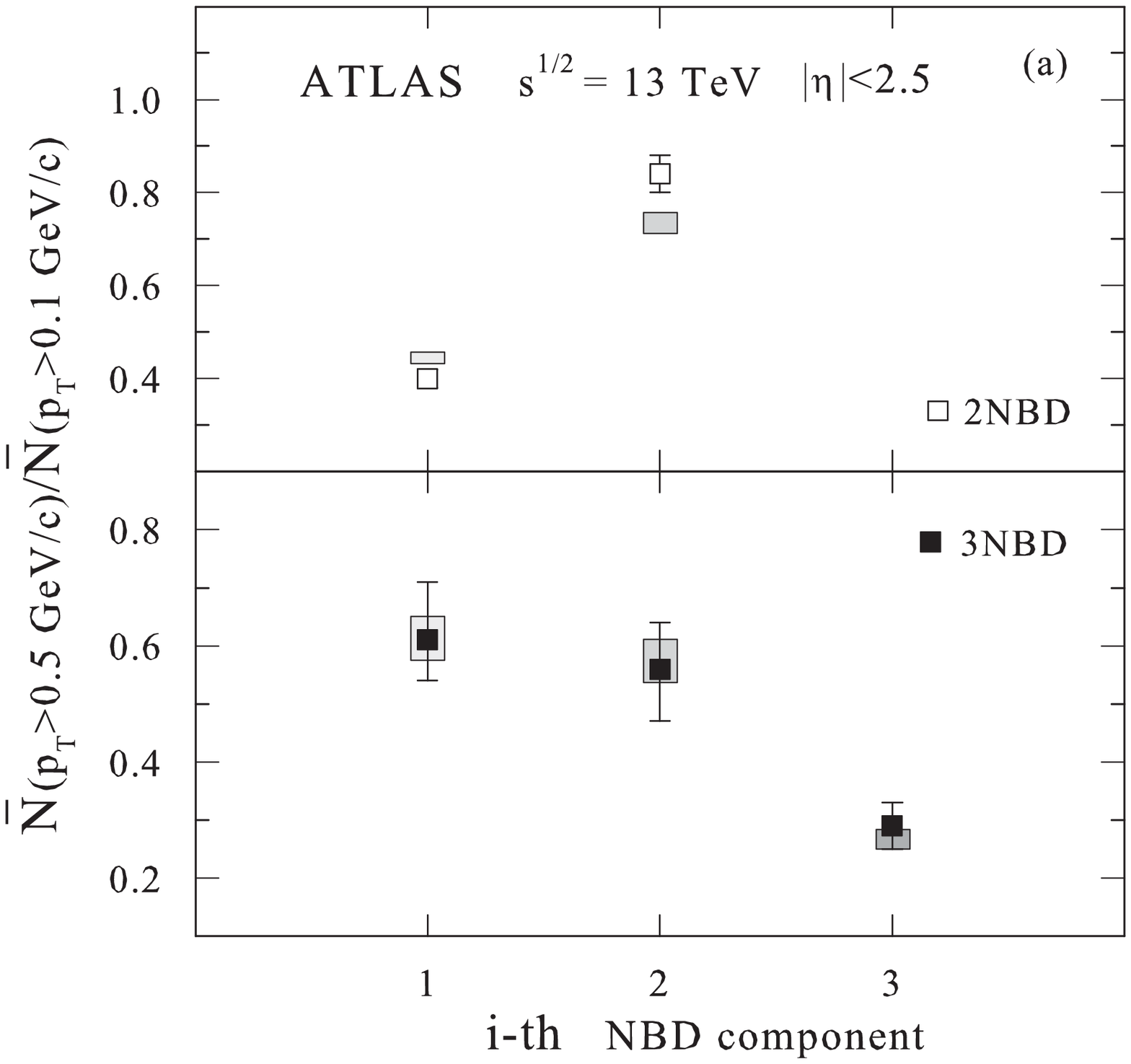}
\vskip -0.5cm
\includegraphics[width=78mm,height=78mm]{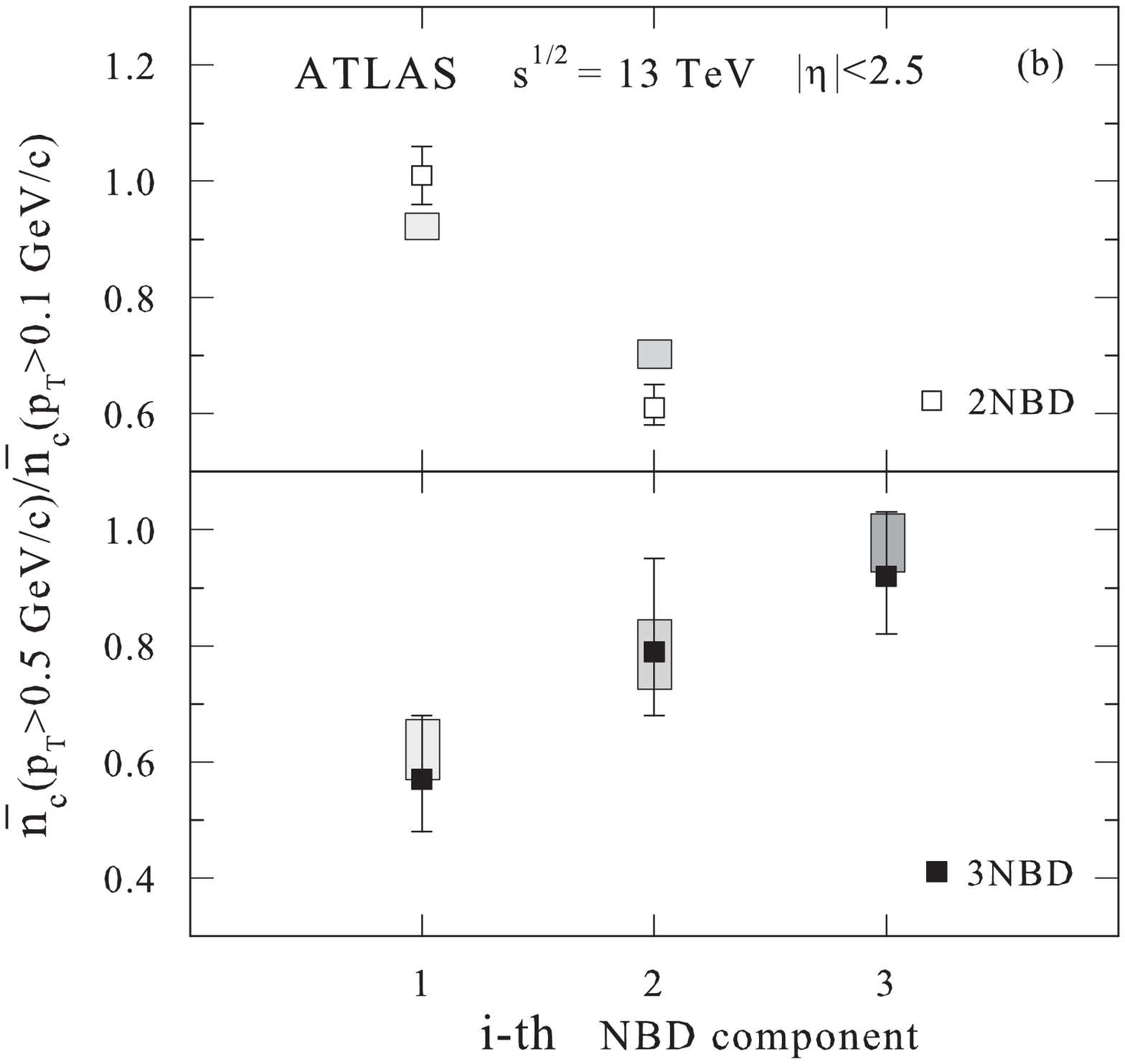}
\vskip -0.5cm
\caption{
Ratios of {\bf a} the average numbers of clans 
and {\bf b} the average numbers of particles per clan
extracted for single NBD components from MDs with the cuts $p_T>$ 500 MeV/c and $p_T>$ 100 MeV/c. 
The upper and lower panels correspond to the respective two- and three-NBD parametrizations
of the ATLAS data \protect\cite{ATLAS3,ATLAS4}
measured in the interval $|\eta|<$ 2.5 at $\sqrt s =$ 13~TeV.
The symbols with error bars and shaded rectangles depict the ratios corresponding to fitted parameters  
obtained by minimization of 
Eqs. (\ref{eq:a1}) and (\ref{eq:a4}), respectively   
}
\label{fig:15}       
\vskip -0.6cm
\end{figure}

We have studied the influence of transverse momentum on clan structure analysis 
using the ATLAS data on MDs \cite{ATLAS1,ATLAS2,ATLAS3,ATLAS4} with different $p_T$ cuts.
Figure \ref{fig:15} shows the ratios of the clan parameters extracted from fits to the data 
with $p_T>$ 500 MeV/c and $p_T>$ 100~ MeV/c at $\sqrt s =$ 13~TeV.
The ratios for the two- and three-NBD superposition are depicted in the upper and lower panels, respectively.
As one can see from Fig. \ref{fig:15}a, 
the average number of clans in the third component shows the greatest suppression with the $p_T$ cut.  
It means that these clans, consisting of very few particles, 
are produced mostly with low transverse momenta. 
For the three-NBD description, the average number of clans of the first component 
$\bar{N}_1$ is reduced the least,  
slightly less than $\bar{N}_2$.
This trend is quite opposite in the two-NBD model, as depicted in the upper panel in Fig. \ref{fig:15}a.

The suppression of the average number of particles per clan with $p_T$ is illustrated in  
Fig. \ref{fig:15}b. 
The value of $\bar{n}_{c3}$ is reduced minimally. 
It is natural because, by definition, a clan contains a minimum of one particle and because  
of the very small number of particles in these clans ($\bar{n}_{c3}\!\sim\!1$). 
At the same time $\bar{n}_{c1}$ is diminished considerably with the harder $p_T$ cut for the three-NBD description. 
The opposite follows from the two-NBD scenario. 
There is a very small suppression of particles inside clans 
of the soft multiplicity component, as depicted for the first NBD in the upper panel in Fig. \ref{fig:15}b.

Let us interpret the results obtained by the description of the ATLAS data on MD with different $p_T$ cuts 
when using weighted superposition of three NBDs.
The relatively small suppression of the average number of clans $\bar{N}_{1}$ 
and considerable reduction of the average number of particles per clan $\bar{n}_{c1}$ 
suggest that clans of the first, dominant component contain particles with a wide spectrum of transverse momenta.
There are many particles in these clans (see Fig. \ref{fig:12}b),
some of them with larger, some of them with smaller $p_T$. 
The imposed cut
$p_T>500$ MeV/c reduces the number of particles $\bar{n}_{c1}$ inside the clans by almost half, 
as those with small $p_T$ are lost. 
At the same time, $\bar{N}_{1}$ is least suppressed with the $p_T$ cut. 
This observation suggests that the large clans of the first multiplicity component 
consisting mostly of soft particles  
must contain also some particles with considerably high $p_T$.
It is therefore reasonable to assume that there are  semi-hard processes with relatively 
large momentum transfer in the dominant class of events.

\begin{figure}
\includegraphics[width=78mm,height=78mm]{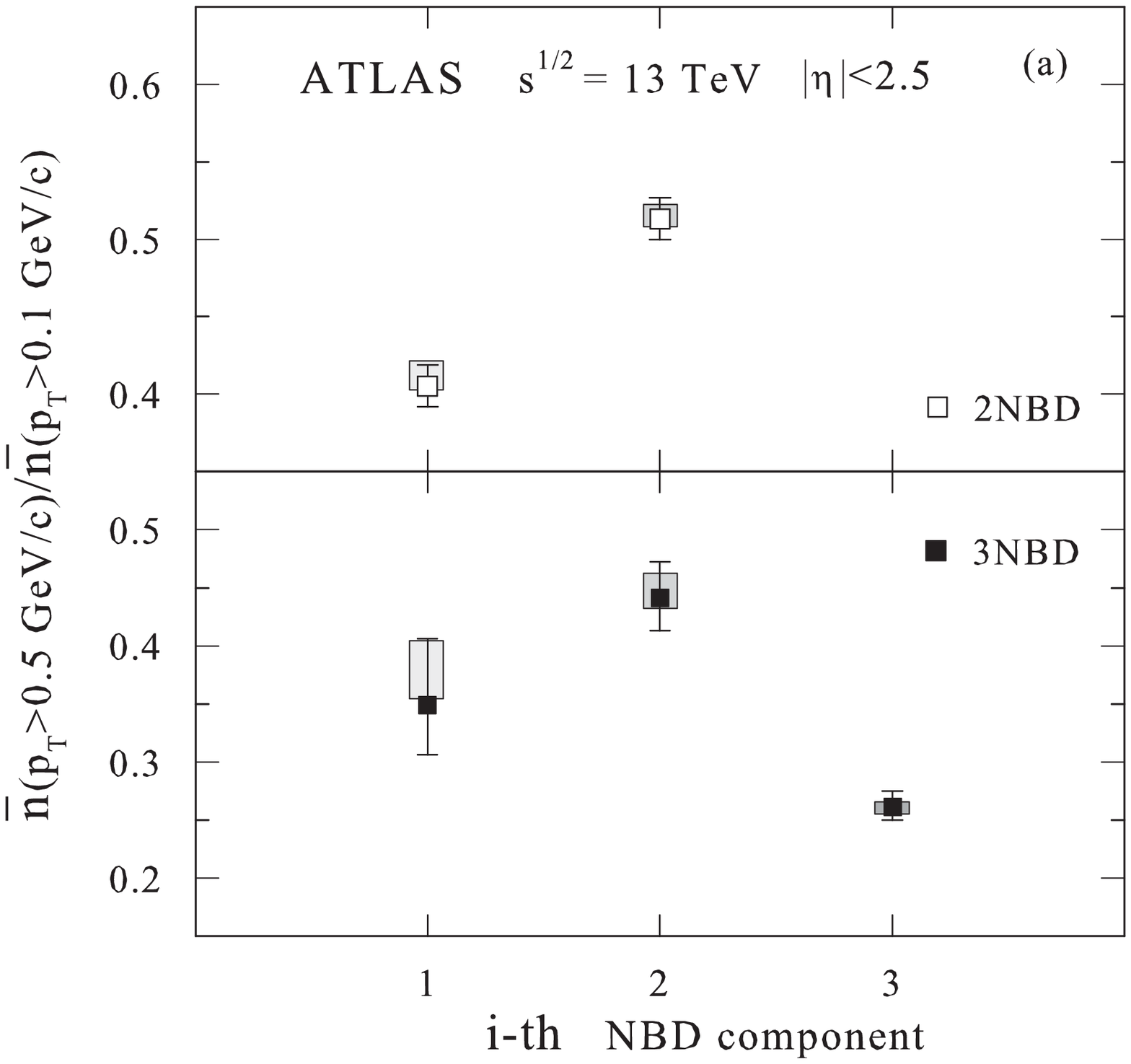}
\vskip -0.6cm
\includegraphics[width=78mm,height=78mm]{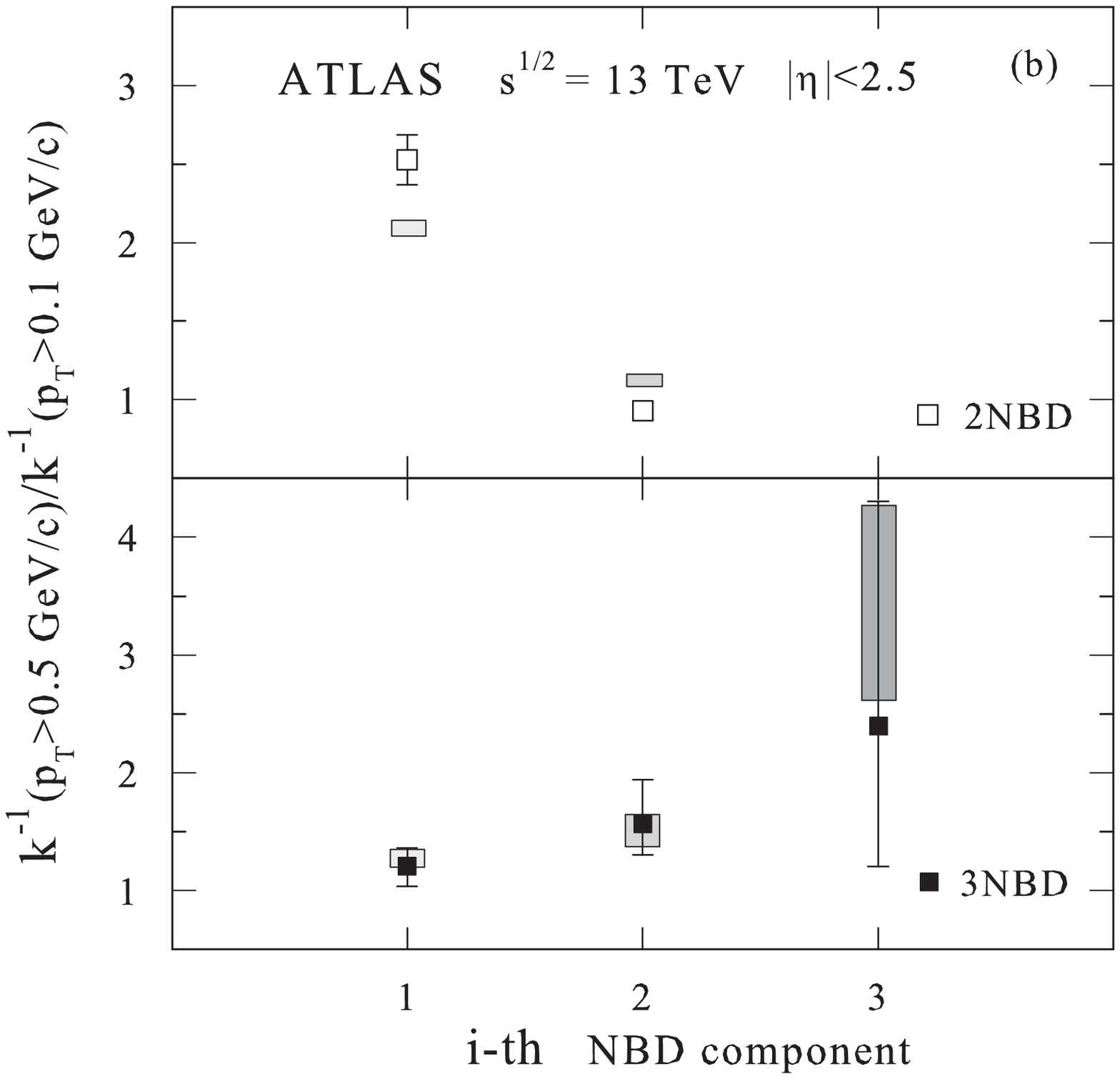}
\vskip -0.6cm
\caption{
Ratios of {\bf a} the average numbers of particles $\bar{n}_i$
and {\bf b} the aggregation parameters $k^{-1}_i$
extracted for single NBD components from MDs with the cuts 
$p_T>$ 500 MeV/c and $p_T>$ 100 MeV/c. 
The upper and lower panels correspond to the respective two- and three-NBD parametrizations
of data \protect\cite{ATLAS3,ATLAS4}
measured by the ATLAS Collaboration in the interval $|\eta|<$ 2.5 at $\sqrt s =$ 13~TeV.
The symbols with error bars and shaded rectangles depict the ratios corresponding to  fitted parameters  
obtained by minimization of 
Eqs. (\ref{eq:a1}) and (\ref{eq:a4}), respectively 
}
\label{fig:16}       
\vskip -0.6cm
\end{figure}

The properties of the clan structure of the second multiplicity component are different.
As depicted in the lower panel in Fig. \ref{fig:15}b,
the number of particles per clan 
$\bar{n}_{c2}$ is less suppressed with the cut $p_T>500$ MeV/c than $\bar{n}_{c1}$.
The relatively small reduction of $\bar{n}_{c2}$ means that fewer particles   
are affected  by the $p_T$ cut.
It suggests that there are fewer particles with low $p_T$ in the clans of the second component 
in comparison with the clans of the first, dominant one.
At the same time 
(see lower panel in Fig. \ref{fig:15}a),
the average number of clans, $\bar{N}_2$, is reduced
similarly and even slightly more than $\bar{N}_1$.
The considerable suppression of $\bar{N}_2$ and small reduction of $\bar{n}_{c2}$ indicate that
the transverse momenta of particles inside clans of the second component 
have a quite narrow distribution concentrated around some $\tilde{p}_T$.
This $\tilde{p}_T$ is higher than the average $<\!\!p_T\!\!>$ of particles of the 
first component, most of which carry low transverse momenta.
Moreover, considering that $\bar{n}_{c2}$ is obviously smaller than  
$\bar{n}_{c1}$ for $p_T>100$ MeV/c at $\sqrt s =$ 13~TeV (see Fig. \ref{fig:12}b), 
one can make the following conclusions.
The clans of the second NBD component contain fewer particles  
than the large clans of the first component.
The particles of the smaller clans carry on average higher $p_T$  
than most of the particles in clans of the dominant component. 
Their transverse momenta are concentrated around some $\tilde{p}_T$.
The average number of these clans, $\bar{N}_{2}$,  increases with energy
and becomes much larger than $\bar{N}_{1}$ at high $\sqrt s$.
The mentioned properties of the clan structure of the second semi-hard component resemble 
some basic features of mini-jets: clustering of particles of close 
transverse momenta with increasing frequency at high energies.

The application of the clan structure analysis to MDs measured by the ATLAS Collaboration  
with different limitations on transverse momenta
reveals properties which differentiate between two- and three-NBD description of the data.
It concerns both the average number of clans and the average number of particles per clan, which are 
reduced with varying intensity when applying the higher $p_T$ cut.
As seen from Fig. \ref{fig:15}, there are opposite trends of the suppression for both models 
in the first and the second multiplicity component. 
In order to clarify what is the cause of such a difference, one can look at the $p_T$ dependence of 
the corresponding NBD parameters.

Figure \ref{fig:16}a shows the ratios of the average number of particles, $\bar{n}_{i}$,
obtained from the fits to the ATLAS data measured in the interval $|\eta|<2.5$
for $p_T>500$~MeV/c and $p_T>100$~MeV/c.
The ratios of the NBD parameters $1/k_i$ extracted from the data are depicted in 
Fig. \ref{fig:16}b.
As concerns the first and the second multiplicity component, the ratios of $\bar{n}_{i}$ follow 
a similar trend for two- and three-NBD parametrization of the data.
The average multiplicities $\bar{n}_{1}$ and $\bar{n}_{2}$ are suppressed almost proportionally for both models.
Contrary to this, the corresponding ratios of the aggregation parameters, $1/k_i$, 
behave completely opposite for parametrization of the ATLAS data with two and three NBDs.
The aggregation of particles into clans increases strongly with the cut $p_T>500$ MeV/c 
for the soft component in the two-NBD model. 
At the same time, the aggregation parameter $1/k_2$ 
changes very little for the semi-hard component.
In the three-NBD model, the aggregation of particles into clans increases slightly with $p_T$ for the first,
dominant component. The aggregation is slightly stronger for the semi-hard component with mini-jets.
As concerns the third NBD, it is hard to tell because of the large error.
The observations indicate that it is just another aggregation of particles into clans, which 
results in different properties of clans in two- and three-NBD models 
used for the description of the ATLAS~data.

\subsection{Discussion} 
\label{sec:G}

A detailed examination of the sensitivity of cumulative combinants to the systematic uncertainties of measurement and to the unfolding procedures applied to raw data requires information on the response matrix for the considered experiments as well as knowledge of the experimental methods used 
by obtaining the final MD.

Though a study of the raw data is not the subject of the present paper, we discuss here an estimation
of errors of the cumulative combinants
based on the published data. 
The combinants are expressible in terms of the first $i$ probability ratios,
$P_n/P_0, n=1,..,i$, which depend on $P_0$. 
For the ATLAS data, the analysed events are required to satisfy special criteria in order to minimize the systematic uncertainties. One of them is the minimal number of charged tracks that depends 
on the particular phase space region. 
The most inclusive phase space region covered by the measurements 
corresponds to the conditions  $n_{ch}\ge 2$ and $p_T > 100$~MeV/c. 
The experimental information on $P_0$ and $P_1$ is missing in these data.
For the CMS data at  $\sqrt s =$ 7~TeV,
the measured value of $P_0$ is very large. 
Due to the experimental difficulties connected with the large error of this probability
and because of the rise of the MDs in the zeroth bin, $P_0$ is usually 
omitted in the NBD fits to the data.  

In order to avoid inaccuracies as to the assumptions with regard to 
the multiplicity probabilities and their errors in the first two bins,
let us consider a modification of combinants based on the relation \ref{eq:r4}.
Instead of the full MD $\{P_0, P_1, P_2,...\}$ one can apply 
this recurrence relation to the truncated sets  
$\{P_2, P_3, P_4,...\}$ or $\{P_1, P_2, P_3,...\}$. 
The corresponding modified quantities $(i\!+\!1){\cal C}_2(i\!+\!1)$ 
and $(i\!+\!1){\cal C}_1(i\!+\!1)$ are defined by the formulas
\begin{equation}
{\cal C}_2(1)=\frac{P_3}{P_2}, \ \ \ \
2{\cal C}_2(2)=2\frac{P_4}{P_2}-{\cal C}_2(1)\frac{P_3}{P_2},...
\label{eq:r8}
\end{equation}
and 
\begin{equation}
{\cal C}_1(1)=\frac{P_2}{P_1}, \ \ \ \
2{\cal C}_1(2)=2\frac{P_3}{P_1}-{\cal C}_1(1)\frac{P_2}{P_1},...,
\label{eq:r9}
\end{equation}
respectively.
The lower index $2$ ($1$) at ${\cal C}$ means that
the truncated probability set begins with $P_2$ ($P_1$).
As illustrated below, the quantities $(i\!+\!1){\cal C}_2(i\!+\!1)$ ($(i\!+\!1){\cal C}_1(i\!+\!1)$)
defined by Eqs. (\ref{eq:r8}) ((\ref{eq:r9})) reveal 
similar oscillating properties to the cumulative combinants calculated by Eq. (\ref{eq:r4})
from a superposition of NBDs  with parameters obtained from the corresponding fits.

\begin{figure}
\includegraphics[width=78mm,height=78mm]{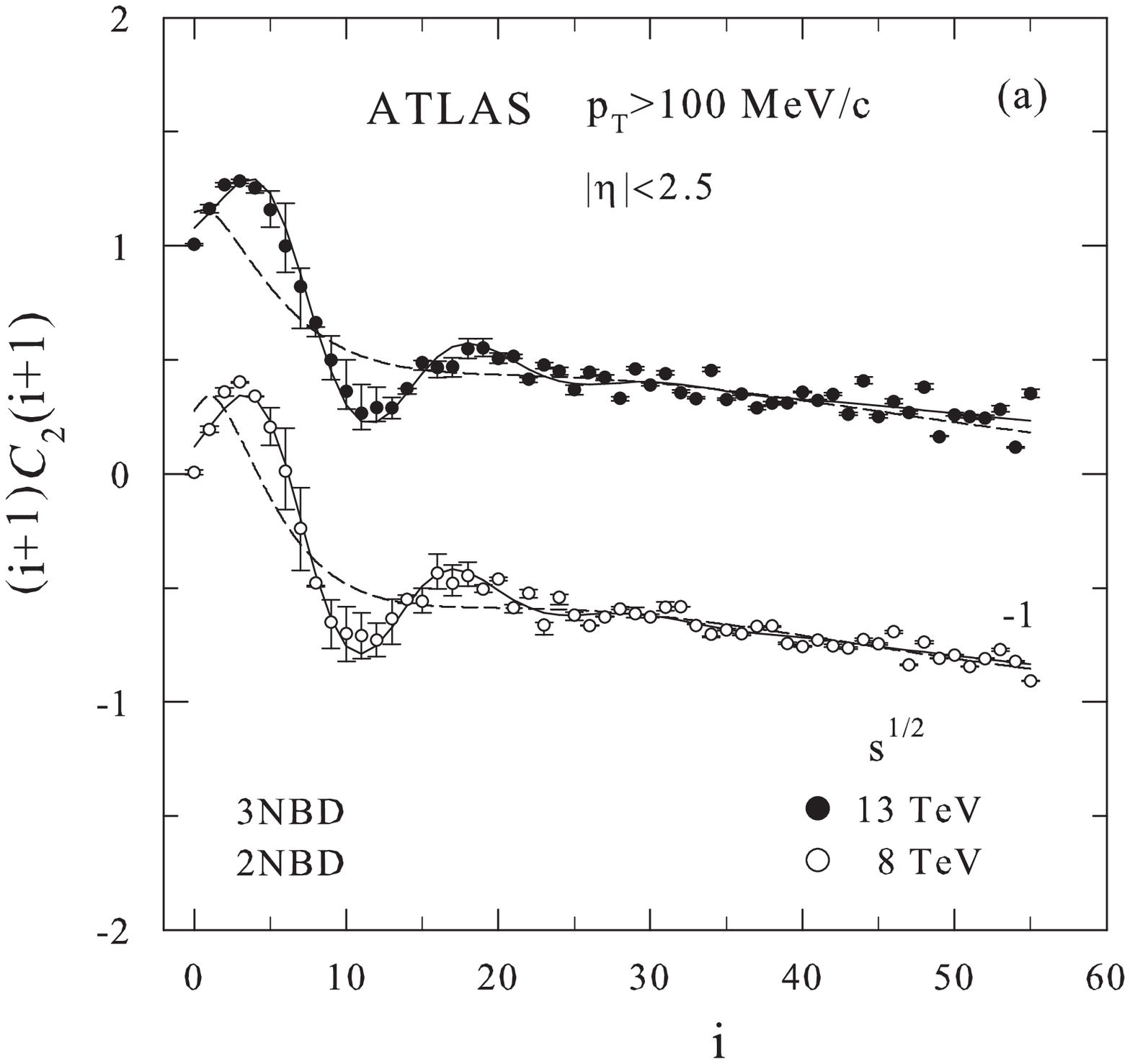}
\vskip -0.6cm
\includegraphics[width=78mm,height=78mm]{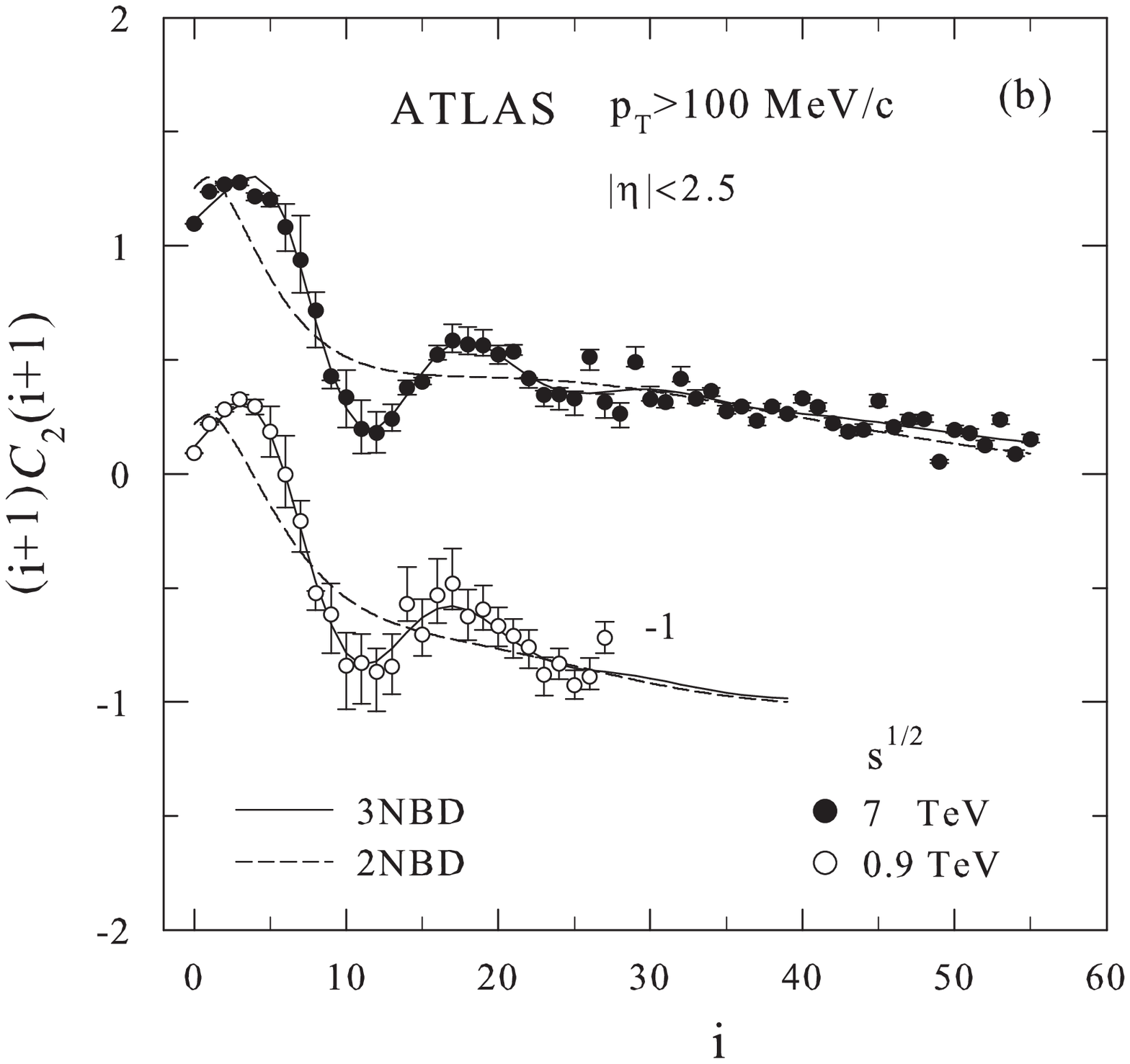}
\vskip -0.6cm
\caption{
The coefficients $(i+1) {\cal C}_2(i+1)$ calculated from 
MDs measured by the ATLAS Collaboration \protect\cite{ATLAS1,ATLAS2,ATLAS4}
in the interval $|\eta|<~2.5$ with $p_T > 100$~MeV/c, $n>1$
at {\bf a} $\sqrt s=$13, 8~TeV 
and {\bf b} $\sqrt s=$ 7, 0.9~TeV.
The error bars correspond to the statistical and the systematic uncertainties summed in quadrature.
The open symbols are shifted by the factor -1. 
The full (dashed) lines are obtained by Eq. (\ref{eq:r8}) from three- (two-) NBD 
fits to the data on MDs using Eq. (\ref{eq:a4})
}
\label{fig:17}       
\vskip -0.6cm
\end{figure}

Figure \ref{fig:17} shows the coefficients $(i\!+\!1){\cal C}_2(i\!+\!1)$ 
in dependence on the rank $i$. 
Their values were calculated from the data on MD ($n_{ch}\ge 2$)
measured by the ATLAS Collaboration in the interval $|\eta|<2.5$
with the transverse momentum cut $p_T>100$~MeV/c
at $\sqrt s=$13, 8, 7 and 0.9~TeV.
The error bars shown in the figure were obtained from experimental errors of the data on 
$P_n$. The latter include both the statistical and the systematic uncertainties summed in quadrature.

In determination of the errors of $(i+1) {\cal C}_2(i+1)$ we have proceeded in the following way
(see Appendix A).
Taking into account the positive correlation between adjacent multiplicity bins and the anti-correlation
between the opposite sides of the MD distribution maximum, we have considered the distributions shifted
to the right and left with respect to the error variation in single multiplicity bins. 
The right shifted distribution is constrained by the upper values of the errors of $P_n$ for $ n > m $  
and by their lower values for $n < m$ where $m$ is the multiplicity with the maximal value of $P_m$.
The left shifted distribution was considered in the analogous way, exchanging mutually 
upper and lower errors at both sides of the maximum.
The right shifted distribution of $P_n$ is responsible for 
the upper error bars of $(i+1) {\cal C}_2(i+1)$ at maxima and the lower error bars near the minimum 
visible in Fig. \ref{fig:17}. The left shifted $P_n$ governs the error bars in the opposite direction.
For higher values of the rank $i$, 
one can see a scattering of points around the curves calculated from three- and two-NBD fits 
to the experimental MDs. 
This corresponds to a statistical spread of the mean values of $P_n$ 
with respect to the fits. As the statistical uncertainties of the data are much smaller than the spread
of the mean values in the ATLAS data, 
the error bars shown in Fig. \ref{fig:17} are practically hidden within the symbols 
for increasing $i$.  

\begin{figure}
\includegraphics[width=78mm,height=78mm]{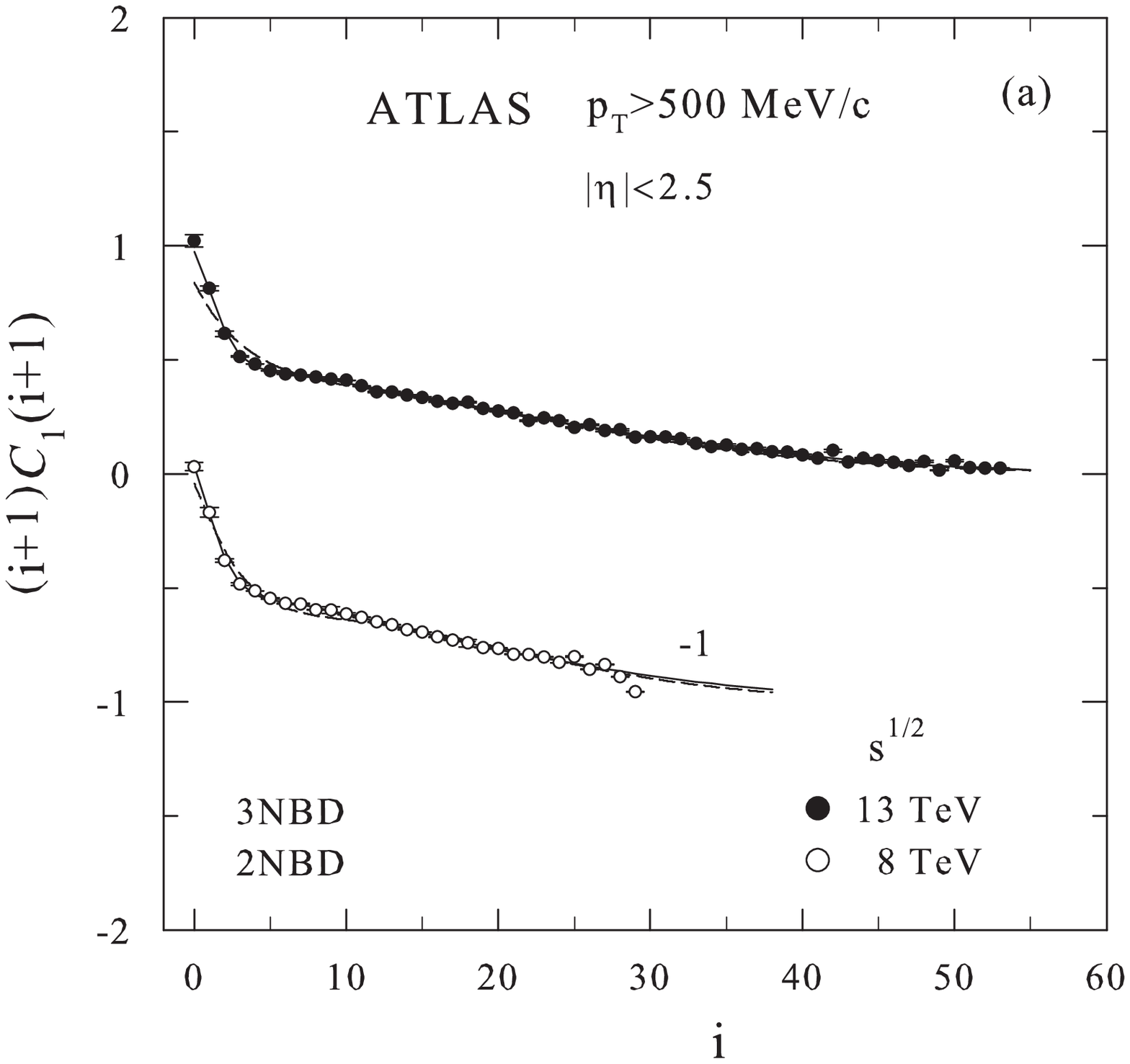}
\vskip -0.6cm
\includegraphics[width=78mm,height=78mm]{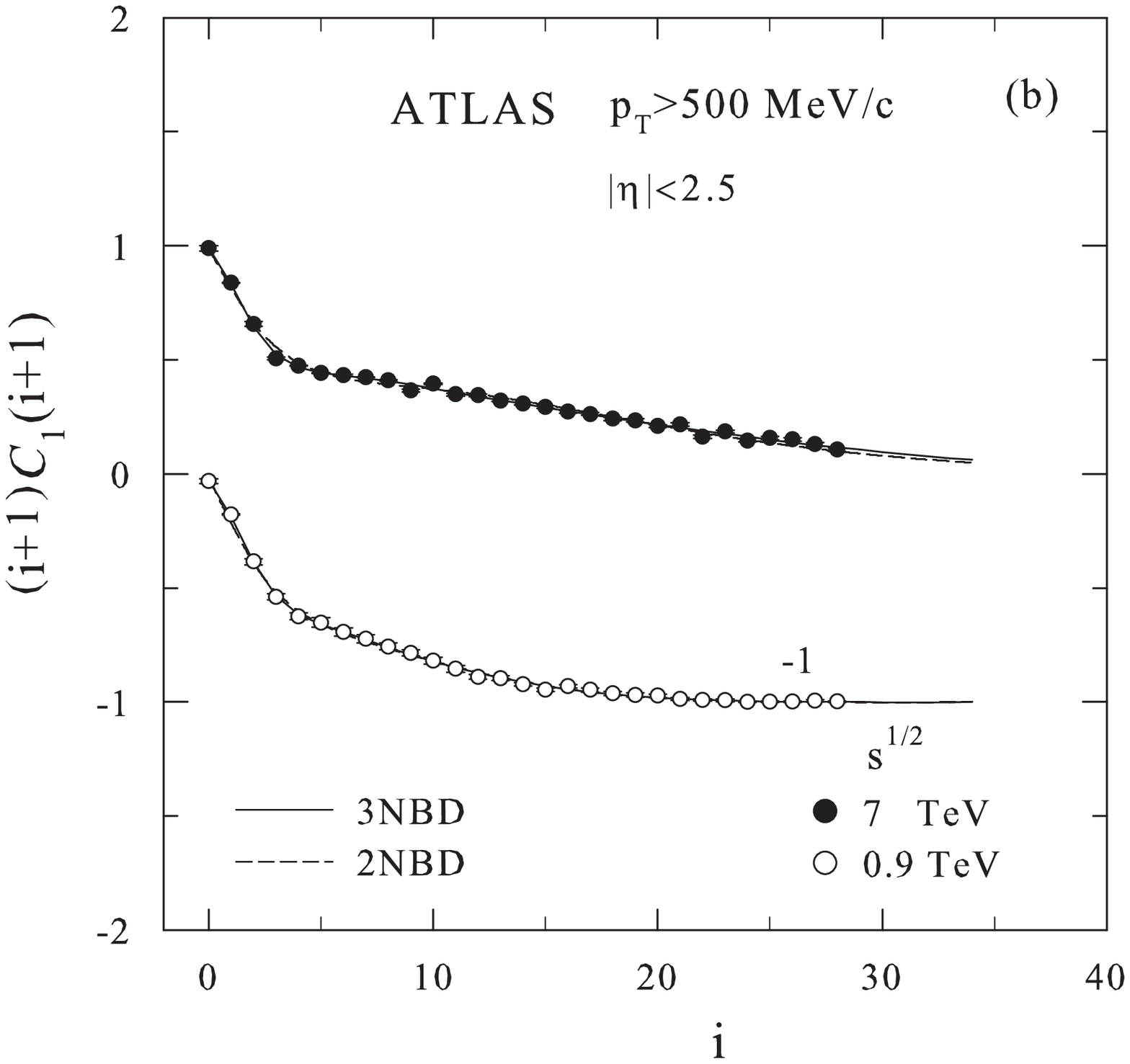}
\vskip -0.6cm
\caption{
{\bf a} The coefficients $(i+1) {\cal C}_1(i+1)$ calculated from 
data on MD measured by the ATLAS Collaboration 
\protect\cite{ATLAS1,ATLAS2,ATLAS3}
in the interval $|\eta|<~2.5$ with $p_T > 500$~MeV/c, $n>0$
at {\bf a} $\sqrt s=$13, 8~TeV 
and {\bf b} $\sqrt s=$ 7, 0.9~TeV.
The error bars correspond to the statistical and the systematic uncertainties summed in quadrature.
The open symbols are shifted by the factor -1. 
The full (dashed) lines are obtained by Eq. (\ref{eq:r9}) from three- (two-) NBD 
fits to the data using Eq. (\ref{eq:a4})
}
\label{fig:18}       
\vskip -0.6cm
\end{figure}
Because of relatively small systematic uncertainties of the measurements,
one can  see a sizeable wavy structure of  
$(i+1) {\cal C}_2(i+1)$ in Fig. \ref{fig:17}, 
which clearly discriminates between
the two- and three-NBD parametrization of the ATLAS data with $p_T>100$ MeV/c. 
The difference is seen in the region of low $i$ and corresponds to a distinct peak near 
the maximum of the measured MD.

\begin{figure*}
\begin{center}
\includegraphics[width=78mm,height=78mm]{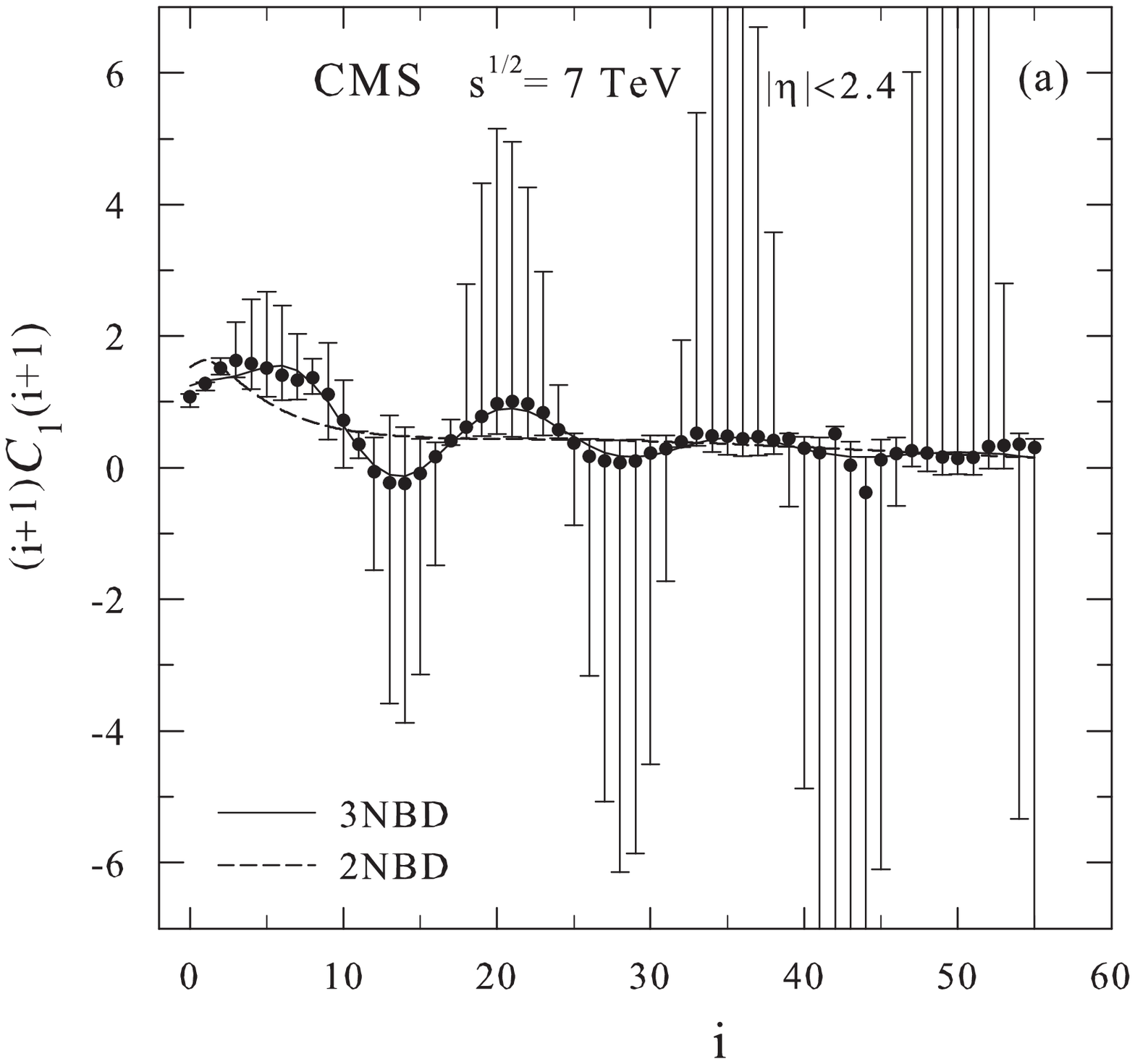}
\includegraphics[width=78mm,height=78mm]{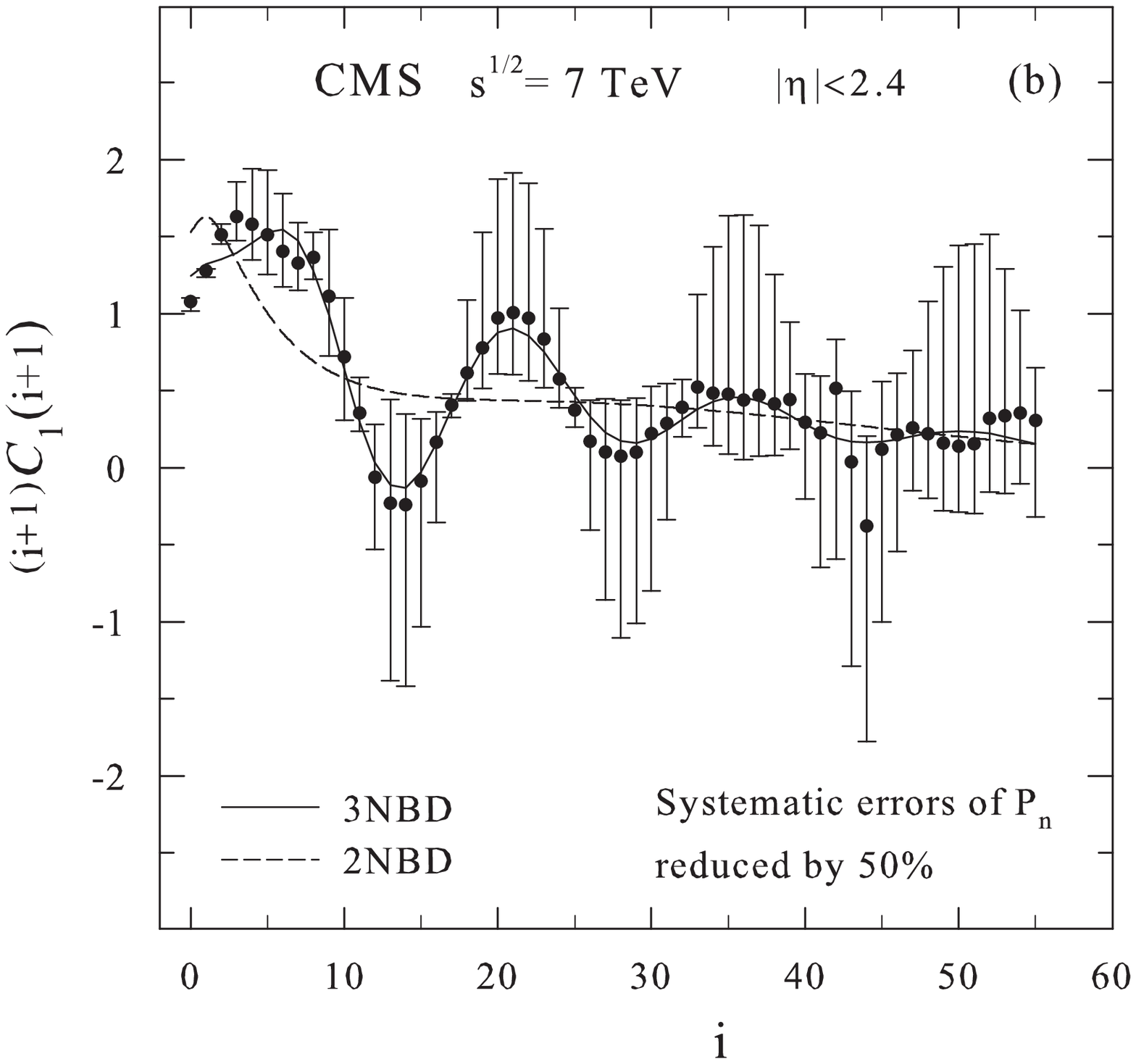}
\end{center} 
\vskip -0.8cm
\caption{
{\bf a} The coefficients $(i+1) {\cal C}_1(i+1)$ calculated from 
data on MD measured by the CMS Collaboration \protect\cite{CMS}
in the interval $|\eta|<2.4$ at $\sqrt s=7$~TeV for $p_T>0$.
The error bars correspond to the statistical and the systematic uncertainties of $P_n$ 
summed in quadrature.
{\bf b} The coefficients $(i+1) {\cal C}_1(i+1)$ calculated from the same data 
with the systematic errors reduced by 50$\%$.
The solid (dashed) lines were calculated from  three- (two-) NBD fits
to the measured MD using Eq. (\ref{eq:r9})
}
\label{fig:19}       
\end{figure*}
\begin{figure*}
\begin{center}
\includegraphics[width=78mm,height=78mm]{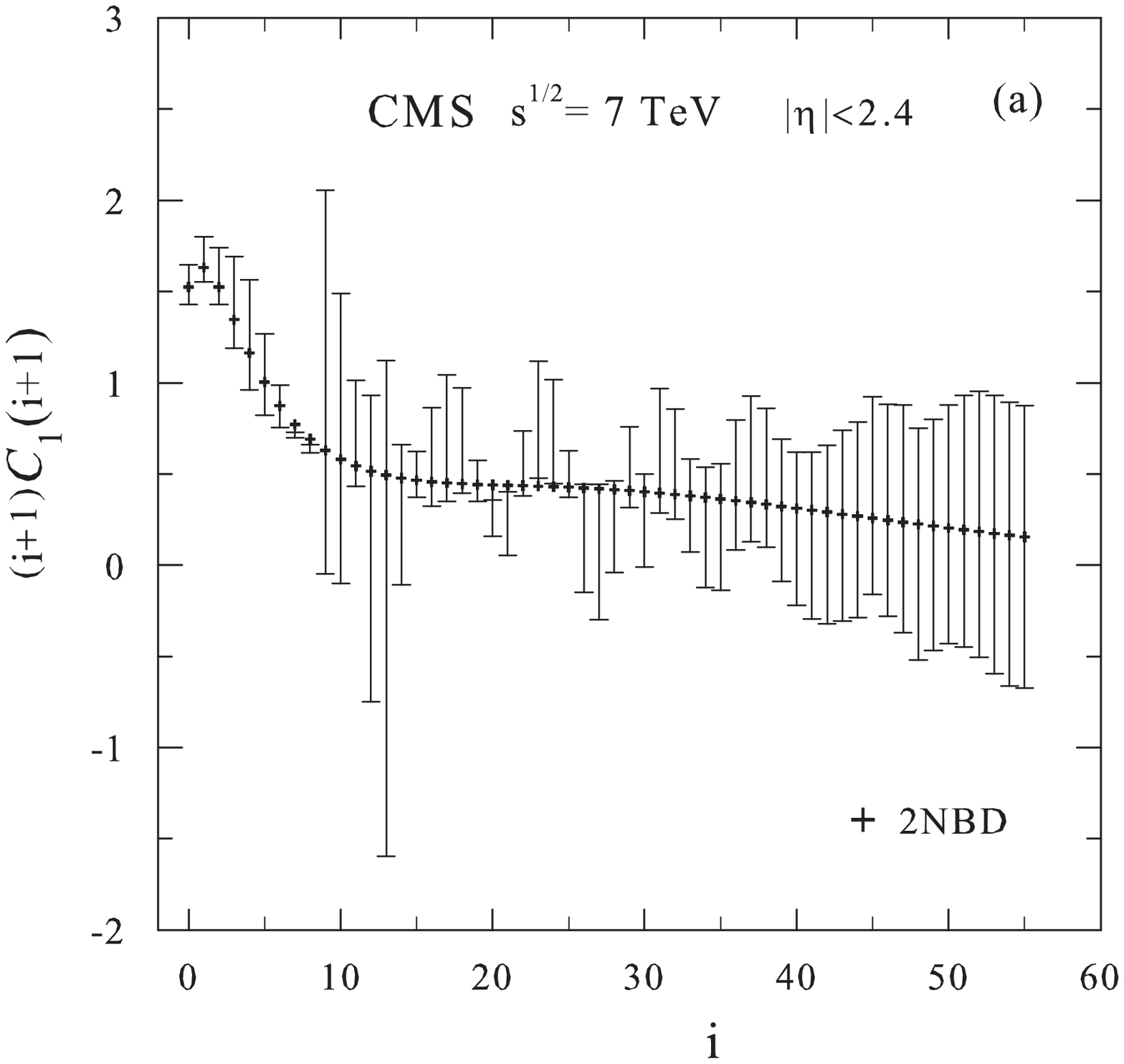}
\includegraphics[width=78mm,height=78mm]{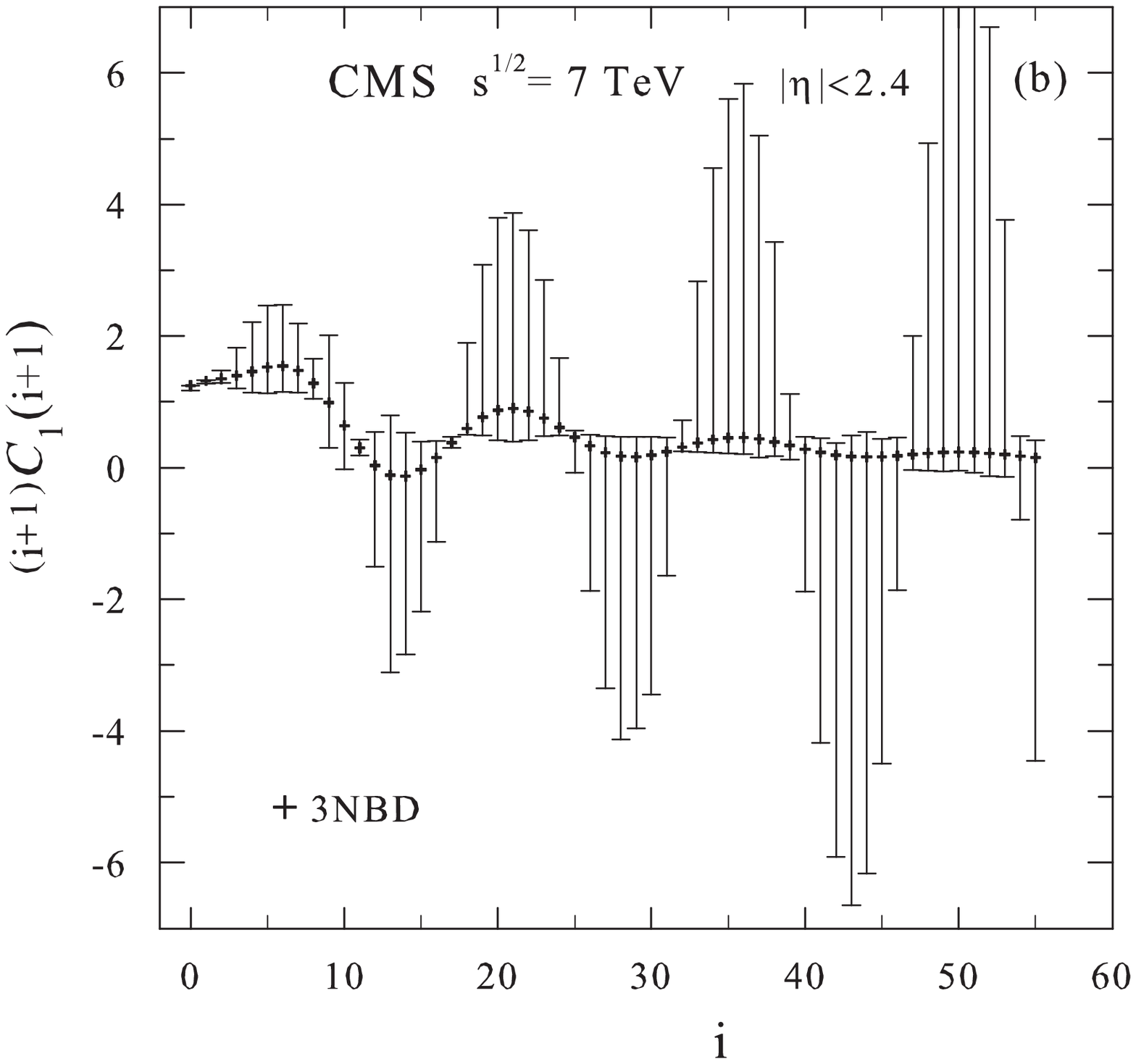}
\end{center} 
\vskip -0.8cm
\caption{
The coefficients $(i+1) {\cal C}_1(i+1)$ calculated from 
a weighted superposition of {\bf a} two and {\bf b} three NBDs 
fitted to the charged-particle MDs 
\protect\cite{CMS}
measured by the CMS Collaboration at $\sqrt s=$7~TeV in the pseudorapidity interval $|\eta|<2.4$ for $p_T>0$. 
The error bars correspond to the systematic and statistical uncertainties 
of $P_n$ summed in quadrature  
}
\label{fig:20}       
\end{figure*}
Figure \ref{fig:18} shows 
the coefficients $(i+1) {\cal C}_1(i+1)$ calculated from data 
on MD ($n_{ch}\ge 1$)
measured by the ATLAS Collaboration in the window $|\eta|<2.5$ with the cut
$p_T > 500$~MeV/c at $\sqrt s=$13, 8, 7 and 0.9~TeV.
The errors of the coefficients were obtained from experimental uncertainties of  
$P_n$ in the same way as in Fig.~\ref{fig:17}.  
Except for a few points at low $i$ in Fig. \ref{fig:18}a, 
the errors are smaller than the size of the displayed symbols.
One can see that there are no oscillations of $(i+1) {\cal C}_1(i+1)$
for the ATLAS data
with $p_T > 500$~MeV/c at all considered energies. 
The quantities calculated from the data on MD reveal  
approximately the same behavior as the cumulative 
combinants obtained from weighted sum of three and two NBDs used to fit the data 
(see Fig. \ref{eq:r8}).
The disappearance of oscillations is connected with 
the small average multiplicity $\bar{n}_3 \sim 3$
of the third NBD component for the data with higher $p_T$ cut 
(cf. the diminishing of the oscillations in
small windows in Fig. 9a where $\bar{n}_3 = 2.7$ for $|\eta| <0.5$).

The situation is somewhat different for the CMS measurements with $p_T>0$ \cite{CMS}  
in sufficiently large pseudorapidity intervals.
Figure \ref{fig:19}a shows the quantities $(i\!+\!1){\cal C}_1(i\!+\!1)$ 
in dependence on the rank $i$ calculated from the $n_{ch}$ distributions 
in the window $|\eta|<2.4$ at $\sqrt s=$7~TeV.
The black symbols correspond to the mean values of $P_n$.   
The error bars were obtained in the same way as 
in the case of the ATLAS data.
\begin{figure*}
\begin{center}
\includegraphics[width=78mm,height=78mm]{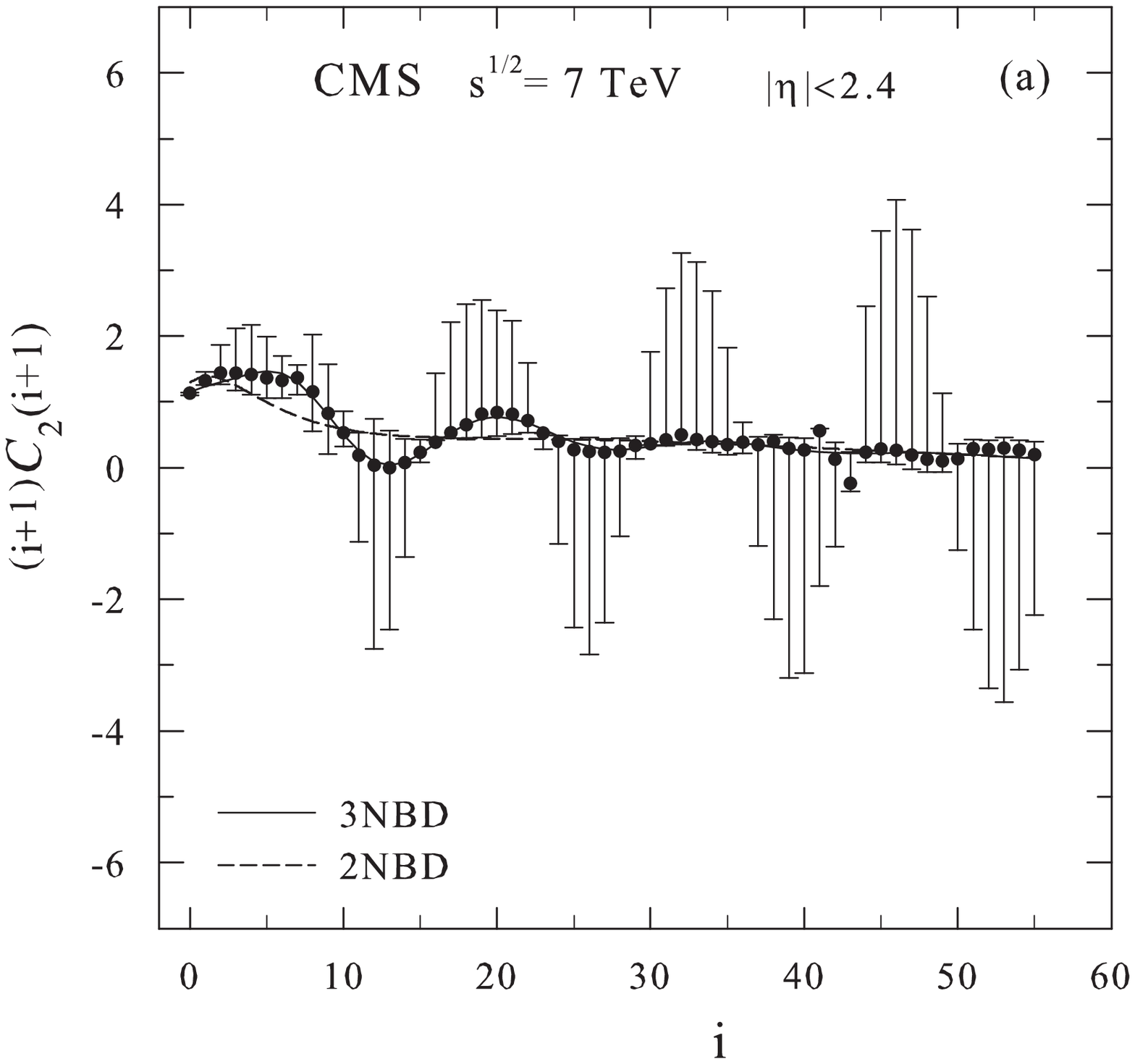}
\includegraphics[width=78mm,height=78mm]{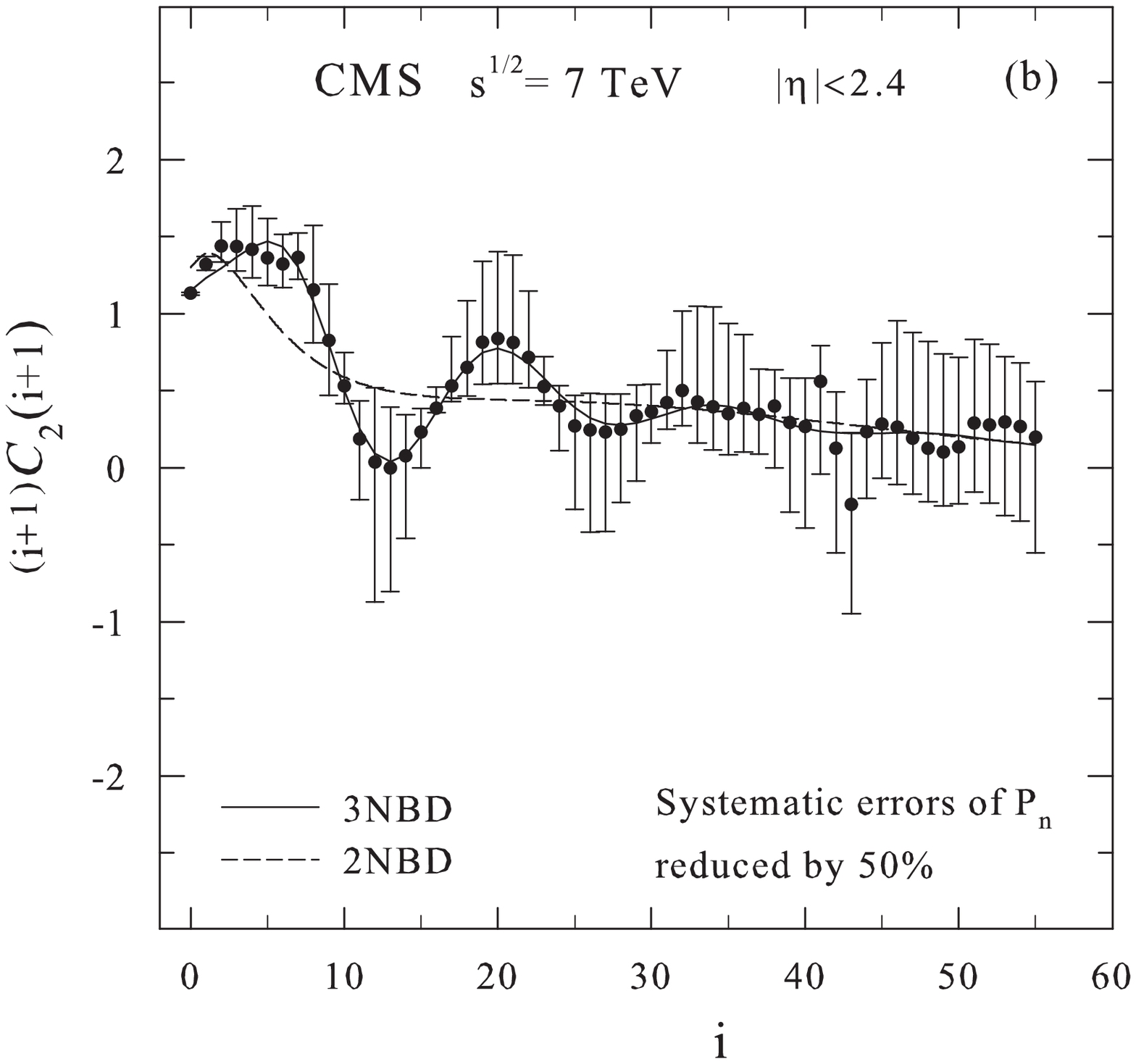}
\end{center} 
\vskip -0.8cm
\caption{
{\bf a} The coefficients $(i+1) {\cal C}_2(i+1)$ calculated from 
data on MD measured by the CMS Collaboration \protect\cite{CMS}
in the interval $|\eta|<2.4$ at $\sqrt s=7$~TeV for $p_T>0$.
The error bars correspond to the statistical and the systematic uncertainties of $P_n$ 
summed in quadrature.
{\bf b} The coefficients $(i+1) {\cal C}_2(i+1)$ calculated from the same data 
with the systematic errors reduced by 50$\%$.
The solid (dashed) lines were calculated from  three- (two-) NBD fits
to the measured MD using Eq. (\ref{eq:r8})
}
\label{fig:21}       
\end{figure*}
The right shifted distribution of $P_n$ results in a large oscillation of errors
that increases strongly in the region of higher rank $i$. 
The right shift determines the upper errors at maxima
and the lower errors at minima of the first oscillations and creates a repeating pattern. 
The left shift of the distribution is mostly responsible for errors in the opposite direction.
One can see from Fig. \ref{fig:19}a that the structure of the error bars agrees with  
the initial wavy character of the black symbols quite well.
The errors grow very rapidly at extrema 
with the increasing number of oscillations. 
Such a behavior is caused by the systematic uncertainties of the CMS data, which are
much larger than in the ATLAS measurements.
The systematic uncertainties govern the errors of residues of the MD with respect to the two-NBD fits \cite{IZ}.
They form a characteristic envelope around the mean values of $P_n$
that follows the fine structure of the residual
mean values quite accurately.

We have examined how a reduction of the systematic uncertainties of MD can affect the information 
on the oscillatory behavior of the coefficients $(i\!+\!1){\cal C}_1(i\!+\!1)$. Assuming that 
the systematic errors of $P_n$ could be reduced by 50$\%$, the oscillating structure 
of cumulative combinants including errors is expected
to be similar to Fig.~\ref{fig:19}b. 
Such an increase in accuracy would help in discriminating between 
two- and three-component description of the data considerably.   
The oscillating pattern depicted in Fig. \ref{fig:19}a can help to discriminate 
between the two- and three-NBD hypotheses also on the basis of the following considerations. 
Let us assume that the distribution of mean values of $P_n$ is described exactly 
by two NBDs, e.g. with parameters quoted in~\cite{IZ}. 
In this model situation we take the
statistical and systematic uncertainties provided by the CMS Collaboration 
and construct the quantities $(i\!+\!1){\cal C}_1(i\!+\!1)$. 
The result for the interval $|\eta|<2.4$ at $\sqrt s=7$~TeV and $p_T>0$ is shown in  Fig.~\ref{fig:20}a.
No characteristic pattern of oscillation can be seen there.
If we assume, however, that the mean values of $P_n$ follow exactly a three-NBD curve,
with parameters taken e.g. from \cite{IZ}, and take the experimental errors from 
the corresponding measurement, the uncertainties of $(i\!+\!1){\cal C}_1(i\!+\!1)$ constructed as above
look like Fig.~\ref{fig:20}b. One can see that the oscillating pattern based on the three-NBD
hypothesis shows a very similar structure, as depicted in Fig.~\ref{fig:19}a. The later corresponds to the modified cumulative combinants and their errors calculated from data published by the CMS Collaboration. 

We examined the sensitivity of the oscillation pattern of errors to the change of 
the truncation condition from $n\ge 1$ to $n\ge 2$.  
For that purpose we used data on MD measured by the CMS Collaboration \cite{CMS}
for $p_T>0$ in the interval $|\eta|<2.4$ at $\sqrt s=7$~TeV  and calculated 
the coefficients $(i\!+\!1){\cal C}_2(i\!+\!1)$.
The result depicted in Fig.~\ref{fig:21} demonstrates
that the error pattern seen  in Fig.~\ref{fig:19} 
is not destroyed with the increase of the truncation point.
A comparison of both figures shows that omitting of $P_1$ from 
the construction of the modified combinants results in a reduction of the amplitude of their errors
and is accompanied by a slight decrease of their periodicity.
Let us note that the structure illustrated in Fig.~\ref{fig:21}b is similar to that
depicted by full symbols in Fig.~\ref{fig:17}b.

A detailed analysis of the cumulative combinants requires study of raw data and 
knowledge of proper response of detectors. The discussed estimation of their uncertainties  
indicates that oscillations of the cumulative combinants seem to be a true
property of $n_{ch}$ distributions. The ATLAS data are most conclusive in that respect.
The CMS data are less decisive as to the oscillations, although some supplementary support 
for such a hypothesis is seen in larger pseudorapidity intervals.

In the following part we discuss some ideas concerning the development of the clan structure 
in the context of the present study of the clan parameters obtained from 
the three-NBD description of the analyzed data. 
The emerging picture
is consistent with a stochastic-physical scenario \cite{IZ_clans} of clans' evolution
involving the ingredients of QCD branching processes.
In this scenario, clans, during their (QCD) evolution, 
can perform collapses onto single particle states. The states evolve further by  
subsequent branching. 
The collapses of clans are most probable in the early stage of their development. 
We consider that such collective behavior  reflects early
neutralization of color configurations and simulates rapid freeze-out of QCD
degrees of freedom inside clans. 
The processes of a clan's collapsing define directions along which the final particles originated from 
their ancestors are spreaded. 
Those clans that experienced (one or more) collapses can be 
identified (under additional conditions) with mini-jets. 
The ``mini-jet clans" are expected to have different properties 
in comparison with clans that evolved without collapses. 
It concerns the transverse momentum distribution of particles which, 
due to collapses, should be concentrated around some $\tilde{p}_T$. 
Moreover, it is natural to expect that the number of clans mini-jets should increase with energy 
and should become larger than the number of clans of which the evolution happened without collapses. 
Due to collapses, clans identified with mini-jets should contain less particles than 
clans that did not suffer any collapse.  

As can be seen from Figs. \ref{fig:11}b and \ref{fig:12}b, 
this picture is in accord with the values of the clan parameters 
extracted from three-NBD description of the ATLAS data.
The first multiplicity component consists of large clans containing many particles.
The average number of these clans is low and constant with energy. 
These properties are in line with the conjecture that the
occurrence of the mini-jets in the dominant component is small.
The second NBD component describes a high-multiplicity tail of the total distribution. 
As usually considered, the component is enriched with mini-jets.
Their symptoms concern clustering of particles into smaller clans with increasing intensity 
at high energies.  
This assumption complies with the clan properties extracted from the experimental data.
First, the clans of the second component contain fewer particles 
in comparison with the clans of the first one ($\bar{n}_{c2}<\bar{n}_{c1}$).
The value of $\bar{n}_{c2}$ increases with energy.
Second, the average number of the smaller clans, $\bar{N}_2$, 
increases with energy and becomes much larger than $\bar{N}_1$ at high $\sqrt{s}$.

The properties of clans in the third component are different.
There is only one or very few particles in these clans,
as illustrated by empty squares in Fig. \ref{fig:12}b.	
It points to little or nearly no branching in such a case.
The parameter $1/k_3$ is small (see Fig. \ref{fig:6}a), 
reflecting a minimal aggregation of particles in comparison with the first and the second  component.
A characteristic feature of the three-NBD fits to the ATLAS data is the approximate constancy 
of the average multiplicity $\bar{n}_{3}$ with energy. 
Moreover, the average number of clans, $\bar{N}_{3}\simeq 9$, 
depends very weakly on $\sqrt s$ as well (see Fig. \ref{fig:11}b).
These observations suggest that there are different mechanisms responsible 
for particle production in the first two components of the MD and the third one 
emerging at low~$n$.

The performed study of the clan parameters indicates that 
the peak around maximum of the MD is influenced by a mechanism which depends weakly on energy and
is characterized by minimum branching and small transverse momenta. 
This is typical for soft particle production.
It is unlikely that the peak is a remnant of diffractive events.
In the central rapidity intervals studied at the LHC, the single-diffractive events 
influence 0- and 1-bin in multiplicity.
The double-diffractive events influence 1- and 2-bin. 
It is therefore hard to assume that the peak at $ \bar{n}_3\simeq 11$ 
in the central region ($ |\eta|<2.5$) is caused by diffraction.
The relevant mechanism could be the production of particles from fragmentation 
of (color) strings stretched between the leading partons of the colliding protons.
The fragmentation of the longitudinally stretched strings is a soft process and results 
in narrow (nearly Poissonian) MD of particles with a limited range of transverse momenta.

The properties of the clan parameters in the first and the second NBD component 
point to a quite different kind of mechanism 
responsible for particle production.
The large amount of branching in both of them can be understood by a two-stage mechanism;
via collisions of constituents of the colliding protons with 
production of massive secondary objects (fireballs, clans' ancestors, etc.) 
in the first stage, and the successive branching of these objects into observable particles 
in the second one.
The events corresponding to the first, dominant multiplicity component
are characterized by the production of a small number of clans containing many particles.
The independence of their number, $\bar{N}_1$, from energy  
corresponds to the occurrence of a small number of mini-jets.
Most particles in the clans carry relatively low transverse momenta. 
However, the weak suppression of $\bar{N}_1$ with $p_T$  indicates 
that the clans contain also particles with relatively high $p_T$.
The above-mentioned properties suggest 
that there are semi-hard processes with a small number of mini-jets followed by
intensive branching and successive production of many low-$p_T$ particles in 
the dominant class of events.
The second multiplicity component, under the tail of the MD, describes production 
of particles in semi-hard processes with extensive production of mini-jets.
The corresponding NBD consists of many clans containing 
less particles on average than in the first component.  
We consider these clans to represent to a certain extent mini-jets, 
i.e. clusters of directed particles of a common ancestor.
Such assumption is supported by the simultaneous increase of the average number of clans $\bar{N}_2$ 
and the average number of particles per clan $\bar{n}_{c2}$ with energy.

\subsection{Summary} 
\label{sec:H}

We have analysed charged-particle MDs  
using new high-statistics data \cite{ATLAS2,ATLAS3,ATLAS4} measured by the ATLAS Collaboration 
in $pp$ collisions at energies $\sqrt s=$8 and 13~TeV. 
The data include measurements in the low-$p_T$ regime ($p_T>100$~MeV/c) and with the 
cut $p_T>500$~MeV/c. 
The analysis extends our previous study \cite{IZ} of the MDs 
obtained at the LHC at lower energies.
The results of our investigations show that the new ATLAS data confirm the existence of a
distinct peak in MD at low multiplicities up to the energy
$\sqrt s=$13~TeV.
The description of the data within a two-component superposition of NBDs is unsatisfactory.
The ATLAS data can be well parametrized by a weighted sum of three NBDs.
The probabilities of the single NBD components show a weak or nearly no energy dependence 
in the most inclusive phase-space region ($p_T>100$~MeV/c).
The third NBD component accounts for the peak-like structure of the experimental MD at low $n$.
Its average multiplicity $\bar{n}_3$ is approximately energy independent and decreases
with the imposed transverse momentum cut.
The parameters $k^{-1}_i$, characterizing widths of the single NBDs, 
demonstrate a clear hierarchy with the index i for the $p_T >$~100~MeV/c data sample.
The third component of the total distribution is 
narrow, well described by the NBD with a large value of $k_3$.
The probability $\alpha_3$ of the low-multiplicity component does not diminish 
when a cut on the transverse momentum is imposed.

We have studied the properties of the cumulative combinants $\langle N\rangle C_i$ calculated  
from a weighted superposition of three- and two-NBD parametrization of the ATLAS and CMS 
data on MDs.
It was demonstrated that the third NBD component at low multiplicities with sufficiently 
large $\bar{n}_3$ can be responsible for the oscillating structure of the combinants.  
The oscillations are clearly visible when they are calculated from the three-NBD fits to 
the ATLAS data measured 
in the pseudorapidity interval $ |\eta|<$2.5 with $p_T >$ 100~MeV/c.
It was demonstrated that the oscillatory behaviour of $\langle N\rangle C_i$ obtained 
from the three-NBD fits \cite{IZ} to the CMS data on MDs
at $\sqrt s=$7~TeV in different pseudorapidity windows reveal similar properties as those found in \cite{WW1}.
They are the decrease of the magnitude and periodicity of the oscillations
with the decreasing window size. The position of the first minimum 
of the oscillations corresponds to the average multiplicity $\bar{n}_3$ of the third NBD component.
The result suggests that 
the average correlation of the system is weaker if just this multiplicity contributes 
to the build-up of the probability $P(n)$ for a higher multiplicity $n$.  
To the contrary, the cumulative combinants $\langle N\rangle C_i$ calculated from a
weighted superposition of two NBDs used to fit the data give no oscillations and
reveal a monotonic decrease only.

We conclude that a detailed parametrization of the LHC data on MD which would give oscillations 
of the cumulative combinants $\langle N\rangle C_i=(i\!+\!1){\cal C}(i\!+\!1)$
requires third component within weighted sum of NBDs.  
This is analogous to the oscillations of the ratio $H_q=K_q/F_q$ 
of the cumulant to the factorial moments which were described by two-NBD fits 
used to account for the shoulder of MDs observed already at lower energies.
The oscillating structures corresponding
to the low- and high-multiplicity regions
cannot be obtained from the two- and one-NBD fits to the experimental data, respectively.

A modification of cumulative combinants based on a truncation of the MD in the first two multiplicity
bins was examined. The modified combinants, $(i\!+\!1){\cal C}_2(i\!+\!1)$ and 
$(i\!+\!1){\cal C}_1(i\!+\!1)$, were obtained from experimental data on MD using 
Eqs. (\ref{eq:r8})  and (\ref{eq:r9}), respectively.
It was illustrated that they reveal similar
properties as the cumulative combinants
calculated by Eq. (\ref{eq:r4}) from a superposition of three NBDs with
the parameters obtained from the corresponding fits.
The results of the analysis of the ATLAS data measured 
in the interval $|\eta| < $2.5 with the transverse
momentum cut $p_T > $100 MeV/c at $\sqrt{s} = $ 13, 8, 7 and 0.9 TeV show that
the quantity $(i\!+\!1){\cal C}_2(i\!+\!1)$ reveals an oscillating structure in the 
region of low $i$.
The wavy structure is clearly visible
because of relatively small systematic uncertainties of the ATLAS measurements.
The oscillation is a consequence of  
the distinct peak at the maximum of the MD. 
The oscillating wave at low $i$ allows to discriminate
between the two- and three-NBD parametrization of the ATLAS data. 

We have computed the quantities $(i\!+\!1){\cal C}_1(i\!+\!1)$ and 
$(i\!+\!1){\cal C}_2(i\!+\!1)$ using data on MD measured by the CMS Collaboration in the 
pseudorapidity interval $|\eta|<$ 2.4 at $\sqrt{s}$=7 TeV. 
The results of the analysis show that 
the CMS data is inconclusive as regards the discussed oscillations
(see Figs. \ref{fig:19}a and \ref{fig:21}a).

We have analyzed the clan structure of the MD of charged particles produced in $pp$ collisions.
The clan parameters were studied in the framework of the 
weighted superposition of three NBDs used to describe 
the ATLAS data on MD in the interval $ |\eta|<$2.5 and $p_T >$ 100~MeV/c.
The examination of the experimental data shows considerable differences  
in the behavior of the average number of clans $\bar{N}_{i}$ and
the average number of particles per clan $\bar{n}_{ci}$ 
in comparison with two-component parametrization of the same data.
The differences concerning the average values of the parameters are summarized as follows.

The three-NBD model gives:
\begin{itemize}
\item[1.]
increasing number of particles per clan $\bar{n}_{c1}$ with $\sqrt s$ 
in the first, dominant component,
\item[2.]
increasing number of clans $\bar{N}_{2}$ with $\sqrt s$ 
in the semi-hard component with mini-jets,
\item[3.]
more particles per clan in the first component 
than in the semi-hard one with mini-jets ($\bar{n}_{c1}>\bar{n}_{c2}$).
\end{itemize}

The two-NBD model gives:
\begin{itemize}
\item[1.]
constant number of particles per clan $\bar{n}_{c1}$ with $\sqrt s$ in the soft component,
\item[2.]
decreasing number of clans $\bar{N}_{2}$ with $\sqrt s$ in the semi-hard component with mini-jets,
\item[3.]
fewer particles per clan in the soft component 
than in the semi-hard one ($\bar{n}_{c1}<\bar{n}_{c2}$).
\end{itemize}
It was shown that the above properties of clans are consistent with the results of the analysis performed
with data on MD measured by the CMS Collaboration in different pseudorapidity intervals at $\sqrt s=$ 7 TeV.
As concerns the third multiplicity component, 
the three-NBD model gives approximately constant value of the average number of clans $\bar{N}_{3}$ 
with energy.  
These clans contain one or very few particles on average.

We have studied the influence of the transverse momentum cut on the clan parameters using the 
ATLAS data with $p_T >$ 500~MeV/c.
The average number of clans and the average number of particles per clan are reduced 
in all three multiplicity components with the $p_T $ cut.
It was shown that the reduction reveals a different behavior from the two-NBD model.
The size of the changes 
suggests that there are semi-hard processes with simultaneous 
production of many low-$p_T$ particles in the first, dominant class of events.
The values of the clans' parameters point to intensive branching and 
a small number of clans mini-jets in these events.
On the other hand, the semi-hard component under the tail of the MD reveals 
the existence of many particle clusters 
with properties expected for mini-jets. 
The transverse momenta of particles in these clusters are on average higher than for
the most particles belonging to the clans of the first component.
The suppression of the clan parameters with the cut $p_T>500$ MeV/c  indicates that 
the transverse  momenta are concentrated around some $\tilde{p}_T$. 
A novel feature is growth of the average number of the clans, $\bar{N}_{2}$, with $\sqrt{s}$. 
As concerns the third multiplicity component, the present study shows that 
it consists mostly of particles with low transverse momenta. 
The biggest suppression of 
$\bar{n}_{3}$ with $p_T $
suggests that there is a soft mechanism responsible for 
particle production under the peak of the MD 
at low $n$.

Based on the performed analysis we argue that there is 
strong evidence of a new component in the MD 
of charged particles measured by the ATLAS Collaboration in 
$pp$ collisions up to the energy $\sqrt s=$13~TeV.
The data are well described by a superposition of three NBD functions in a wide range of $\sqrt s$.  
The third component at low $n$ gives oscillations of the cumulative combinants 
$\langle N\rangle C_i=(i\!+\!1){\cal C}(i\!+\!1)$
observed for the first time in~\cite{WW1}.
The analysis of the clan properties applied to three-NBD fits to the ATLAS data
on multiplicities reveals new features in the LHC energy domain. 
Further study of the clan characteristics and the fine structure of the MD with future high quality data 
is a challenging problem, which can give important information on 
production mechanism and correlation structures  
of the multi-particle states created in high energy proton-proton collisions.

\begin{acknowledgements}
The investigations were supported by the  institutional support 
RVO61389005 and 
by the grant LG 15052 of the Ministry of Education of the Czech Republic.   
\end{acknowledgements}

\newpage
\setcounter{equation}{0}
\begin{Aeqno}

\appendix 
\section*{Appendix A}

Here we summarize the details of the least-squares criteria and 
the conditions used in the minimization procedure.
The value of $\chi^2$ was computed as  
\begin{equation}
\chi^2=\sum\limits_{m=1}^{M}
\frac{(P^{ex}_m-P^{ph}_m)^2}{\sigma^2_m}
\label{eq:a1}
\end{equation} 
where $M$ is the number of bins considered in the experiment.
The bins are defined using the information accessible in the following form:
the probability $P^{ex}_m$ in the point $n_m$ and the bin interval $(a_m,b_m)$. 
In general, the values of $n_m$ are not given as integers.
In the case when the $m$th bin consists of a single multiplicity $n$, 
$n_m=n$ and $P^{ph}_m = P(n)$ with $P(n)$ given by (\ref{eq:r2}).
For the bins containing more multiplicities, such as in the tails of the distributions,  
$n_m\neq n$ and we define 
\begin{equation}
P^{ph}_m= (1-\Delta)P([n_m]) + \Delta P([n_m]+1)
\label{eq:a2}
\end{equation} 
with $\Delta\equiv n_m-[n_m]$.
All values of  $P^{ph}_m$ are then renormalized to fulfill the condition
\begin{equation}
\sum\limits_{m=1}^{M} P^{ph}_m =  
\sum\limits_{m=1}^{M} P^{ex}_m .
\label{eq:a3}
\end{equation}  
This ensures the same normalization of the phenomenological and experimental distributions over the fitted 
range of bins.  
The value of $\chi^2$ is computed according to (\ref{eq:a1}) from the normalized distributions in each iteration step.
The errors $ \sigma_m $ are calculated as quadratic sums of the statistical and systematic 
uncertainties. 
For the asymmetric systematic errors we used 
$\sigma_m^{+}$ if $ P^{ph}_m>P^{ex}_m$ and $\sigma_m^{-}$ if $ P^{ph}_m<P^{ex}_m$.
The results of the analysis with the CERN-MINUIT program are given 
in Tables~\ref{tab:1} - \ref{tab:4}. 
The obtained values of $\chi^2/dof$ are quoted in the separate rows. 
The number of degrees of freedom, $dof$, 
is stated explicitly 
as the number of experimental points minus the number of free parameters.

In order to estimate how covariations in the data on MD can affect the results obtained
by minimization of the expression (\ref{eq:a1}), we have performed calculations using
the generalized formula
\begin{equation}
\widetilde{\chi}^2=\sum\limits_{n,m=1}^{M}
(P^{ex}_n\!-\!P^{ph}_n)V_{n,m}(P^{ex}_m\!-\!P^{ph}_m) .
\label{eq:a4}
\end{equation} 
The weight matrix $V$ is the inverse of the covariance error matrix $\Sigma=V^{-1}$.
The covariance matrix of the charged-particle MD takes into account correlations between 
multiplicities as well correlations introduced by corrections applied on the data.

The published MDs are obtained from data which suffer from low track 
reconstruction efficiencies $\varepsilon_{trk}$, resolution effects etc.,
so that experimental errors in the resulting MDs are strongly correlated
across adjacent multiplicity bins. 
The track efficiency uncertainty is dominant one for the $n_{ch}$ distributions.
It results in positive correlations between the adjacent multiplicity bins and 
in anti-correlation between the low- and high-multiplicity tails.
The variations of $\varepsilon_{trk}$ induce shifts of the maximum of the MD which, together with 
a proper normalization, increases the right tail of the distribution with a simultaneous decrease
of the left tail and vice versa.
As the leading uncertainty is due to the track efficiency systematics,
the shifts correspond to the up/down systematic uncertainties of $P_n$.
Because systematic errors largely exceed the statistical ones, the effect of
covariations between the latter is likely to
be negligible compared to the systematic covariations.
Taking all this into account, 
given the asymmetry of the systematic errors, we write
the covariance matrix in the following way.
Its diagonal part is constructed in the form  
\begin{equation}
\Sigma=
\begin{array}{lc}
\frac{1}{2}(\sigma^+_i\!\sigma^+_j\!+\!\sigma^-_i\!\sigma^-_j)g_ig_j
\!+\!\sigma^{st}_i \sigma^{st}_j 
\ \ \ \ \ \ \ i\!=\!j \ . \\
\end{array}
\label{eq:a5}
\end{equation} 
The off diagonal part ($i\!\ne\!j$) is considered as follows 
\begin{equation}
\Sigma\!=\!
\left\{ \begin{array}{lc}
\ \ \frac{1}{2}(\sigma^+_i\!\sigma^+_j\!+\!\sigma^-_i\!\sigma^-_j)\rho_{i,j},
\ \ (i\!-\!\!m)(j\!-\!\!m)\!\ge\!0\! \\ \\
\!-\!\frac{1}{2}(\sigma^+_i\!\sigma^-_j\!+\!\sigma^-_i\!\sigma^+_j)\rho_{i,j},
\ \ (i\!-\!\!m)(j\!-\!\!m)\!<\!0\! \ .\\
\end{array}
\right . 
\label{eq:a6}
\end{equation} 
The symbols $\sigma^{st}_i$ and $\sigma^+_i$, $\sigma^-_i>0$ mean the statistical and the
systematic (up and down) errors in the $i$th bin, respectively. The value of $m$ denotes multiplicity, 
which corresponds to the maximum of the measured distribution. The correlation factors 
$\rho_{i,j}=\rho_{c}g_ig_j$ are governed by the overall correlation coefficient $\rho_{c}$.
As the track efficiency uncertainty is dominant one for the $n_{ch}$ distributions 
and because it results in positive correlations between the adjacent multiplicities,
we consider $\rho_{c}>0$. 
Due to the alternating signs of $\Sigma$ in different multiplicity quadrants defined in  (\ref{eq:a6}),
we introduce the functions 
\begin{equation}
g_i = 1-\textup{exp}\left[-\frac{(i\!-\!m)^2}{\Delta^2}
\right],
\label{eq:a7}
\end{equation}
which model the necessary interpolation of 
the shifted distributions in the vicinity of the point $m$.
We used $\Delta=1$ so that $g_i$ has a perceptible impact on the nearest neighbouring 
bins next to this point only. 
The definition of the matrix $\Sigma$ satisfies the two requirements:
positive covariations between adjacent multiplicity bins, 
and long-range negative covariations 
between bins on the opposite sides of the distribution maximum.

The considered structure of the covariance matrix leads to a certain limitation  
on the parameter $\rho_{c}$. 
The restriction follows from 
the requirement of the positive definiteness of the matrix~$\Sigma$ and 
from data on bin-to-bin errors in single measurements. 
The extra strong correlations ($\rho_{c}\rightarrow 1$) do not meet 
the criteria for valid correlation matrices \cite{CoV}. 
Our calculations were performed 
with $80\%$ correlations ($\rho_{c}=0.8$) for the ATLAS data  
with $p_T>$ 100 MeV/c and for the CMS data at $\sqrt s=$7~TeV.
We have used a smaller value of $\rho_{c}=0.6$
for the ATLAS data with the cut $p_T>$ 500 MeV/c. 
This is motivated by an increase of the 
track reconstruction efficiency parameter $\varepsilon_{trk}\simeq 0.75$ \cite{ATLAS3}
for the $p_T$-cut data, which results in effective decrease of correlations between 
multiplicity bins.
The change of $\rho_{c}$ affects the uncertainties of the obtained parameters but their values
remain within the specified errors. 
We have checked the positive definiteness 
of the corresponding error matrices $\Sigma$
in all considered cases.
The error matrices were inverted
to obtain the weight matrices $V$.

The parameters of the weighted superposition of the three NBDs  obtained   
by  minimization of Eq. (\ref{eq:a4}) 
are shown in Tables~\ref{tab:5} and~\ref{tab:6}.
The shaded rectangles depicted in the respective figures represent the parameter values 
within the stated errors.
The CMS data measured in the intervals $|\eta|<\eta_c=$ 0.5, 1.0 and 1.5 were fitted 
with the fixed value of $k_3=\infty$.
Except for some results of the fit to the ATLAS data with $p_T>500$~MeV/c at $\sqrt{s}=7$~TeV,  
one can see reasonable agreement between the parameters 
obtained  by  Eq. (\ref{eq:a1}) (full symbols with error bars) and by Eq. (\ref{eq:a4}) 
(shaded rectangles) for all fittings when using the superposition of three NBDs. 
The fits with the three-component function (\ref{eq:r2}) seem to be  
rather stable with respect to covariations of the systematic uncertainties reflected 
by Eqs. (\ref{eq:a5}) and (\ref{eq:a6})
for the error matrix $\Sigma$.

\begin{table}
\caption{
The parameters of the three-NBD superposition   
extracted from fits to data on MDs \protect\cite{ATLAS1,ATLAS2,ATLAS4} measured by the ATLAS
Collaboration in the window $|\eta|<2.5$ with the cut $p_T>100$~MeV/c, $n>1$.
The parameter values were obtained by minimization of Eq. (\ref{eq:a4})
}
\label{tab:5}       
\begin{tabular*}{\columnwidth}{@{}l@{\extracolsep{\fill}}lll@{}} 
\hline\noalign{\smallskip}
$\sqrt{s}$ \ \ \ \ \ \ \  {\it i} &  $\ \ \ \alpha_i$  & \ \ \ $\bar{n}_i$ & \ \ \ $k_i$  \\             
\noalign{\smallskip}\hline\noalign{\smallskip}

13 TeV \  $\!$ 
1 &  0.772$^{+0.015}_{-0.016}$  &  27.2$^{+0.7  }_{-0.7  }$ &\ \,1.30$^{+0.02 }_{-0.02 }$   \\ \noalign{\smallskip} \ \ \ \ \ \ \ \ \ \ \ \
2 &  0.143$^{+0.014}_{-0.012}$  &  72.7$^{+1.0  }_{-1.1  }$ &\  \,6.0$^{+0.3  }_{-0.3  }$   \\ \noalign{\smallskip} \ \ \ \ \ \ \ \ \ \ \ \
3 &  0.085$^{+0.002}_{-0.003}$  & 11.14$^{+0.04 }_{-0.04 }$ &    18.2$^{+1.6  }_{-1.4  }$   \\ \noalign{\smallskip}
      & \multicolumn{3}{c}{     $\widetilde{\chi}^2/dof$ = 88.0/(86-8)}  \\
\noalign{\smallskip}\hline\noalign{\smallskip}

8 TeV \ \ \ 
1 &  0.762$^{+0.019}_{-0.022}$  &  24.2$^{+0.7  }_{-0.8  }$ &\ \,1.46$^{+0.02 }_{-0.02 }$   \\ \noalign{\smallskip}  \ \ \ \ \ \ \ \ \ \ \ \
2 &  0.149$^{+0.018}_{-0.015}$  &  62.5$^{+0.9  }_{-0.9  }$ &\  \,6.0$^{+0.3  }_{-0.3  }$   \\ \noalign{\smallskip}  \ \ \ \ \ \ \ \ \ \ \ \
3 &  0.089$^{+0.004}_{-0.004}$  &  10.84$^{+0.06}_{-0.06 }$ &    19.4$^{+2.7  }_{-2.2  }$   \\ \noalign{\smallskip}
     & \multicolumn{3}{c}{     $\widetilde{\chi}^2/dof$ = 56.9/(85-8)}  \\
\noalign{\smallskip}\hline\noalign{\smallskip}

7 TeV \ \ \ 
1 &  0.745$^{+0.026}_{-0.031}$  &  21.9$^{+0.9  }_{-1.1  }$ &\ \,1.51$^{+0.04}_{-0.03 }$   \\ \noalign{\smallskip}  \ \ \ \ \ \ \ \ \ \ \ \
2 &  0.179$^{+0.027}_{-0.022}$  &  57.3$^{+1.2  }_{-1.4  }$ &\  \,5.8$^{+0.4 }_{-0.3  }$   \\ \noalign{\smallskip}  \ \ \ \ \ \ \ \ \ \ \ \
3 &  0.076$^{+0.004}_{-0.004}$  &  11.2$^{+0.1  }_{-0.1  }$ &     27.$^{+7.  }_{-5.   }$   \\ \noalign{\smallskip}
     & \multicolumn{3}{c}{     $\widetilde{\chi}^2/dof$ = 33.7/(85-8)}  \\
\noalign{\smallskip}\hline\noalign{\smallskip}

0.9 TeV \ 
1 &  0.52$^{+0.13 }_{-0.15 }$  &  12.1$^{+1.5  }_{-2.6  }$ &\ \,1.63$^{+0.15 }_{-0.19 }$   \\ \noalign{\smallskip}  \ \ \ \ \ \ \ \ \ \ \ \
2 &  0.33$^{+0.11 }_{-0.09 }$  &  26.5$^{+1.6  }_{-1.5  }$ &\  \,4.6$^{+0.6  }_{-0.4  }$   \\ \noalign{\smallskip}  \ \ \ \ \ \ \ \ \ \ \ \
3 &  0.15$^{+0.04 }_{-0.04 }$  &  10.9$^{+0.2  }_{-0.2  }$ &     16.$^{+10.  }_{-5.   }$   \\ \noalign{\smallskip}
     & \multicolumn{3}{c}{     $\widetilde{\chi}^2/dof$ = 19.2/(51-8)}  \\
\noalign{\smallskip}\hline\noalign{\smallskip}

\end{tabular*}
\end{table}

As indicated in Figs. \ref{fig:11}a and \ref{fig:12}a, the performed estimation of 
covariations has great impact on two-NBD parametrization of the ATLAS data. 
The fits with two NBDs to the data with the cut $p_T>$ 100 MeV/c
give large values of 
$\widetilde{\chi}^2/dof$ = 776/81, $\widetilde{\chi}^2/dof$ = 660/80,
$\widetilde{\chi}^2/dof$ = 417/80, $\widetilde{\chi}^2/dof$ = 145/46 
at $\sqrt{s}=$ 13, 8, 7, and 0.9 TeV, respectively.
The two-NBD fits to the data with the cut $p_T>$ 500 MeV/c give
$\widetilde{\chi}^2/dof$ = 1143/76, $\widetilde{\chi}^2/dof$ = 174/34,
$\widetilde{\chi}^2/dof$ = 211/34, $\widetilde{\chi}^2/dof$ =~15/29 
at 
$\sqrt{s}=$ 
13, 8, 7, and 0.9 TeV, respectively. 
Except for the last one, all these numbers are unsatisfactory.
The quoted values of $\widetilde{\chi}^2$  confirm the conclusion that the
two-NBD hypothesis does not reproduce the ATLAS data well.

\begin{table}[t!]
\caption{
The parameters of the three-NBD superposition   
extracted from fits to data on  MDs \protect\cite{ATLAS1,ATLAS2,ATLAS3} measured by the ATLAS
Collaboration in the window $|\eta|<2.5$ with the cut $p_T>500$~MeV/c, $n>0$.
The parameter values were obtained by minimization of Eq. (\ref{eq:a4})
}
\label{tab:6}       
\begin{tabular*}{\columnwidth}{@{}l@{\extracolsep{\fill}}lll@{}} 
\hline\noalign{\smallskip}
$\sqrt{s}$ \ \ \ \ \ \ \ {\it i} &  $\ \ \ \alpha_i$  & \ \ \ $\bar{n}_i$ & \ \ \ $k_i$  \\             
\noalign{\smallskip}\hline\noalign{\smallskip}

13 TeV \ $\!$
1 &  0.714$^{+0.028 }_{-0.034 }$ & 10.3$^{+0.4  }_{-0.4  }$ &\ \ 1.02$^{+0.06 }_{-0.04 }$   \\ \noalign{\smallskip} \ \ \ \ \ \ \ \ \ \ \ \
2 &  0.170$^{+0.020 }_{-0.016 }$  &  32.5$^{+0.6  }_{-0.6  }$ &\ \ 4.0$^{+0.2 }_{-0.2 }$   \\ \noalign{\smallskip} \ \ \ \ \ \ \ \ \ \ \ \
3 &  0.116$^{+0.014 }_{-0.012 }$  &\, 2.91$^{+0.03 }_{-0.05 }$ &\ \  5.4$^{+1.0  }_{-0.8  }$   \\ \noalign{\smallskip}
      & \multicolumn{3}{c}{     $\widetilde{\chi}^2/dof$ = 84.0/(81-8)}  \\
\noalign{\smallskip}\hline\noalign{\smallskip}

8 TeV \ \ \ 
1 &  0.68$^{+0.09}_{-0.17}$  &  10.2$^{+1.3 }_{-1.3 }$ &\ \ 1.31$^{+0.53}_{-0.14 }$   \\ 
\noalign{\smallskip}  \ \ \ \ \ \ \ \ \ \ \ \
2 &  0.15$^{+0.07}_{-0.05}$  &  29.3$^{+1.9 }_{-2.0 }$ &\ \ 4.6$^{+0.7  }_{-0.5  }$   \\ 
\noalign{\smallskip}  \ \ \ \ \ \ \ \ \ \ \ \
3 &  0.17$^{+0.10}_{-0.04}$  &\, 2.8$^{+0.1 }_{-0.1 }$ &\ \ 4.5$^{+2.1  }_{-1.8  }$   \\ 
\noalign{\smallskip}
     & \multicolumn{3}{c}{     $\widetilde{\chi}^2/dof$ = 13.1/(39-8)}  \\
\noalign{\smallskip}\hline\noalign{\smallskip}

7 TeV \ \ \ 
1 &  0.737$^{+0.039}_{-0.057}$  &\, 8.9$^{+0.6  }_{-0.7  }$ &\ \, 1.09$^{+0.05 }_{-0.03 }$   \\ \noalign{\smallskip}  \ \ \ \ \ \ \ \ \ \ \ \
2 &  0.155$^{+0.034}_{-0.024}$  &  26.6$^{+0.7  }_{-0.9  }$ &\ \, 4.5$^{+0.3 }_{-0.3 }$   \\ \noalign{\smallskip}  \ \ \ \ \ \ \ \ \ \ \ \
3 &  0.108$^{+0.023}_{-0.015}$  &\, 3.0$^{+0.1  }_{-0.1  }$ &\ \, 7.1$^{+3.5  }_{-2.2  }$   \\ \noalign{\smallskip}
     & \multicolumn{3}{c}{     $\widetilde{\chi}^2/dof$ = 23.0/(39-8)}  \\
\noalign{\smallskip}\hline\noalign{\smallskip}

0.9 TeV $\!$
1 &  0.83$^{+0.05 }_{-0.09 }$  &\, 5.0$^{+0.3  }_{-0.6  }$ &\ \, 1.37$^{+0.12 }_{-0.06 }$   \\ \noalign{\smallskip}  \ \ \ \ \ \ \ \ \ \ \ \
2 &  0.11$^{+0.08 }_{-0.04 }$  &  13.9$^{+1.0  }_{-1.2  }$ &\ \, 5.9$^{+1.7  }_{-1.1  }$   \\ \noalign{\smallskip}  \ \ \ \ \ \ \ \ \ \ \ \
3 &  0.06$^{+0.01 }_{-0.01 }$  &\, 3.1$^{+0.2 }_{-0.2  }$ &\  \, $\infty$   \\ 
\noalign{\smallskip}
     & \multicolumn{3}{c}{     $\widetilde{\chi}^2/dof$ = 9.0/(34-7)}  \\
\noalign{\smallskip}\hline\noalign{\smallskip}

\end{tabular*}
\end{table}

The CMS measurement at the energy $\sqrt{s}=$ 7 TeV is less critical to the mentioned hypothesis.
Using Eq. (\ref{eq:a4}), two-NBD fits to the data 
give small values of    
$\widetilde{\chi}^2/dof$ = 9/35,  $\widetilde{\chi}^2/dof$ = 8/64,
$\widetilde{\chi}^2/dof$ = 28/89, $\widetilde{\chi}^2/dof$ = 53/109, 
$\widetilde{\chi}^2/dof$ = 44/121 
in the intervals $|\eta|<0.5$, $|\eta|<1.0$, $|\eta|<1.5$, $|\eta|<2.0$, $|\eta|<2.4$,
respectively.
This is due to the much larger systematic uncertainties as compared with the ATLAS data.
However, in regard to the residual analysis \cite{IZ} with respect to a parametrization 
of the CMS data by two NBDs 
and in the light of a similar analysis of the ATLAS data over a wide range of $\sqrt{s}$= 0.9 - 13 TeV, 
one has to look at these numbers relatively.
The systematic emergence of a peak in the residues, visible in the larger 
pseudorapidity windows, can be described by a superposition of three NBDs, 
yielding 
$\widetilde{\chi}^2/dof$ = 7/35,  $\widetilde{\chi}^2/dof$ = 3/64,
$\widetilde{\chi}^2/dof$ = 7/89, $\widetilde{\chi}^2/dof$ = 10/109, 
$\widetilde{\chi}^2/dof$ = 7/121
in the intervals $|\eta|<0.5$, $|\eta|<1.0$, $|\eta|<1.5$, $|\eta|<2.0$, $|\eta|<2.4$,
respectively.

\end{Aeqno}

\end{document}